\documentclass[twocolumn,traditabstract,longauth]{aa}

\usepackage{graphicx}
\usepackage{epsf}

\usepackage[breaklinks, colorlinks, citecolor=blue]{hyperref}  
\usepackage{natbib}
\bibpunct{(}{)}{;}{a}{}{,} 

\def\setsymbol#1#2{\expandafter\def\csname #1\endcsname{#2}}
\def\getsymbol#1{\csname #1\endcsname}

\def\Planck{\textit{Planck}}

\def\all2013resultspapers{\nocite{planck2013-p01, planck2013-p02, planck2013-p02a, planck2013-p02d, planck2013-p02b, planck2013-p03, planck2013-p03c, planck2013-p03f, planck2013-p03d, planck2013-p03e, planck2013-p06, planck2013-p03a, planck2013-pip88, planck2013-p08, planck2013-p11, planck2013-p12, planck2013-p13, planck2013-p14, planck2013-p15, planck2013-p05b, planck2013-p17, planck2013-p09, planck2013-p09a, planck2013-p20, planck2013-p19, planck2013-pipaberration, planck2013-p05, planck2013-p05a, planck2013-pip56, planck2013-p06b, planck2013-p01a}}

\newbox\tablebox    \newdimen\tablewidth
\def\leaderfil{\leaders\hbox to 5pt{\hss.\hss}\hfil}
\def\endPlancktable{\tablewidth=\columnwidth 
    $$\hss\copy\tablebox\hss$$
    \vskip-\lastskip\vskip -2pt}
\def\endPlancktablewide{\tablewidth=\textwidth 
    $$\hss\copy\tablebox\hss$$
    \vskip-\lastskip\vskip -2pt}
\def\tablenote#1 #2\par{\begingroup \parindent=0.8em
    \abovedisplayshortskip=0pt\belowdisplayshortskip=0pt
    \noindent
    $$\hss\vbox{\hsize\tablewidth \hangindent=\parindent \hangafter=1 \noindent
    \hbox to \parindent{$^#1$\hss}\strut#2\strut\par}\hss$$
    \endgroup}
\def\doubleline{\vskip 3pt\hrule \vskip 1.5pt \hrule \vskip 5pt}

\def\L2{\ifmmode L_2\else $L_2$\fi}

\def\DeltaT{\ifmmode \Delta T\else $\Delta T$\fi}
\def\deltat{\ifmmode \Delta t\else $\Delta t$\fi}
\def\fknee{\ifmmode f_{\rm knee}\else $f_{\rm knee}$\fi}
\def\Fmax{\ifmmode F_{\rm max}\else $F_{\rm max}$\fi}
\def\solar{\ifmmode{\rm M}_{\mathord\odot}\else${\rm M}_{\mathord\odot}$\fi}
\def\Msolar{\ifmmode{\rm M}_{\mathord\odot}\else${\rm M}_{\mathord\odot}$\fi}
\def\Lsolar{\ifmmode{\rm L}_{\mathord\odot}\else${\rm L}_{\mathord\odot}$\fi}

\def\inv{\ifmmode^{-1}\else$^{-1}$\fi}
\def\mo{\ifmmode^{-1}\else$^{-1}$\fi}
\def\sup#1{\ifmmode ^{\rm #1}\else $^{\rm #1}$\fi}
\def\expo#1{\ifmmode \times 10^{#1}\else $\times 10^{#1}$\fi}
\def\,{\thinspace}
\def\lsim{\mathrel{\raise .4ex\hbox{\rlap{$<$}\lower 1.2ex\hbox{$\sim$}}}}
\def\gsim{\mathrel{\raise .4ex\hbox{\rlap{$>$}\lower 1.2ex\hbox{$\sim$}}}}

\def\simprop{\mathrel{\raise .4ex\hbox{\rlap{$\propto$}\lower 1.2ex\hbox{$\sim$}}}}
\def\deg{\ifmmode^\circ\else$^\circ$\fi}
\def\pdeg{\ifmmode $\setbox0=\hbox{$^{\circ}$}\rlap{\hskip.11\wd0 .}$^{\circ}
          \else \setbox0=\hbox{$^{\circ}$}\rlap{\hskip.11\wd0 .}$^{\circ}$\fi}
\def\arcs{\ifmmode {^{\scriptstyle\prime\prime}}
          \else $^{\scriptstyle\prime\prime}$\fi}
\def\arcm{\ifmmode {^{\scriptstyle\prime}}
          \else $^{\scriptstyle\prime}$\fi}
\newdimen\sa  \newdimen\sb
\def\parcs{\sa=.07em \sb=.03em
     \ifmmode \hbox{\rlap{.}}^{\scriptstyle\prime\kern -\sb\prime}\hbox{\kern -\sa}
     \else \rlap{.}$^{\scriptstyle\prime\kern -\sb\prime}$\kern -\sa\fi}
\def\parcm{\sa=.08em \sb=.03em
     \ifmmode \hbox{\rlap{.}\kern\sa}^{\scriptstyle\prime}\hbox{\kern-\sb}
     \else \rlap{.}\kern\sa$^{\scriptstyle\prime}$\kern-\sb\fi}
\def\ra[#1 #2 #3.#4]{#1\sup{h}#2\sup{m}#3\sup{s}\llap.#4}
\def\dec[#1 #2 #3.#4]{#1\deg#2\arcm#3\arcs\llap.#4}
\def\deco[#1 #2 #3]{#1\deg#2\arcm#3\arcs}
\def\rra[#1 #2]{#1\sup{h}#2\sup{m}}

\def\dots{\relax\ifmmode \ldots\else $\ldots$\fi}
\def\WHzsr{\ifmmode $W\,Hz\mo\,sr\mo$\else W\,Hz\mo\,sr\mo\fi}
\def\mHz{\ifmmode $\,mHz$\else \,mHz\fi}
\def\GHz{\ifmmode $\,GHz$\else \,GHz\fi}
\def\mKs{\ifmmode $\,mK\,s$^{1/2}\else \,mK\,s$^{1/2}$\fi}
\def\muKs{\ifmmode \,\mu$K\,s$^{1/2}\else \,$\mu$K\,s$^{1/2}$\fi}
\def\muKRJs{\ifmmode \,\mu$K$_{\rm RJ}$\,s$^{1/2}\else \,$\mu$K$_{\rm RJ}$\,s$^{1/2}$\fi}
\def\muKHz{\ifmmode \,\mu$K\,Hz$^{-1/2}\else \,$\mu$K\,Hz$^{-1/2}$\fi}
\def\MJysr{\ifmmode \,$MJy\,sr\mo$\else \,MJy\,sr\mo\fi}
\def\MJysrmK{\ifmmode \,$MJy\,sr\mo$\,mK$_{\rm CMB}\mo\else \,MJy\,sr\mo\,mK$_{\rm CMB}\mo$\fi}
\def\microns{\ifmmode \,\mu$m$\else \,$\mu$m\fi}
\def\micron{\microns}
\def\muK{\ifmmode \,\mu$K$\else \,$\mu$\hbox{K}\fi}
\def\microK{\ifmmode \,\mu$K$\else \,$\mu$\hbox{K}\fi}
\def\muW{\ifmmode \,\mu$W$\else \,$\mu$\hbox{W}\fi}
\def\kms{\ifmmode $\,km\,s$^{-1}\else \,km\,s$^{-1}$\fi}
\def\kmsMpc{\ifmmode $\,\kms\,Mpc\mo$\else \,\kms\,Mpc\mo\fi}
\setsymbol{LFI:center:frequency:70GHz:units}{70.3\,GHz}
\setsymbol{LFI:center:frequency:44GHz:units}{44.1\,GHz}
\setsymbol{LFI:center:frequency:30GHz:units}{28.5\,GHz}

\setsymbol{LFI:center:frequency:70GHz}{70.3}
\setsymbol{LFI:center:frequency:44GHz}{44.1}
\setsymbol{LFI:center:frequency:30GHz}{28.5}

\setsymbol{LFI:center:frequency:LFI18:Rad:M:units}{71.7\GHz}
\setsymbol{LFI:center:frequency:LFI19:Rad:M:units}{67.5\GHz}
\setsymbol{LFI:center:frequency:LFI20:Rad:M:units}{69.2\GHz}
\setsymbol{LFI:center:frequency:LFI21:Rad:M:units}{70.4\GHz}
\setsymbol{LFI:center:frequency:LFI22:Rad:M:units}{71.5\GHz}
\setsymbol{LFI:center:frequency:LFI23:Rad:M:units}{70.8\GHz}
\setsymbol{LFI:center:frequency:LFI24:Rad:M:units}{44.4\GHz}
\setsymbol{LFI:center:frequency:LFI25:Rad:M:units}{44.0\GHz}
\setsymbol{LFI:center:frequency:LFI26:Rad:M:units}{43.9\GHz}
\setsymbol{LFI:center:frequency:LFI27:Rad:M:units}{28.3\GHz}
\setsymbol{LFI:center:frequency:LFI28:Rad:M:units}{28.8\GHz}
\setsymbol{LFI:center:frequency:LFI18:Rad:S:units}{70.1\GHz}
\setsymbol{LFI:center:frequency:LFI19:Rad:S:units}{69.6\GHz}
\setsymbol{LFI:center:frequency:LFI20:Rad:S:units}{69.5\GHz}
\setsymbol{LFI:center:frequency:LFI21:Rad:S:units}{69.5\GHz}
\setsymbol{LFI:center:frequency:LFI22:Rad:S:units}{72.8\GHz}
\setsymbol{LFI:center:frequency:LFI23:Rad:S:units}{71.3\GHz}
\setsymbol{LFI:center:frequency:LFI24:Rad:S:units}{44.1\GHz}
\setsymbol{LFI:center:frequency:LFI25:Rad:S:units}{44.1\GHz}
\setsymbol{LFI:center:frequency:LFI26:Rad:S:units}{44.1\GHz}
\setsymbol{LFI:center:frequency:LFI27:Rad:S:units}{28.5\GHz}
\setsymbol{LFI:center:frequency:LFI28:Rad:S:units}{28.2\GHz}

\setsymbol{LFI:center:frequency:LFI18:Rad:M}{71.7}
\setsymbol{LFI:center:frequency:LFI19:Rad:M}{67.5}
\setsymbol{LFI:center:frequency:LFI20:Rad:M}{69.2}
\setsymbol{LFI:center:frequency:LFI21:Rad:M}{70.4}
\setsymbol{LFI:center:frequency:LFI22:Rad:M}{71.5}
\setsymbol{LFI:center:frequency:LFI23:Rad:M}{70.8}
\setsymbol{LFI:center:frequency:LFI24:Rad:M}{44.4}
\setsymbol{LFI:center:frequency:LFI25:Rad:M}{44.0}
\setsymbol{LFI:center:frequency:LFI26:Rad:M}{43.9}
\setsymbol{LFI:center:frequency:LFI27:Rad:M}{28.3}
\setsymbol{LFI:center:frequency:LFI28:Rad:M}{28.8}
\setsymbol{LFI:center:frequency:LFI18:Rad:S}{70.1}
\setsymbol{LFI:center:frequency:LFI19:Rad:S}{69.6}
\setsymbol{LFI:center:frequency:LFI20:Rad:S}{69.5}
\setsymbol{LFI:center:frequency:LFI21:Rad:S}{69.5}
\setsymbol{LFI:center:frequency:LFI22:Rad:S}{72.8}
\setsymbol{LFI:center:frequency:LFI23:Rad:S}{71.3}
\setsymbol{LFI:center:frequency:LFI24:Rad:S}{44.1}
\setsymbol{LFI:center:frequency:LFI25:Rad:S}{44.1}
\setsymbol{LFI:center:frequency:LFI26:Rad:S}{44.1}
\setsymbol{LFI:center:frequency:LFI27:Rad:S}{28.5}
\setsymbol{LFI:center:frequency:LFI28:Rad:S}{28.2}

\setsymbol{LFI:white:noise:sensitivity:70GHz:units}{134.7\muKs}
\setsymbol{LFI:white:noise:sensitivity:44GHz:units}{164.7\muKs}
\setsymbol{LFI:white:noise:sensitivity:30GHz:units}{143.4\muKs}

\setsymbol{LFI:white:noise:sensitivity:70GHz}{134.7}
\setsymbol{LFI:white:noise:sensitivity:44GHz}{164.7}
\setsymbol{LFI:white:noise:sensitivity:30GHz}{143.4}

\setsymbol{LFI:white:noise:sensitivity:LFI18:Rad:M:units}{512.0\muKs}
\setsymbol{LFI:white:noise:sensitivity:LFI19:Rad:M:units}{581.4\muKs}
\setsymbol{LFI:white:noise:sensitivity:LFI20:Rad:M:units}{590.8\muKs}
\setsymbol{LFI:white:noise:sensitivity:LFI21:Rad:M:units}{455.2\muKs}
\setsymbol{LFI:white:noise:sensitivity:LFI22:Rad:M:units}{492.0\muKs}
\setsymbol{LFI:white:noise:sensitivity:LFI23:Rad:M:units}{507.7\muKs}
\setsymbol{LFI:white:noise:sensitivity:LFI24:Rad:M:units}{462.2\muKs}
\setsymbol{LFI:white:noise:sensitivity:LFI25:Rad:M:units}{413.6\muKs}
\setsymbol{LFI:white:noise:sensitivity:LFI26:Rad:M:units}{478.6\muKs}
\setsymbol{LFI:white:noise:sensitivity:LFI27:Rad:M:units}{277.7\muKs}
\setsymbol{LFI:white:noise:sensitivity:LFI28:Rad:M:units}{312.3\muKs}
\setsymbol{LFI:white:noise:sensitivity:LFI18:Rad:S:units}{465.7\muKs}
\setsymbol{LFI:white:noise:sensitivity:LFI19:Rad:S:units}{555.6\muKs}
\setsymbol{LFI:white:noise:sensitivity:LFI20:Rad:S:units}{623.2\muKs}
\setsymbol{LFI:white:noise:sensitivity:LFI21:Rad:S:units}{564.1\muKs}
\setsymbol{LFI:white:noise:sensitivity:LFI22:Rad:S:units}{534.4\muKs}
\setsymbol{LFI:white:noise:sensitivity:LFI23:Rad:S:units}{542.4\muKs}
\setsymbol{LFI:white:noise:sensitivity:LFI24:Rad:S:units}{399.2\muKs}
\setsymbol{LFI:white:noise:sensitivity:LFI25:Rad:S:units}{392.6\muKs}
\setsymbol{LFI:white:noise:sensitivity:LFI26:Rad:S:units}{418.6\muKs}
\setsymbol{LFI:white:noise:sensitivity:LFI27:Rad:S:units}{302.9\muKs}
\setsymbol{LFI:white:noise:sensitivity:LFI28:Rad:S:units}{285.3\muKs}

\setsymbol{LFI:white:noise:sensitivity:LFI18:Rad:M}{512.0}
\setsymbol{LFI:white:noise:sensitivity:LFI19:Rad:M}{581.4}
\setsymbol{LFI:white:noise:sensitivity:LFI20:Rad:M}{590.8}
\setsymbol{LFI:white:noise:sensitivity:LFI21:Rad:M}{455.2}
\setsymbol{LFI:white:noise:sensitivity:LFI22:Rad:M}{492.0}
\setsymbol{LFI:white:noise:sensitivity:LFI23:Rad:M}{507.7}
\setsymbol{LFI:white:noise:sensitivity:LFI24:Rad:M}{462.2}
\setsymbol{LFI:white:noise:sensitivity:LFI25:Rad:M}{413.6}
\setsymbol{LFI:white:noise:sensitivity:LFI26:Rad:M}{478.6}
\setsymbol{LFI:white:noise:sensitivity:LFI27:Rad:M}{277.7}
\setsymbol{LFI:white:noise:sensitivity:LFI28:Rad:M}{312.3}
\setsymbol{LFI:white:noise:sensitivity:LFI18:Rad:S}{465.7}
\setsymbol{LFI:white:noise:sensitivity:LFI19:Rad:S}{555.6}
\setsymbol{LFI:white:noise:sensitivity:LFI20:Rad:S}{623.2}
\setsymbol{LFI:white:noise:sensitivity:LFI21:Rad:S}{564.1}
\setsymbol{LFI:white:noise:sensitivity:LFI22:Rad:S}{534.4}
\setsymbol{LFI:white:noise:sensitivity:LFI23:Rad:S}{542.4}
\setsymbol{LFI:white:noise:sensitivity:LFI24:Rad:S}{399.2}
\setsymbol{LFI:white:noise:sensitivity:LFI25:Rad:S}{392.6}
\setsymbol{LFI:white:noise:sensitivity:LFI26:Rad:S}{418.6}
\setsymbol{LFI:white:noise:sensitivity:LFI27:Rad:S}{302.9}
\setsymbol{LFI:white:noise:sensitivity:LFI28:Rad:S}{285.3}

\setsymbol{LFI:knee:frequency:70GHz:units}{29.5\mHz}
\setsymbol{LFI:knee:frequency:44GHz:units}{56.2\mHz}
\setsymbol{LFI:knee:frequency:30GHz:units}{113.7\mHz}

\setsymbol{LFI:knee:frequency:70GHz}{29.5}
\setsymbol{LFI:knee:frequency:44GHz}{56.2}
\setsymbol{LFI:knee:frequency:30GHz}{113.7}

\setsymbol{LFI:knee:frequency:LFI18:Rad:M:units}{16.3\mHz}
\setsymbol{LFI:knee:frequency:LFI19:Rad:M:units}{15.1\mHz}
\setsymbol{LFI:knee:frequency:LFI20:Rad:M:units}{18.7\mHz}
\setsymbol{LFI:knee:frequency:LFI21:Rad:M:units}{37.2\mHz}
\setsymbol{LFI:knee:frequency:LFI22:Rad:M:units}{12.7\mHz}
\setsymbol{LFI:knee:frequency:LFI23:Rad:M:units}{34.6\mHz}
\setsymbol{LFI:knee:frequency:LFI24:Rad:M:units}{46.2\mHz}
\setsymbol{LFI:knee:frequency:LFI25:Rad:M:units}{24.9\mHz}
\setsymbol{LFI:knee:frequency:LFI26:Rad:M:units}{67.6\mHz}
\setsymbol{LFI:knee:frequency:LFI27:Rad:M:units}{187.4\mHz}
\setsymbol{LFI:knee:frequency:LFI28:Rad:M:units}{122.2\mHz}
\setsymbol{LFI:knee:frequency:LFI18:Rad:S:units}{17.7\mHz}
\setsymbol{LFI:knee:frequency:LFI19:Rad:S:units}{22.0\mHz}
\setsymbol{LFI:knee:frequency:LFI20:Rad:S:units}{8.7\mHz}
\setsymbol{LFI:knee:frequency:LFI21:Rad:S:units}{25.9\mHz}
\setsymbol{LFI:knee:frequency:LFI22:Rad:S:units}{15.8\mHz}
\setsymbol{LFI:knee:frequency:LFI23:Rad:S:units}{129.8\mHz}
\setsymbol{LFI:knee:frequency:LFI24:Rad:S:units}{100.9\mHz}
\setsymbol{LFI:knee:frequency:LFI25:Rad:S:units}{38.9\mHz}
\setsymbol{LFI:knee:frequency:LFI26:Rad:S:units}{58.9\mHz}
\setsymbol{LFI:knee:frequency:LFI27:Rad:S:units}{104.4\mHz}
\setsymbol{LFI:knee:frequency:LFI28:Rad:S:units}{40.7\mHz}

\setsymbol{LFI:knee:frequency:LFI18:Rad:M}{16.3}
\setsymbol{LFI:knee:frequency:LFI19:Rad:M}{15.1}
\setsymbol{LFI:knee:frequency:LFI20:Rad:M}{18.7}
\setsymbol{LFI:knee:frequency:LFI21:Rad:M}{37.2}
\setsymbol{LFI:knee:frequency:LFI22:Rad:M}{12.7}
\setsymbol{LFI:knee:frequency:LFI23:Rad:M}{34.6}
\setsymbol{LFI:knee:frequency:LFI24:Rad:M}{46.2}
\setsymbol{LFI:knee:frequency:LFI25:Rad:M}{24.9}
\setsymbol{LFI:knee:frequency:LFI26:Rad:M}{67.6}
\setsymbol{LFI:knee:frequency:LFI27:Rad:M}{187.4}
\setsymbol{LFI:knee:frequency:LFI28:Rad:M}{122.2}
\setsymbol{LFI:knee:frequency:LFI18:Rad:S}{17.7}
\setsymbol{LFI:knee:frequency:LFI19:Rad:S}{22.0}
\setsymbol{LFI:knee:frequency:LFI20:Rad:S}{8.7}
\setsymbol{LFI:knee:frequency:LFI21:Rad:S}{25.9}
\setsymbol{LFI:knee:frequency:LFI22:Rad:S}{15.8}
\setsymbol{LFI:knee:frequency:LFI23:Rad:S}{129.8}
\setsymbol{LFI:knee:frequency:LFI24:Rad:S}{100.9}
\setsymbol{LFI:knee:frequency:LFI25:Rad:S}{38.9}
\setsymbol{LFI:knee:frequency:LFI26:Rad:S}{58.9}
\setsymbol{LFI:knee:frequency:LFI27:Rad:S}{104.4}
\setsymbol{LFI:knee:frequency:LFI28:Rad:S}{40.7}

\setsymbol{LFI:slope:70GHz:units}{$-1.03$\mHz}
\setsymbol{LFI:slope:44GHz:units}{$-0.89$\mHz}
\setsymbol{LFI:slope:30GHz:units}{$-0.87$\mHz}

\setsymbol{LFI:slope:70GHz}{$-1.03$}
\setsymbol{LFI:slope:44GHz}{$-0.89$}
\setsymbol{LFI:slope:30GHz}{$-0.87$}

\setsymbol{LFI:slope:LFI18:Rad:M:units}{$-1.04$\mHz}
\setsymbol{LFI:slope:LFI19:Rad:M:units}{$-1.09$\mHz}
\setsymbol{LFI:slope:LFI20:Rad:M:units}{$-0.69$\mHz}
\setsymbol{LFI:slope:LFI21:Rad:M:units}{$-1.56$\mHz}
\setsymbol{LFI:slope:LFI22:Rad:M:units}{$-1.01$\mHz}
\setsymbol{LFI:slope:LFI23:Rad:M:units}{$-0.96$\mHz}
\setsymbol{LFI:slope:LFI24:Rad:M:units}{$-0.83$\mHz}
\setsymbol{LFI:slope:LFI25:Rad:M:units}{$-0.91$\mHz}
\setsymbol{LFI:slope:LFI26:Rad:M:units}{$-0.95$\mHz}
\setsymbol{LFI:slope:LFI27:Rad:M:units}{$-0.87$\mHz}
\setsymbol{LFI:slope:LFI28:Rad:M:units}{$-0.88$\mHz}
\setsymbol{LFI:slope:LFI18:Rad:S:units}{$-1.15$\mHz}
\setsymbol{LFI:slope:LFI19:Rad:S:units}{$-1.00$\mHz}
\setsymbol{LFI:slope:LFI20:Rad:S:units}{$-0.95$\mHz}
\setsymbol{LFI:slope:LFI21:Rad:S:units}{$-0.92$\mHz}
\setsymbol{LFI:slope:LFI22:Rad:S:units}{$-1.01$\mHz}
\setsymbol{LFI:slope:LFI23:Rad:S:units}{$-0.95$\mHz}
\setsymbol{LFI:slope:LFI24:Rad:S:units}{$-0.73$\mHz}
\setsymbol{LFI:slope:LFI25:Rad:S:units}{$-1.16$\mHz}
\setsymbol{LFI:slope:LFI26:Rad:S:units}{$-0.79$\mHz}
\setsymbol{LFI:slope:LFI27:Rad:S:units}{$-0.82$\mHz}
\setsymbol{LFI:slope:LFI28:Rad:S:units}{$-0.91$\mHz}

\setsymbol{LFI:slope:LFI18:Rad:M}{$-1.04$}
\setsymbol{LFI:slope:LFI19:Rad:M}{$-1.09$}
\setsymbol{LFI:slope:LFI20:Rad:M}{$-0.69$}
\setsymbol{LFI:slope:LFI21:Rad:M}{$-1.56$}
\setsymbol{LFI:slope:LFI22:Rad:M}{$-1.01$}
\setsymbol{LFI:slope:LFI23:Rad:M}{$-0.96$}
\setsymbol{LFI:slope:LFI24:Rad:M}{$-0.83$}
\setsymbol{LFI:slope:LFI25:Rad:M}{$-0.91$}
\setsymbol{LFI:slope:LFI26:Rad:M}{$-0.95$}
\setsymbol{LFI:slope:LFI27:Rad:M}{$-0.87$}
\setsymbol{LFI:slope:LFI28:Rad:M}{$-0.88$}
\setsymbol{LFI:slope:LFI18:Rad:S}{$-1.15$}
\setsymbol{LFI:slope:LFI19:Rad:S}{$-1.00$}
\setsymbol{LFI:slope:LFI20:Rad:S}{$-0.95$}
\setsymbol{LFI:slope:LFI21:Rad:S}{$-0.92$}
\setsymbol{LFI:slope:LFI22:Rad:S}{$-1.01$}
\setsymbol{LFI:slope:LFI23:Rad:S}{$-0.95$}
\setsymbol{LFI:slope:LFI24:Rad:S}{$-0.73$}
\setsymbol{LFI:slope:LFI25:Rad:S}{$-1.16$}
\setsymbol{LFI:slope:LFI26:Rad:S}{$-0.79$}
\setsymbol{LFI:slope:LFI27:Rad:S}{$-0.82$}
\setsymbol{LFI:slope:LFI28:Rad:S}{$-0.91$}

\setsymbol{LFI:FWHM:70GHz:units}{13\parcm01}
\setsymbol{LFI:FWHM:44GHz:units}{27\parcm92}
\setsymbol{LFI:FWHM:30GHz:units}{32\parcm65}

\setsymbol{LFI:FWHM:70GHz}{13.01}
\setsymbol{LFI:FWHM:44GHz}{27.92}
\setsymbol{LFI:FWHM:30GHz}{32.65}

\setsymbol{LFI:FWHM:LFI18:units}{13\parcm39}
\setsymbol{LFI:FWHM:LFI19:units}{13\parcm01}
\setsymbol{LFI:FWHM:LFI20:units}{12\parcm75}
\setsymbol{LFI:FWHM:LFI21:units}{12\parcm74}
\setsymbol{LFI:FWHM:LFI22:units}{12\parcm87}
\setsymbol{LFI:FWHM:LFI23:units}{13\parcm27}
\setsymbol{LFI:FWHM:LFI24:units}{22\parcm98}
\setsymbol{LFI:FWHM:LFI25:units}{30\parcm46}
\setsymbol{LFI:FWHM:LFI26:units}{30\parcm31}
\setsymbol{LFI:FWHM:LFI27:units}{32\parcm65}
\setsymbol{LFI:FWHM:LFI28:units}{32\parcm66}

\setsymbol{LFI:FWHM:LFI18}{13.39}
\setsymbol{LFI:FWHM:LFI19}{13.01}
\setsymbol{LFI:FWHM:LFI20}{12.75}
\setsymbol{LFI:FWHM:LFI21}{12.74}
\setsymbol{LFI:FWHM:LFI22}{12.87}
\setsymbol{LFI:FWHM:LFI23}{13.27}
\setsymbol{LFI:FWHM:LFI24}{22.98}
\setsymbol{LFI:FWHM:LFI25}{30.46}
\setsymbol{LFI:FWHM:LFI26}{30.31}
\setsymbol{LFI:FWHM:LFI27}{32.65}
\setsymbol{LFI:FWHM:LFI28}{32.66}

\setsymbol{LFI:FWHM:uncertainty:LFI18:units}{0.170\arcm}
\setsymbol{LFI:FWHM:uncertainty:LFI19:units}{0.174\arcm}
\setsymbol{LFI:FWHM:uncertainty:LFI20:units}{0.170\arcm}
\setsymbol{LFI:FWHM:uncertainty:LFI21:units}{0.156\arcm}
\setsymbol{LFI:FWHM:uncertainty:LFI22:units}{0.164\arcm}
\setsymbol{LFI:FWHM:uncertainty:LFI23:units}{0.171\arcm}
\setsymbol{LFI:FWHM:uncertainty:LFI24:units}{0.652\arcm}
\setsymbol{LFI:FWHM:uncertainty:LFI25:units}{1.075\arcm}
\setsymbol{LFI:FWHM:uncertainty:LFI26:units}{1.131\arcm}
\setsymbol{LFI:FWHM:uncertainty:LFI27:units}{1.266\arcm}
\setsymbol{LFI:FWHM:uncertainty:LFI28:units}{1.287\arcm}

\setsymbol{LFI:FWHM:uncertainty:LFI18}{0.170}
\setsymbol{LFI:FWHM:uncertainty:LFI19}{0.174}
\setsymbol{LFI:FWHM:uncertainty:LFI20}{0.170}
\setsymbol{LFI:FWHM:uncertainty:LFI21}{0.156}
\setsymbol{LFI:FWHM:uncertainty:LFI22}{0.164}
\setsymbol{LFI:FWHM:uncertainty:LFI23}{0.171}
\setsymbol{LFI:FWHM:uncertainty:LFI24}{0.652}
\setsymbol{LFI:FWHM:uncertainty:LFI25}{1.075}
\setsymbol{LFI:FWHM:uncertainty:LFI26}{1.131}
\setsymbol{LFI:FWHM:uncertainty:LFI27}{1.266}
\setsymbol{LFI:FWHM:uncertainty:LFI28}{1.287}

\setsymbol{HFI:center:frequency:100GHz:units}{100\,GHz}
\setsymbol{HFI:center:frequency:143GHz:units}{143\,GHz}
\setsymbol{HFI:center:frequency:217GHz:units}{217\,GHz}
\setsymbol{HFI:center:frequency:353GHz:units}{353\,GHz}
\setsymbol{HFI:center:frequency:545GHz:units}{545\,GHz}
\setsymbol{HFI:center:frequency:857GHz:units}{857\,GHz}

\setsymbol{HFI:center:frequency:100GHz}{100}
\setsymbol{HFI:center:frequency:143GHz}{143}
\setsymbol{HFI:center:frequency:217GHz}{217}
\setsymbol{HFI:center:frequency:353GHz}{353}
\setsymbol{HFI:center:frequency:545GHz}{545}
\setsymbol{HFI:center:frequency:857GHz}{857}

\setsymbol{HFI:Ndetectors:100GHz}{8}
\setsymbol{HFI:Ndetectors:143GHz}{11}
\setsymbol{HFI:Ndetectors:217GHz}{12}
\setsymbol{HFI:Ndetectors:353GHz}{12}
\setsymbol{HFI:Ndetectors:545GHz}{3}
\setsymbol{HFI:Ndetectors:857GHz}{4}

\setsymbol{HFI:FWHM:Maps:100GHz:units}{9\parcm88}
\setsymbol{HFI:FWHM:Maps:143GHz:units}{7\parcm18}
\setsymbol{HFI:FWHM:Maps:217GHz:units}{4\parcm87}
\setsymbol{HFI:FWHM:Maps:353GHz:units}{4\parcm65}
\setsymbol{HFI:FWHM:Maps:545GHz:units}{4\parcm72}
\setsymbol{HFI:FWHM:Maps:857GHz:units}{4\parcm39}
\setsymbol{HFI:FWHM:Maps:100GHz}{9.88}
\setsymbol{HFI:FWHM:Maps:143GHz}{7.18}
\setsymbol{HFI:FWHM:Maps:217GHz}{4.87}
\setsymbol{HFI:FWHM:Maps:353GHz}{4.65}
\setsymbol{HFI:FWHM:Maps:545GHz}{4.72}
\setsymbol{HFI:FWHM:Maps:857GHz}{4.39}

\setsymbol{HFI:beam:ellipticity:Maps:100GHz}{1.15}
\setsymbol{HFI:beam:ellipticity:Maps:143GHz}{1.01}
\setsymbol{HFI:beam:ellipticity:Maps:217GHz}{1.06}
\setsymbol{HFI:beam:ellipticity:Maps:353GHz}{1.05}
\setsymbol{HFI:beam:ellipticity:Maps:545GHz}{1.14}
\setsymbol{HFI:beam:ellipticity:Maps:857GHz}{1.19}

\setsymbol{HFI:FWHM:Mars:100GHz:units}{9\parcm37}
\setsymbol{HFI:FWHM:Mars:143GHz:units}{7\parcm04}
\setsymbol{HFI:FWHM:Mars:217GHz:units}{4\parcm68}
\setsymbol{HFI:FWHM:Mars:353GHz:units}{4\parcm43}
\setsymbol{HFI:FWHM:Mars:545GHz:units}{3\parcm80}
\setsymbol{HFI:FWHM:Mars:857GHz:units}{3\parcm67}

\setsymbol{HFI:FWHM:Mars:100GHz}{9.37}
\setsymbol{HFI:FWHM:Mars:143GHz}{7.04}
\setsymbol{HFI:FWHM:Mars:217GHz}{4.68}
\setsymbol{HFI:FWHM:Mars:353GHz}{4.43}
\setsymbol{HFI:FWHM:Mars:545GHz}{3.80}
\setsymbol{HFI:FWHM:Mars:857GHz}{3.67}

\setsymbol{HFI:beam:ellipticity:Mars:100GHz}{1.18}
\setsymbol{HFI:beam:ellipticity:Mars:143GHz}{1.03}
\setsymbol{HFI:beam:ellipticity:Mars:217GHz}{1.14}
\setsymbol{HFI:beam:ellipticity:Mars:353GHz}{1.09}
\setsymbol{HFI:beam:ellipticity:Mars:545GHz}{1.25}
\setsymbol{HFI:beam:ellipticity:Mars:857GHz}{1.03}

\setsymbol{HFI:CMB:relative:calibration:100GHz}{$\lsim 1\%$}
\setsymbol{HFI:CMB:relative:calibration:143GHz}{$\lsim 1\%$}
\setsymbol{HFI:CMB:relative:calibration:217GHz}{$\lsim 1\%$}
\setsymbol{HFI:CMB:relative:calibration:353GHz}{$\lsim 1\%$}
\setsymbol{HFI:CMB:relative:calibration:545GHz}{}
\setsymbol{HFI:CMB:relative:calibration:857GHz}{}

\setsymbol{HFI:CMB:absolute:calibration:100GHz}{$\lsim 2\%$}
\setsymbol{HFI:CMB:absolute:calibration:143GHz}{$\lsim 2\%$}
\setsymbol{HFI:CMB:absolute:calibration:217GHz}{$\lsim 2\%$}
\setsymbol{HFI:CMB:absolute:calibration:353GHz}{$\lsim 2\%$}
\setsymbol{HFI:CMB:absolute:calibration:545GHz}{}
\setsymbol{HFI:CMB:absolute:calibration:857GHz}{}

\setsymbol{HFI:FIRAS:gain:calibration:accuracy:statistical:100GHz}{}
\setsymbol{HFI:FIRAS:gain:calibration:accuracy:statistical:143GHz}{}
\setsymbol{HFI:FIRAS:gain:calibration:accuracy:statistical:217GHz}{}
\setsymbol{HFI:FIRAS:gain:calibration:accuracy:statistical:353GHz}{2.5\%}
\setsymbol{HFI:FIRAS:gain:calibration:accuracy:statistical:545GHz}{1\%}
\setsymbol{HFI:FIRAS:gain:calibration:accuracy:statistical:857GHz}{0.5\%}

\setsymbol{HFI:FIRAS:gain:calibration:accuracy:systematic:100GHz}{}
\setsymbol{HFI:FIRAS:gain:calibration:accuracy:systematic:143GHz}{}
\setsymbol{HFI:FIRAS:gain:calibration:accuracy:systematic:217GHz}{}
\setsymbol{HFI:FIRAS:gain:calibration:accuracy:systematic:353GHz}{}
\setsymbol{HFI:FIRAS:gain:calibration:accuracy:systematic:545GHz}{7\%}
\setsymbol{HFI:FIRAS:gain:calibration:accuracy:systematic:857GHz}{7\%}

\setsymbol{HFI:FIRAS:zero:point:accuracy:100GHz:units}{0.8\MJysr}
\setsymbol{HFI:FIRAS:zero:point:accuracy:143GHz:units}{}
\setsymbol{HFI:FIRAS:zero:point:accuracy:217GHz:units}{}
\setsymbol{HFI:FIRAS:zero:point:accuracy:353GHz:units}{1.4\MJysr}
\setsymbol{HFI:FIRAS:zero:point:accuracy:545GHz:units}{2.2\MJysr}
\setsymbol{HFI:FIRAS:zero:point:accuracy:857GHz:units}{1.7\MJysr}

\setsymbol{HFI:FIRAS:zero:point:accuracy:100GHz}{0.8}
\setsymbol{HFI:FIRAS:zero:point:accuracy:143GHz}{}
\setsymbol{HFI:FIRAS:zero:point:accuracy:217GHz}{}
\setsymbol{HFI:FIRAS:zero:point:accuracy:353GHz}{1.4}
\setsymbol{HFI:FIRAS:zero:point:accuracy:545GHz}{2.2}
\setsymbol{HFI:FIRAS:zero:point:accuracy:857GHz}{1.7}

\setsymbol{HFI:unit:conversion:100GHz:units}{0.2415\MJysrmK}
\setsymbol{HFI:unit:conversion:143GHz:units}{0.3694\MJysrmK}
\setsymbol{HFI:unit:conversion:217GHz:units}{0.4811\MJysrmK}
\setsymbol{HFI:unit:conversion:353GHz:units}{0.2883\MJysrmK}
\setsymbol{HFI:unit:conversion:545GHz:units}{0.05826\MJysrmK}
\setsymbol{HFI:unit:conversion:857GHz:units}{0.002238\MJysrmK}

\setsymbol{HFI:unit:conversion:100GHz}{0.2415}
\setsymbol{HFI:unit:conversion:143GHz}{0.3694}
\setsymbol{HFI:unit:conversion:217GHz}{0.4811}
\setsymbol{HFI:unit:conversion:353GHz}{0.2883}
\setsymbol{HFI:unit:conversion:545GHz}{0.05826}
\setsymbol{HFI:unit:conversion:857GHz}{0.002238}

\setsymbol{HFI:colour:correction:alpha=-2:V1.01:100GHz}{0.9893}
\setsymbol{HFI:colour:correction:alpha=-2:V1.01:143GHz}{0.9759}
\setsymbol{HFI:colour:correction:alpha=-2:V1.01:217GHz}{1.0007}
\setsymbol{HFI:colour:correction:alpha=-2:V1.01:353GHz}{1.0028}
\setsymbol{HFI:colour:correction:alpha=-2:V1.01:545GHz}{1.0019}
\setsymbol{HFI:colour:correction:alpha=-2:V1.01:857GHz}{0.9889}

\setsymbol{HFI:colour:correction:alpha=0:V1.01:100GHz}{1.0008}
\setsymbol{HFI:colour:correction:alpha=0:V1.01:143GHz}{1.0148}
\setsymbol{HFI:colour:correction:alpha=0:V1.01:217GHz}{0.9909}
\setsymbol{HFI:colour:correction:alpha=0:V1.01:353GHz}{0.9888}
\setsymbol{HFI:colour:correction:alpha=0:V1.01:545GHz}{0.9878}
\setsymbol{HFI:colour:correction:alpha=0:V1.01:857GHz}{1.0014}

\providecommand{\sorthelp}[1]{}

\def\bfc{}
\def\bfcc{}

\begin{document}
%
\author{\small
Planck Collaboration: M.~Arnaud\inst{66}
\and
M.~Ashdown\inst{62, 5}
\and
F.~Atrio-Barandela\inst{17}
\and
J.~Aumont\inst{53}
\and
C.~Baccigalupi\inst{81}
\and
A.~J.~Banday\inst{87, 10}
\and
R.~B.~Barreiro\inst{58}
\and
E.~Battaner\inst{89, 90}
\and
K.~Benabed\inst{54, 86}
\and
A.~Benoit-L\'{e}vy\inst{22, 54, 86}
% ...
}
\institute{\small
APC, AstroParticule et Cosmologie, Universit\'{e} Paris Diderot, CNRS/IN2P3, CEA/lrfu, Observatoire de Paris, Sorbonne Paris Cit\'{e}, 10, rue Alice Domon et L\'{e}onie Duquet, 75205 Paris Cedex 13, France\goodbreak
\and
Aalto University Mets\"{a}hovi Radio Observatory and Dept of Radio Science and Engineering, P.O. Box 13000, FI-00076 AALTO, Finland\goodbreak
\and
African Institute for Mathematical Sciences, 6-8 Melrose Road, Muizenberg, Cape Town, South Africa\goodbreak
\and
Agenzia Spaziale Italiana Science Data Center, Via del Politecnico snc, 00133, Roma, Italy\goodbreak
\and
Astrophysics Group, Cavendish Laboratory, University of Cambridge, J J Thomson Avenue, Cambridge CB3 0HE, U.K.\goodbreak
\and
Astrophysics \& Cosmology Research Unit, School of Mathematics, Statistics \& Computer Science, University of KwaZulu-Natal, Westville Campus, Private Bag X54001, Durban 4000, South Africa\goodbreak
\and
Atacama Large Millimeter/submillimeter Array, ALMA Santiago Central Offices, Alonso de Cordova 3107, Vitacura, Casilla 763 0355, Santiago, Chile\goodbreak
\and
CGEE, SCS Qd 9, Lote C, Torre C, 4$^{\circ}$ andar, Ed. Parque Cidade Corporate, CEP 70308-200, Bras\'{i}lia, DF,Ê Brazil\goodbreak
\and
CITA, University of Toronto, 60 St. George St., Toronto, ON M5S 3H8, Canada\goodbreak
\and
CNRS, IRAP, 9 Av. colonel Roche, BP 44346, F-31028 Toulouse cedex 4, France\goodbreak
\and
California Institute of Technology, Pasadena, California, U.S.A.\goodbreak
\and
Centro de Estudios de F\'{i}sica del Cosmos de Arag\'{o}n (CEFCA), Plaza San Juan, 1, planta 2, E-44001, Teruel, Spain\goodbreak
\and
Computational Cosmology Center, Lawrence Berkeley National Laboratory, Berkeley, California, U.S.A.\goodbreak
\and
DSM/Irfu/SPP, CEA-Saclay, F-91191 Gif-sur-Yvette Cedex, France\goodbreak
\and
DTU Space, National Space Institute, Technical University of Denmark, Elektrovej 327, DK-2800 Kgs. Lyngby, Denmark\goodbreak
\and
D\'{e}partement de Physique Th\'{e}orique, Universit\'{e} de Gen\`{e}ve, 24, Quai E. Ansermet,1211 Gen\`{e}ve 4, Switzerland\goodbreak
\and
Departamento de F\'{\i}sica Fundamental, Facultad de Ciencias, Universidad de Salamanca, 37008 Salamanca, Spain\goodbreak
\and
Departamento de F\'{\i}sica, Universidad de Oviedo, Avda. Calvo Sotelo s/n, Oviedo, Spain\goodbreak
\and
Department of Astrophysics/IMAPP, Radboud University Nijmegen, P.O. Box 9010, 6500 GL Nijmegen, The Netherlands\goodbreak
\and
Department of Physics \& Astronomy, University of British Columbia, 6224 Agricultural Road, Vancouver, British Columbia, Canada\goodbreak
\and
Department of Physics and Astronomy, Dana and David Dornsife College of Letter, Arts and Sciences, University of Southern California, Los Angeles, CA 90089, U.S.A.\goodbreak
\and
Department of Physics and Astronomy, University College London, London WC1E 6BT, U.K.\goodbreak
\and
Department of Physics, Florida State University, Keen Physics Building, 77 Chieftan Way, Tallahassee, Florida, U.S.A.\goodbreak
\and
Department of Physics, Gustaf H\"{a}llstr\"{o}min katu 2a, University of Helsinki, Helsinki, Finland\goodbreak
\and
Department of Physics, Princeton University, Princeton, New Jersey, U.S.A.\goodbreak
\and
Department of Physics, University of California, Santa Barbara, California, U.S.A.\goodbreak
\and
Dipartimento di Fisica e Astronomia G. Galilei, Universit\`{a} degli Studi di Padova, via Marzolo 8, 35131 Padova, Italy\goodbreak
\and
Dipartimento di Fisica e Scienze della Terra, Universit\`{a} di Ferrara, Via Saragat 1, 44122 Ferrara, Italy\goodbreak
\and
Dipartimento di Fisica, Universit\`{a} La Sapienza, P. le A. Moro 2, Roma, Italy\goodbreak
\and
Dipartimento di Fisica, Universit\`{a} degli Studi di Milano, Via Celoria, 16, Milano, Italy\goodbreak
\and
Dipartimento di Fisica, Universit\`{a} degli Studi di Trieste, via A. Valerio 2, Trieste, Italy\goodbreak
\and
Dipartimento di Fisica, Universit\`{a} di Roma Tor Vergata, Via della Ricerca Scientifica, 1, Roma, Italy\goodbreak
\and
Discovery Center, Niels Bohr Institute, Blegdamsvej 17, Copenhagen, Denmark\goodbreak
\and
Discovery Center, Niels Bohr Institute, Copenhagen University, Blegdamsvej 17, Copenhagen, Denmark\goodbreak
\and
European Southern Observatory, ESO Vitacura, Alonso de Cordova 3107, Vitacura, Casilla 19001, Santiago, Chile\goodbreak
\and
European Space Agency, ESAC, Planck Science Office, Camino bajo del Castillo, s/n, Urbanizaci\'{o}n Villafranca del Castillo, Villanueva de la Ca\~{n}ada, Madrid, Spain\goodbreak
\and
European Space Agency, ESTEC, Keplerlaan 1, 2201 AZ Noordwijk, The Netherlands\goodbreak
\and
Facolt\`{a} di Ingegneria, Universit\`{a} degli Studi e-Campus, Via Isimbardi 10, Novedrate (CO), 22060, Italy\goodbreak
\and
Gran Sasso Science Institute, INFN, viale F. Crispi 7, 67100 L'Aquila, Italy\goodbreak
\and
Helsinki Institute of Physics, Gustaf H\"{a}llstr\"{o}min katu 2, University of Helsinki, Helsinki, Finland\goodbreak
\and
INAF - Osservatorio Astrofisico di Catania, Via S. Sofia 78, Catania, Italy\goodbreak
\and
INAF - Osservatorio Astronomico di Padova, Vicolo dell'Osservatorio 5, Padova, Italy\goodbreak
\and
INAF - Osservatorio Astronomico di Roma, via di Frascati 33, Monte Porzio Catone, Italy\goodbreak
\and
INAF - Osservatorio Astronomico di Trieste, Via G.B. Tiepolo 11, Trieste, Italy\goodbreak
\and
INAF/IASF Bologna, Via Gobetti 101, Bologna, Italy\goodbreak
\and
INAF/IASF Milano, Via E. Bassini 15, Milano, Italy\goodbreak
\and
INFN, Sezione di Bologna, Via Irnerio 46, I-40126, Bologna, Italy\goodbreak
\and
INFN, Sezione di Roma 1, Universit\`{a} di Roma Sapienza, Piazzale Aldo Moro 2, 00185, Roma, Italy\goodbreak
\and
INFN/National Institute for Nuclear Physics, Via Valerio 2, I-34127 Trieste, Italy\goodbreak
\and
IPAG: Institut de Plan\'{e}tologie et d'Astrophysique de Grenoble, Universit\'{e} Grenoble Alpes, IPAG, F-38000 Grenoble, France, CNRS, IPAG, F-38000 Grenoble, France\goodbreak
\and
Imperial College London, Astrophysics group, Blackett Laboratory, Prince Consort Road, London, SW7 2AZ, U.K.\goodbreak
\and
Infrared Processing and Analysis Center, California Institute of Technology, Pasadena, CA 91125, U.S.A.\goodbreak
\and
Institut d'Astrophysique Spatiale, CNRS (UMR8617) Universit\'{e} Paris-Sud 11, B\^{a}timent 121, Orsay, France\goodbreak
\and
Institut d'Astrophysique de Paris, CNRS (UMR7095), 98 bis Boulevard Arago, F-75014, Paris, France\goodbreak
\and
Institute for Space Sciences, Bucharest-Magurale, Romania\goodbreak
\and
Institute of Astronomy, University of Cambridge, Madingley Road, Cambridge CB3 0HA, U.K.\goodbreak
\and
Institute of Theoretical Astrophysics, University of Oslo, Blindern, Oslo, Norway\goodbreak
\and
Instituto de F\'{\i}sica de Cantabria (CSIC-Universidad de Cantabria), Avda. de los Castros s/n, Santander, Spain\goodbreak
\and
Istituto Nazionale di Fisica Nucleare, Sezione di Padova, via Marzolo 8, I-35131 Padova, Italy\goodbreak
\and
Jet Propulsion Laboratory, California Institute of Technology, 4800 Oak Grove Drive, Pasadena, California, U.S.A.\goodbreak
\and
Jodrell Bank Centre for Astrophysics, Alan Turing Building, School of Physics and Astronomy, The University of Manchester, Oxford Road, Manchester, M13 9PL, U.K.\goodbreak
\and
Kavli Institute for Cosmology Cambridge, Madingley Road, Cambridge, CB3 0HA, U.K.\goodbreak
\and
Kazan Federal University, 18 Kremlyovskaya St., Kazan, 420008, Russia\goodbreak
\and
LAL, Universit\'{e} Paris-Sud, CNRS/IN2P3, Orsay, France\goodbreak
\and
LERMA, CNRS, Observatoire de Paris, 61 Avenue de l'Observatoire, Paris, France\goodbreak
\and
Laboratoire AIM, IRFU/Service d'Astrophysique - CEA/DSM - CNRS - Universit\'{e} Paris Diderot, B\^{a}t. 709, CEA-Saclay, F-91191 Gif-sur-Yvette Cedex, France\goodbreak
\and
Laboratoire Traitement et Communication de l'Information, CNRS (UMR 5141) and T\'{e}l\'{e}com ParisTech, 46 rue Barrault F-75634 Paris Cedex 13, France\goodbreak
\and
Laboratoire de Physique Subatomique et Cosmologie, Universit\'{e} Grenoble-Alpes, CNRS/IN2P3, 53, rue des Martyrs, 38026 Grenoble Cedex, France\goodbreak
\and
Laboratoire de Physique Th\'{e}orique, Universit\'{e} Paris-Sud 11 \& CNRS, B\^{a}timent 210, 91405 Orsay, France\goodbreak
\and
Lawrence Berkeley National Laboratory, Berkeley, California, U.S.A.\goodbreak
\and
Lebedev Physical Institute of the Russian Academy of Sciences, Astro Space Centre, 84/32 Profsoyuznaya st., Moscow, GSP-7, 117997, Russia\goodbreak
\and
Max-Planck-Institut f\"{u}r Astrophysik, Karl-Schwarzschild-Str. 1, 85741 Garching, Germany\goodbreak
\and
Max-Planck-Institut f\"{u}r Radioastronomie, Auf dem H\"{u}gel 69, 53121 Bonn, Germany\goodbreak
\and
National Radio Astronomy Observatory, 520 Edgemont Road, Charlottesville VA 22903-2475, U.S.A.\goodbreak
\and
National University of Ireland, Department of Experimental Physics, Maynooth, Co. Kildare, Ireland\goodbreak
\and
Nicolaus Copernicus Astronomical Center, Bartycka 18, 00-716 Warsaw, Poland\goodbreak
\and
Niels Bohr Institute, Blegdamsvej 17, Copenhagen, Denmark\goodbreak
\and
Niels Bohr Institute, Copenhagen University, Blegdamsvej 17, Copenhagen, Denmark\goodbreak
\and
Optical Science Laboratory, University College London, Gower Street, London, U.K.\goodbreak
\and
SETI Institute and SOFIA Science Center, NASA Ames Research Center, MS 211-3, Mountain View, CA 94035, U.S.A.\goodbreak
\and
SISSA, Astrophysics Sector, via Bonomea 265, 34136, Trieste, Italy\goodbreak
\and
School of Physics and Astronomy, Cardiff University, Queens Buildings, The Parade, Cardiff, CF24 3AA, U.K.\goodbreak
\and
Sorbonne Universit\'{e}-UPMC, UMR7095, Institut d'Astrophysique de Paris, 98 bis Boulevard Arago, F-75014, Paris, France\goodbreak
\and
Space Sciences Laboratory, University of California, Berkeley, California, U.S.A.\goodbreak
\and
Special Astrophysical Observatory, Russian Academy of Sciences, Nizhnij Arkhyz, Zelenchukskiy region, Karachai-Cherkessian Republic, 369167, Russia\goodbreak
\and
UPMC Univ Paris 06, UMR7095, 98 bis Boulevard Arago, F-75014, Paris, France\goodbreak
\and
Universit\'{e} de Toulouse, UPS-OMP, IRAP, F-31028 Toulouse cedex 4, France\goodbreak
\and
Universities Space Research Association, Stratospheric Observatory for Infrared Astronomy, MS 232-11, Moffett Field, CA 94035, U.S.A.\goodbreak
\and
University of Granada, Departamento de F\'{\i}sica Te\'{o}rica y del Cosmos, Facultad de Ciencias, Granada, Spain\goodbreak
\and
University of Granada, Instituto Carlos I de F\'{\i}sica Te\'{o}rica y Computacional, Granada, Spain\goodbreak
% ...
}

\def\planck{\textit{Planck}}
\def\Planck{\textit{Planck}}

\title{\Planck\ intermediate results. XXXI. Microwave survey of Galactic supernova remnants}
\authorrunning{\planck\ Collaboration}
\titlerunning{\planck\ Supernova Remnant Survey}

\date{Received                ; accepted               }

\def\ndetected{16}

\abstract{
The all-sky \planck\ survey in 9 frequency bands was used to search for emission from all 274 known Galactic supernova remnants.
Of these, \ndetected\ were detected in at least two \planck\ frequencies. The radio-through-microwave spectral energy distributions were
compiled to determine the emission mechanism for microwave emission. 
In only one case, IC\,443, is there high-frequency emission clearly from dust associated with the supernova remnant.
In all cases, the low-frequency emission is from synchrotron radiation.
As predicted for a population of relativistic particles with energy distribution that extends continuously to high energies, 
a single power law is evident for many sources, including the Crab and PKS\,1209-51/52. A decrease in flux density relative to the extrapolation of radio emission is evident in several sources. Their spectral energy distributions can be approximated as broken power laws, 
$S_\nu\propto\nu^{-\alpha}$, with the spectral index, $\alpha$, increasing 
 by 0.5--1 above a break frequency in the range 10--60\,GHz. The break could be due to synchrotron losses.
}


\maketitle

\section {Introduction}

Supernovae leave behind remnants that take on several forms \citep{mccraywang96}. The core of the star that explodes is either flung out with the rest of the ejecta into the surrounding medium (in the case of a Type I supernova), or it survives as a neutron star or black hole (in the case of a Type II supernova). The neutron stars eject relativistic particles from jets, making them visible as pulsars and powering wind nebulae. These wind nebulae are sometimes called `plerions', and are exemplified by the Crab Nebula. Ejecta from the stellar explosion are only visible for young ($\sim 10^3$\,yr) supernova remnants, before they are mixed with surrounding interstellar or residual circumstellar material; such objects are exemplified by the historical Tycho and Kepler supernova remnants and Cassiopeia~A.
{\bfc Type~I supernovae from white dwarf deflagration leave behind supernova remnants, because their blast waves propagate into the interstellar medium.}
Most supernova remnants are from older explosions, and the material being observed is interstellar (and circumstellar in some cases) material shocked by the supernova blast waves. The magnetic field of the medium is enhanced in the compressed post-shock gas, and charged particles are accelerated to relativistic speeds, generating copious synchrotron emission that is the hallmark of a supernova remnant at radio frequencies.

At the highest radio frequencies and in the microwave, supernova remnants may transition from synchrotron emission to other mechanisms.
The synchrotron brightness decreases as frequencies increase, and free-free emission (with its flatter spectrum) and dust emission (with its steeply rising spectrum) will gain prominence. Dipole radiation from spinning dust grains could possibly contribute \citep{scaife07}. Because the synchrotron radiation 
itself is an energy loss mechanism, the electrons decrease in energy over time, and relatively fewer higher-energy electrons should exist as the remnants age;
therefore, the synchrotron emission will diminish at higher frequencies.
For these reasons, a survey of supernova remnants at microwave frequencies could reveal some keys to the evolution of relativistic particles as they are produced and injected into the interstellar medium, as well as potentially unveiling new emission mechanisms that can trace the nature of the older supernova remnants.

\section{Observations}

\subsection{Properties of the \planck\ survey}

\planck\footnote{\Planck\ (\url{http://www.esa.int/Planck}) is a project of the 
European Space Agency  (ESA) with instruments provided by two scientific 
consortia funded by ESA member states and led by Principal Investigators 
from France and Italy, telescope reflectors provided through a collaboration 
between ESA and a scientific consortium led and funded by Denmark, and 
additional contributions from NASA (USA).} \citep{tauber2010a, Planck2011-1.1} is the third generation
space mission to measure the an\-isotropy of the cosmic microwave
background (CMB).  It observed the sky in nine frequency bands
covering 30--857\,GHz with high sensitivity and angular resolution
from 31\arcm\ to 5\arcm.  The Low Frequency Instrument (LFI;
\citealt{Mandolesi2010, Bersanelli2010, Planck2011-1.4}) covers the
30, 44, and 70\,GHz bands with amplifiers cooled to 20\,\hbox{K}.  The
High Frequency Instrument (HFI; \citealt{Lamarre2010, Planck2011-1.5})
covers the 100, 143, 217, 353, 545, and 857\,GHz bands with bolometers
cooled to 0.1\,\hbox{K}.  A combination of
radiative cooling and three mechanical coolers produces the
temperatures needed for the detectors and optics
\citep{Planck2011-1.3}.  Two data processing centres (DPCs) check and
calibrate the data and make maps of the sky \citep{Planck2011-1.7,
  Planck2011-1.6, planck2013-p02b, planck2013-p03f}.  \planck's sensitivity, angular resolution, and
frequency coverage make it a powerful instrument for Galactic and
extragalactic astrophysics as well as cosmology.  Early astrophysics
results are given in Planck Collaboration VIII--XXVI 2011, based on
data taken between 13~August 2009 and 7~June 2010.  Intermediate
astrophysics results are being presented in a series of papers
based on data taken between 13~August 2009 and 27~November 2010.

Relevant properties of \planck\ are summarized for each frequency band in Table~\ref{planckprop}. The effective beam shapes vary across the sky and with data selection.  Details are given in \cite{planck2013-p02d} and \cite{planck2013-p03c}. Average beam sizes are given in Table~\ref{planckprop}. Calibration of the brightness scale was achieved by measuring the amplitude of the dipole of the cosmic microwave background (CMB) radiation, which has a known spatial and spectral form; this calibration was used for the seven lower frequency bands (30--353\,GHz). At the two high frequencies (545 and 857\,GHz), the CMB signal is relatively low, so calibration was performed using measurements of Uranus and Neptune. The accuracy of the calibration is given in Table~\ref{planckprop}; 
{\bfc the \Planck\ calibration is precise enough that it is not an issue for the results discussed in this paper.}

The \planck\ image products are at {\tt Healpix} {\bfc \citep{gorski05}} $N_{\rm side}=2048$ for frequencies 100--857\,GHz and $N_{\rm side}=1024$ for frequencies 30--70\,GHz, in units of CMB thermodynamic temperature up to 353\,GHz.  For astronomical use, the temperature units are converted to flux density per pixel by multiplication by the factor given in the last column of Table~\ref{planckprop}. 
At 545 and 857\,GHz, the maps are provided in MJy~sr$^{-1}$ units, and the scaling simply reflects the pixel size.

\begin{table}
\caption{Planck survey properties}
\label{planckprop}
\tiny
\setbox\tablebox=\vbox{
\newdimen\digitwidth
\setbox0=\hbox{\rm 0}
\digitwidth=\wd0
\catcode`*=\active
\def*{\kern\digitwidth}
\newdimen\signwidth
\setbox0=\hbox{+}
\signwidth=\wd0
\catcode`!=\active
\def!{\kern\signwidth}
\halign{\hbox to 0.7in{#\leaderfil}\tabskip=2em&
    \hfil#\hfil\tabskip=1em& 
    \hfil#\hfil\tabskip=2em& 
    \hfil#\hfil\tabskip=1em& 
    \hfil#\hfil\tabskip=0pt\cr
\noalign{\doubleline}
\omit&Center&Beam&Calibration&Units\cr
\omit&Frequency&FWHM\rlap{$^{\rm a}$}&Accuracy\rlap{$^{\rm b}$}&Factor\rlap{$^{\rm c}$}\cr
\omit\hfil Band\hfil&[GHz]&[arcmin]&[\%]&[Jy\,pix\mo]\cr
\noalign{\vskip 3pt\hrule\vskip 5pt}
*30&  *28.5&           32.3 & 0.25&           *27.00*\cr
*44&  *44.1&           27.1 & 0.25&           *56.56*\cr
*70&  *70.3&           13.3 & 0.25&           128.4**\cr
100&100\phantom{.}*&   *9.66& 0.6*&           *59.72*\cr
143&143\phantom{.}*&   *7.27& 0.5*&           *95.09*\cr
217&217\phantom{.}*&   *5.01& 0.7*&           121.2**\cr
353&353\phantom{.}*&   *4.86& 2.5*&           *74.63*\cr
545&545\phantom{.}*&   *4.84& 5\phantom{.}**& **0.250\cr
857&857\phantom{.}*&   *4.63& 5\phantom{.}**& **0.250\cr
\noalign{\vskip 5pt\hrule\vskip 3pt}}}
\endPlancktable
\tablenote {{\rm a}} Beams are from \citep{planck2013-p03c} for HFI and \citep{planck2013-p02d} for LFI\par
\tablenote {{\rm b}} Calibration is from \citep{planck2013-p03f} for HFI and \citep{planck2013-p02b} for LFI\par
\tablenote {{\rm c}} Pixel sizes refer to the all-sky Healpix maps and are 11.8$'$ for frequencies 30--70 GHz and 2.9$'$ for frequencies 100-857 GHz.\par
\end{table}

\subsection{Flux density measurements}

{\bfcc Flux densities were measured using circular aperture photometry.
Other approaches to measuring the source fluxes were explored and could be pursued by future investigators. This includes model fitting (e.g. Gaussian or other source shape motivated by morphology seen at other wavelengths convolved with the beam) or non-circular aperture photometry (e.g. drawing a shape around the source and hand-selecting a background). We experimented with both methods and found they were highly dependent upon the choices made. The circular-aperture method used in this paper has the advantages of being symmetric about the source center (hence, eliminating all linear gradients in the background) as well as being objective about the shape of the source (hence, independent of assumptions about the microwave emitting region).
}

Because of the wide range of \Planck\ beam sizes and the comparable size of the supernova remnants (SNRs), we took care to adjust the aperture sizes as a function of frequency and to scale the results to a flux density scale.  For each SNR, the source size in the maps was taken to be the combination of the intrinsic source size from the Green catalogue \citep{greencatalog}, 
$\theta_{\rm SNR}$ in Table~\ref{sourcetab}, and the {\it Planck} beam size, 
$\theta_{\rm b}$ in Table~\ref{planckprop}:
\begin{equation}
\theta_{\rm s} = \sqrt{\theta_{\rm SNR}^2 + \theta_{\rm b}^2}.
\end{equation}
The flux densities were measured using standard aperture photometry, with the aperture size centred on each target with a diameter scaled to  
$\theta_{\rm ap} = 1.5 \theta_{\rm s}$.
The background was determined in an annulus of inner and outer radii $1.5\theta_{\rm ap}$ and $2\theta_{ap}$.  To correct for loss of flux density outside of the aperture, an aperture correction as predicted for a Gaussian flux density distribution with size $\theta_s$ was applied to each measurement:
\begin{equation}
f_{\rm A} = \frac{1}{f(\theta_{\rm ap})-\left[f(\theta_{\rm out})-f(\theta_{\rm in})\right]}
\label{eq:fap}
\end{equation}
where the flux density enclosed within a given aperture is
\begin{equation}
f(\theta)=1-e^{-4\ln 2 (\theta/\theta_{\rm s})^2}
\end{equation}
and $\theta_{\rm in}$ and $\theta_{\rm out}$ are the inner and outer radii of annulus within which the background is measured. For the well-resolved sources (diameter 50\,\% larger than the beam FWHM), no aperture correction was applied.  Aperture corrections are typically 1.5 for the compact sources and by
definition unity for the large sources. The use of a Gaussian source model for the intermediate cases (source comparable to beam) is not strictly appropriate for SNRs, which are often limb-brightened (shell-like), so the aperture corrections are only good to about 20\,\%.  

{\bfcc The uncertainties for the flux densities are a root-sum-square of the calibration uncertainty (from Table~\ref{planckprop}) and the propagated statistical uncertainties.
The statistical uncertainties} (technically, appropriate for white uncorrelated noise only) take into account the number of pixels in the on-source aperture and background annulus and using the robust standard deviation within the background annulus to estimate the pixel-to-pixel noise for each source.
The measurements are made using the native {\tt Healpix} maps, by searching for pixels within the appropriate apertures and background annuli and
calculating the sums and robust medians, respectively. This procedure avoids the need of generating extra projections and maintains the native
pixelization of the survey.

\begin{figure}
\includegraphics[width=8.9cm]{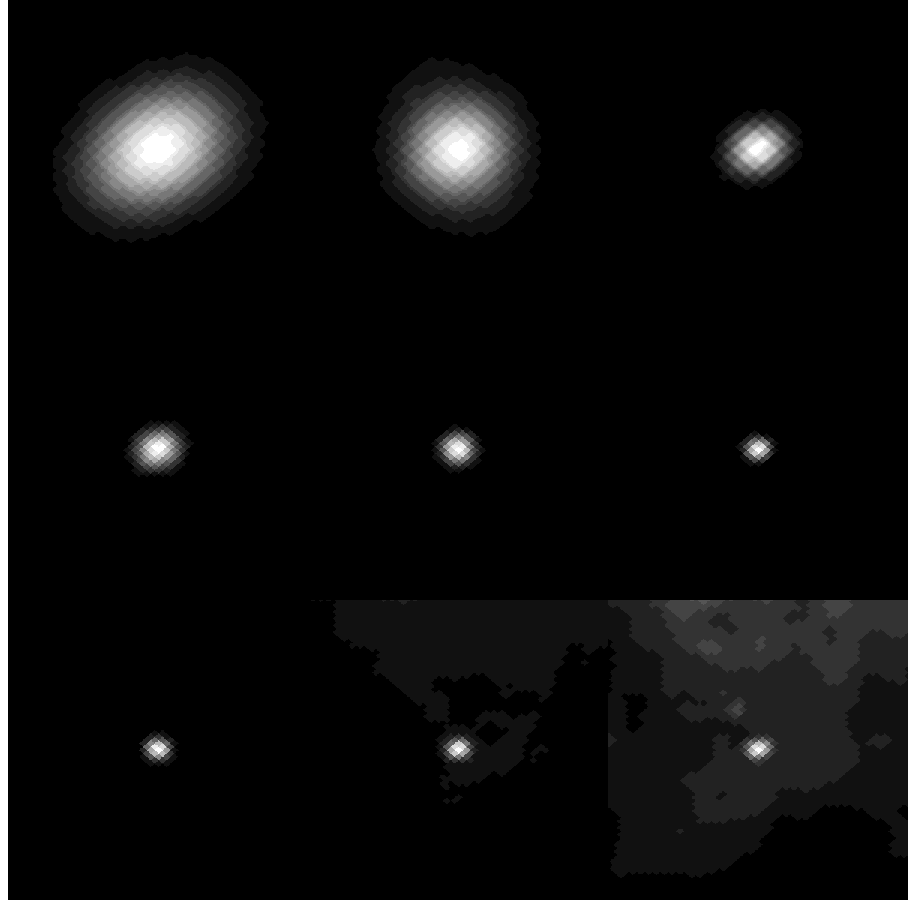}
\vskip 0.5em
\includegraphics[width=9cm]{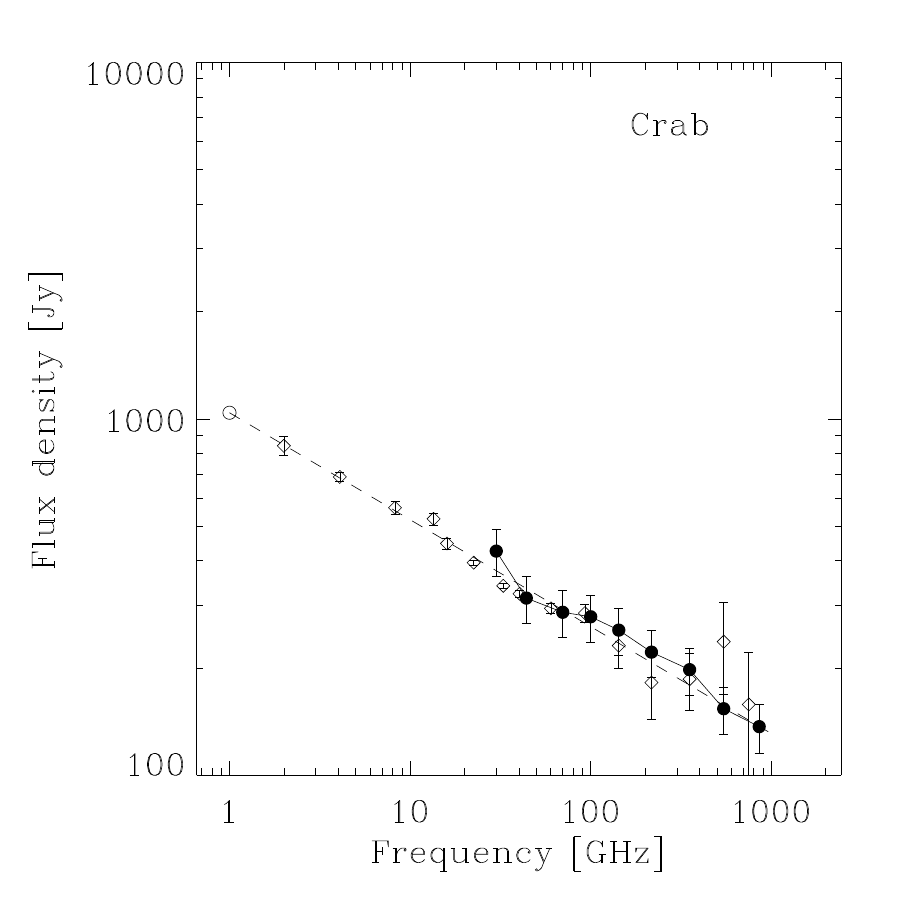}
\caption{
{\it Top\/}:  \Planck\ Images of the Crab nebula at frequencies increasing from 30\,GHz at top left to 857\,GHz at bottom right.
Each image is 100\arcm\ on a side, with the three lowest-frequency images smoothed by 7$'$. 
{\it Bottom\/}: Microwave spectral energy distribution of the Crab Nebula. Filled circles are \Planck\ measurements with 3$\sigma$ error bars. 
The open symbol at 1\,GHz and the dashed line emanating from it are the flux 
and power law with spectral index from the Green catalogue. 
Open diamonds are from the compilation of flux densities by \citet{macias10}. 
}
\label{crabsed}
\end{figure}

{\bfc We verify the flux calibration scale by using the procedure described above on the well-measured Crab Nebula.}
The flux densities are compared to previous measurements in Fig.~\ref{crabsed}. The good agreement 
verifies the measurement procedure used for this survey, at least for a compact source.
The \Planck\ Early Release Compact Source Catalogue  {\bfc \citep[hereafter ERCSC]{planck2011-1.10}} flux density measurements 
for the Crab are lower
than those determined in this paper, due to the source
being marginally resolved by \planck\ at high frequencies. 

Table~\ref{sourcetab} lists the basic properties of SNRs that were detected by \planck.
Table~\ref{fluxtab} summarises the \planck\ flux-density measurements for detected SNRs.
To be considered a detection, each SNR must have a statistically significant flux density measurement (flux density greater than 3.5 times 
the statistical uncertainty in the aperture photometry measurement) and be evident by eye for at 
least two \planck\ frequencies. Inspection of the higher-frequency images allows for identification of interstellar foregrounds for the supernova
remnants, which are almost all best-detected at the lowest frequencies. A large fraction of the SNRs are located close to the
Galactic plane and are smaller than the low-frequency \planck\ beam. Essentially none of those targets could be detected with \planck\ due to 
confusion from surrounding \ion{H}{ii} regions. Dust and free-free emission from \ion{H}{ii} regions makes them extremely bright at far-infrared wavelengths
and moderately bright at radio frequencies. Supernova remnants in the Galactic plane would only be separable from \ion{H}{ii} regions using a multi-frequency 
approach and angular resolution significantly higher than \planck.
More detailed results from the flux measurements are provided in the Appendix.

\begin{table*}
\begingroup
\newdimen\tblskip \tblskip=5pt
\caption{Properties of detected supernova remnants.}
\label{sourcetab}
\nointerlineskip
\vskip -3mm
\footnotesize
\setbox\tablebox=\vbox{
   \newdimen\digitwidth 
   \setbox0=\hbox{\rm 0} 
   \digitwidth=\wd0 
   \catcode`*=\active 
   \def*{\kern\digitwidth}
   \newdimen\signwidth 
   \setbox0=\hbox{+} 
   \signwidth=\wd0 
   \catcode`!=\active 
   \def!{\kern\signwidth}
\halign{#\hfil\tabskip=1em&
   \hbox to 1.5in{#\leaderfil}\tabskip=0.2em&
   \hfil#\hfil\tabskip=1em&
   \hfil#\hfil&
   \hfil#\hfil&
   #\hfil\tabskip=0pt\cr
\noalign{\doubleline}
\multispan2\hfil Supernova remnant\hfil & Angular \rlap{$^{\rm a}$} & $S_{\rm 1\,GHz}$\rlap{$^{\rm a}$}&Spectral\rlap{$^{\rm a}$}&\omit {\bfc Notes}\rlap{$^{\rm b}$}\cr
\multispan2\hfil   \hfil  & Diameter ($'$) & (Jy) & Index
\hfil\cr
\noalign{\vskip 3pt\hrule\vskip 5pt}
G21.5-0.9&       G21.5-0.9&  **4&	***7&  0.00& $<10^3$ yr, pulsar\cr
G34.7-0.4&           W\,44&  *35& *230&  0.37& $2\times 10^4$ yr, dense\cr
G69.0+2.7&         CTB\,80&  *80& *120& \dots& $10^5$ yr, pulsar, plerion\cr
G74.0-8.5&     Cygnus Loop&  230& *210& \dots& mature\cr
G89.0+4.7&          HB\,21&  120& *220&  0.38& 5000 yr, dense\cr
G111.7-2.1&         Cas\,A&  **5& 2720&  0.77& 330 yr, ejecta, CCO\cr
G120.1+1.4&          Tycho&  **8& **56&  0.65& young, Type Ia\cr
G130.7+3.1&         3C\,58&  **9& **33&  0.07& 830 yr, plerion, pulsar\cr
G184.6-5.8&           Crab&  **7& 1040&  0.30& 958 yr, plerion, pulsar\cr
G189.1+3.0&        IC\,443&  *40& *160&  0.36& $3\times 10^4$ yr, dense ISM, CCO\cr
G260.4-3.4&       Puppis\,A&  *60& *130&  0.50& CCO, 3700 yr\cr
G263.9-3.3&           Vela&  255& 1750& \dots& $10^4$ yr\cr
G296.5+10.0&PKS\,1209-51/52&  *90& **48&  0.5*& $\sim 10^4$ yr\cr
G315.4-2.3&        RCW\,86&  *42& **49&  0.60& 1800 yr\cr
G326.3-1.8&MSH\,15-5{\it 6}&  *38& *145& \dots\cr
G327.6+14.6&      SN\,1006&  *30& **19&  0.60& 1008 yr, Type Ia\cr
\noalign{\vskip 5pt\hrule\vskip 3pt}}}
\endPlancktablewide
\tablenote {{\rm a}} Flux densities at 1\,GHz, radio spectral indices, and angular diameters are from \citet{greencatalog}. For cases of unknown or strongly-spatially-variable spectral index, we put `\dots' in the table and use a nominal value of 0.5 in the figures for illustration purposes only.\par
\tablenote {{\rm b}} Notes on the properties of the SNR, gleaned from the \citet{greencatalog} catalogue and references therein, as well as updates from the online version
\url{http://www.mrao.cam.ac.uk/projects/surveys/snrs/}. {\it Plerion\/}: radio emission is centre-filled. {\it Pulsar\/}: a pulsar is associated with the SNR. {\it CCO}: a central compact object (presumably a neutron star) is associated with the supernova remnant from high-energy images. {\it Dense ISM\/}: the supernova remnant is interacting with a dense interstellar medium. {\it X\/}yr: estimated age of the supernova remnant. {\it Mature\/}: age likely greater than $10^5$ yr. {\it Type X\/}: the supernova that produced this supernova remnant was of type $X$.\par
\endgroup
\end{table*}

\def\Tiny{ \font\Tinyfont = cmr10 at 7.5pt \relax  \Tinyfont}

\begin{table*}
\begingroup
\newdimen\tblskip \tblskip=5pt
\caption{\Planck\ integrated flux density measurements of supernova remnants.}
\label{fluxtab}
\nointerlineskip
\vskip -5mm
\scriptsize
\setbox\tablebox=\vbox{
   \newdimen\digitwidth 
   \setbox0=\hbox{\rm 0} 
   \digitwidth=\wd0 
   \catcode`*=\active 
   \def*{\kern\digitwidth}
   \newdimen\signwidth 
   \setbox0=\hbox{+} 
   \signwidth=\wd0 
   \catcode`!=\active 
   \def!{\kern\signwidth}
\halign{\hbox to 0.8in{#\leaderfil}\tabskip=1em&
   \hfil#\hfil&
   \hfil#\hfil&
   \hfil#\hfil&
   \hfil#\hfil&
   \hfil#\hfil&
   \hfil#\hfil&
   \hfil#\hfil&
   \hfil#\hfil&
   \hfil#\hfil\tabskip=0pt\cr
\noalign{\doubleline}
\omit&\multispan9\hfil\sc Flux density [Jy]\hfil\cr
\noalign{\vskip -3pt}
\omit&\multispan9\hrulefill\cr
\omit\hfil\sc Source\hfil&30\,GHz&44\,GHz&70\,GHz&100\,GHz&143\,GHz&217\,GHz&353\,GHz&545\,GHz&857\,GHz\cr
\noalign{\vskip 3pt\hrule\vskip 5pt}
G21.5-0.9&                \dots&       \dots&    $**4.3\pm*0.6$& $**2.7\pm*0.5$& $**3.0\pm*0.4$&          \dots&          \dots&          \dots&            \dots\cr
W\,44&           $121\pm*8$& $*68\pm*5$& $*30\pm*5$&          \dots&          \dots&          \dots&          \dots&          \dots&            \dots\cr
CTB\,80&         $*12.2\pm*1.7$& $**3.7\pm*1.2$&          \dots& \dots&          \dots&          \dots&          \dots&          \dots&            \dots\cr
Cygnus Loop&     $*24.9\pm*1.7$&          \dots&          \dots&          \dots&          \dots&          \dots&          \dots&          \dots&            \dots\cr
HB\,21&          $*14.7\pm*1.7$& $*11.9\pm*1.3$& \dots&          \dots&          \dots&          \dots&          \dots&          \dots&            \dots\cr
Cas\,A&          $228\pm11$& $140\pm*7$& $109\pm*6$& $*94\pm*5$& $*73\pm*4$& $*62\pm*4$& $*52\pm*7$&          \dots&            \dots\cr
Tycho&           $**8.1\pm*0.4$& $**5.2\pm*0.3$& $**4.4\pm*0.3$& $**4.0\pm*0.2$& $**3.5\pm*0.2$&\dots&          \dots&          \dots&            \dots\cr
3C\,58&          $*22.2\pm*2.2$& $*16.4\pm*1.6$& $*14.2\pm*1.4$& $*12.7\pm*1.3$& $*10.8\pm*1.1$& $**8.4\pm*0.8$& $**4.8\pm*0.7$&          \dots&            \dots\cr
Crab&            $425\pm21$& $314\pm16$& $287\pm14$& $278\pm14$& $255\pm13$& $222\pm11$& $198\pm 10$& $154\pm 7.7$& $*137\pm**7$\cr
IC\,443&         $*56\pm*3$& $*39\pm*3$& $*20.5\pm*1.2$& $*17.3\pm*0.9$& $*11.1\pm*0.8$& $*56\pm10$& $*229\pm 30$ & $*720\pm 100$& $2004\pm 300$\cr
Puppis\,A&       $*32.0\pm*1.7$& $*21.9\pm*1.2$& $*13.6\pm*1.1$& $**4.5\pm*0.8$&          \dots&          \dots&          \dots&          \dots&            \dots\cr
Vela&            $488\pm25$& $345\pm17$& $289\pm15$&          \dots&          \dots&          \dots&          \dots&          \dots&            \dots\cr
PKS\,1209&       $**9.3\pm*0.5$& $**7.0\pm*0.6$& $**5.5\pm*1.1$& $**4.0\pm*0.4$& \dots&          \dots&          \dots&          \dots&            \dots\cr
RC\,W86&         $**3.7\pm*1.0$& $**4.8\pm*0.7$& $**3.3\pm*0.6$&          \dots&          \dots&          \dots&          \dots&          \dots&            \dots\cr
MSH\,15-5{\it 6}&$*56\pm*4$& $*39\pm*3$& $*22.4\pm*1.5$& $*20.6\pm*1.7$&      $*12.8\pm*2.7$     \dots&          \dots&          \dots&          \dots&            \dots\cr
SN1006&          $**3.2\pm*0.2$& $**2.2\pm*0.3$& $**1.0\pm*0.3$& $**0.7\pm*0.1$&          \dots&          \dots&          \dots&          \dots&            \dots\cr
\noalign{\vskip 5pt\hrule\vskip 3pt}}}
\endPlancktablewide
\endgroup
\end{table*}

As a test of the quality of the results and robustness to contamination from the CMB, we performed the flux density measurements using the total intensity maps as well as the
CMB-subtracted maps. Differences greater than $1\sigma$ were seen only for the largest SNRs. Of the measurements in Table~\ref{fluxtab}, only the following flux densities were affected at
the 2-$\sigma$ or greater level:  Cygnus Loop (44 and 70\,GHz), HB\,21 (70\,GHz), and Vela (70\,GHz).

\section{Results and discussion of individual objects}

The images and spectral energy distributions (SEDs) of detected SNRs are summarized in the following subsections.
A goal of the survey is to determine whether new emission mechanisms or changes in the radio emission mechanisms are detected
in the microwave range.
Therefore, for each target, the 1\,GHz radio flux density \citep{greencatalog} was used to  extrapolate to  microwave frequencies using a power law.
{\bfc The extrapolation illustrates the expected \planck\ flux densities, if synchrotron radiation is the 
sole source of emission and the high-energy particles have a power-law energy distribution.}
The \citet{greencatalog} SNR catalogue was compiled from an extensive and continuously updated literature search.
The radio spectral indices are gleaned from that same compendium.  They represent a fit to the flux densities from 0.4 to 5\,GHz, where available, and are
the value $\alpha$ in a SED $S_\nu=S_\nu({\rm 1\,GHz}) \nu_{\rm GHz}^{-\alpha}$. The spectral indices can be quite uncertain in
some cases, where observations with very different observing techniques are combined (in particular, interferometric and single-dish).
The typical spectral index for synchrotron emission from SNRs is $\alpha\sim 0.4-0.8$.
On theoretical grounds \citep{reynolds11}, it has been shown that for a shock that compresses the gas by a factor $r$, the spectral index of the 
synchrotron emission from the relativistic electrons in the compressed magnetic field has index $\alpha=3/(2r-2)$.
For a strong adiabatic shock, $r=4$, so the expected spectral index is $\alpha=0.5$, similar to the typical observed value for SNRs.
Shallower spectra are seen towards pulsar wind nebulae, where the relativistic particles are freshly injected.

Where there was a positive deviation relative to the radio power law, there would be an indication of a different emission mechanism.
Free-free emission has a radio spectral index $\alpha\sim 0.1$ that is only weakly dependent upon the electron temperature. 
Ionized gas near massive star-forming regions dominates the microwave emission from the Galactic plane and is a primary source of confusion
for the survey presented here. Small SNRs  in the Galactic plane are essentially impossible to detect with \Planck\ due to this confusion.
Larger SNRs can still be identified because their morphology can be recognized by comparison to lower-frequency radio images.

At higher \planck\ frequencies, thermal emission from dust grains dominates the sky brightness. 
The thermal dust emission can be produced in massive star forming regions (just like the free-free
emission), as well as in lower-mass star forming regions, where cold dust in molecular clouds contributes. 
Confusion due to interstellar dust makes identification of SNRs
at the higher \planck\ frequencies essentially impossible without a detailed study of individual cases, and even then the results will require confirmation. For the present
survey we only measure four SNRs at frequencies 353\,GHz and higher; in all cases the targets are bright and compact, making them distinguishable from unrelated emission.

\subsection{G21.5-0.9}

\begin{figure}
\includegraphics[width=9cm]{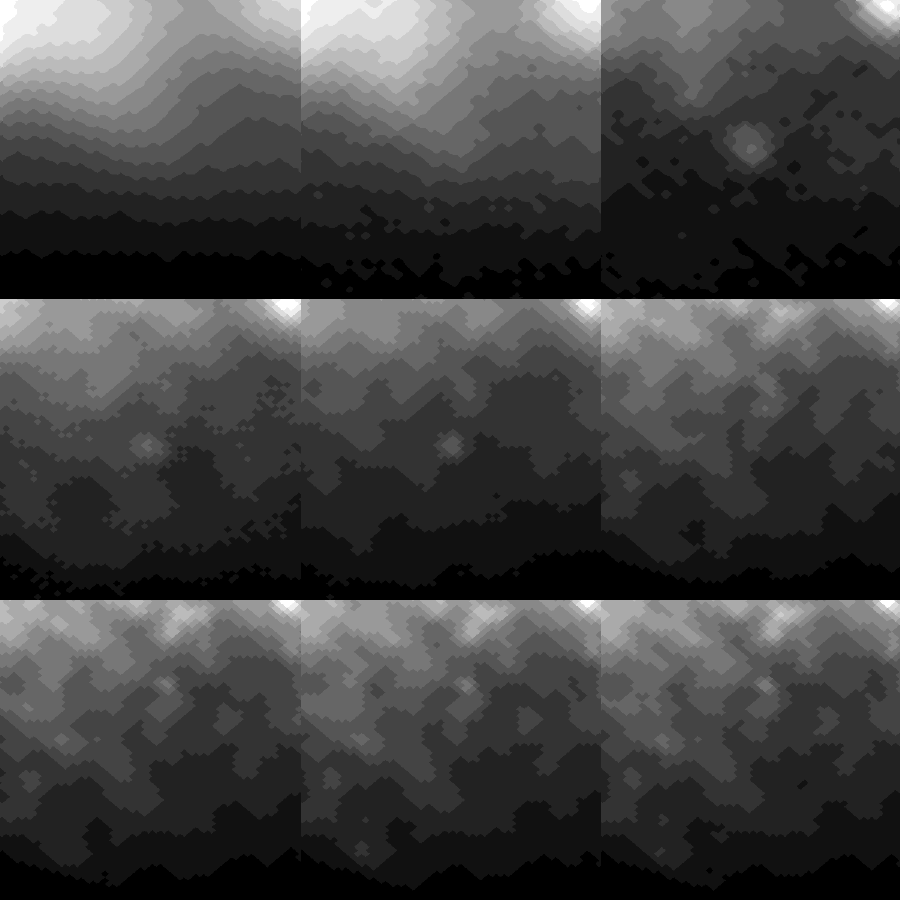}
\vskip 0.5em
\includegraphics[width=9cm]{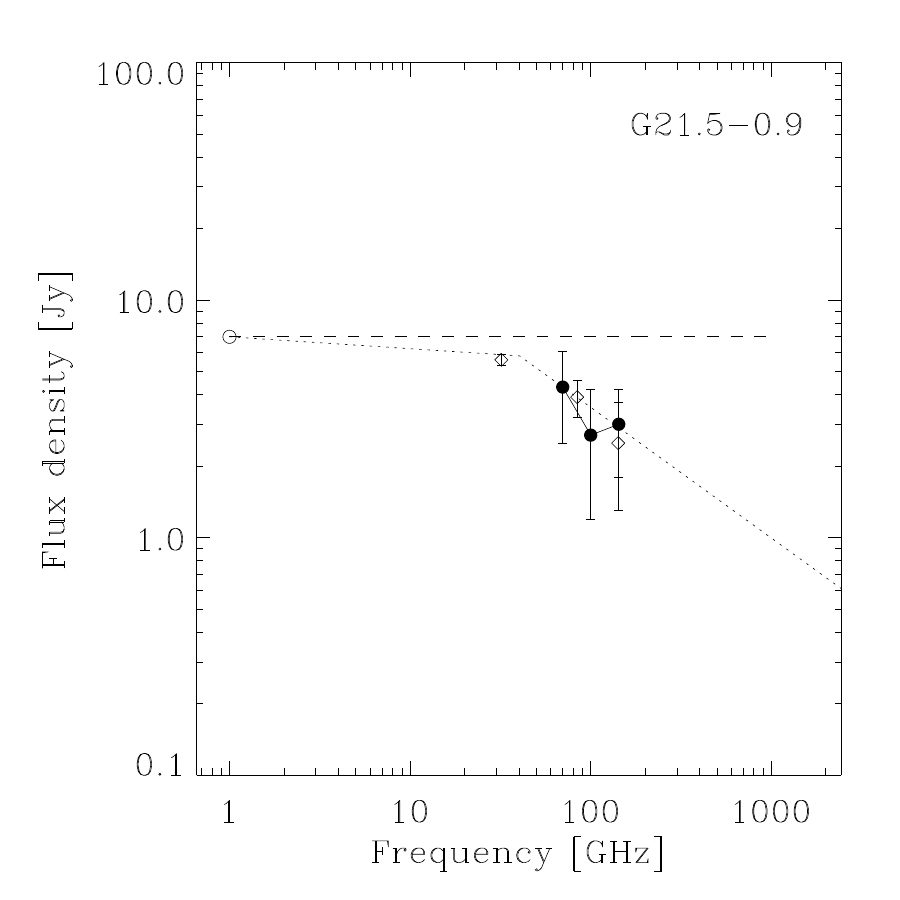}
\caption{
{\it Top\/}:  Images of the G21.5-0.9  environment at the nine \Planck\ frequencies (Table 1), increasing from 30\,GHz at top left to 857\,GHz at bottom right.  Each image is 100\arcm\ on a side{\bfc , with galactic coordinate orientation.}
{\it Bottom\/}: Microwave SED of  G21.5-0.9. Filled circles are \Planck\ measurements from this paper, with 3$\sigma$ error bars. The open symbol at 1\,GHz is from the Green catalogue. The dashed line emanating from the 1\,GHz flux density is a flat spectrum, for illustration. The dotted line is a broken power law, with a 
break frequency at 40\,GHz, a spectral index of 0.05 at lower frequencies and 0.55 at higher frequencies.
Open diamonds show prior radio flux densities from \citet{morsi87} and \citet{salter89}.
}
\label{g21sed}
\end{figure}

G21.5-0.9 is a supernova remnant and pulsar wind nebula, powered by the recently-discovered PSR J1833-1034  \citep{gupta05,camilo06}, with an estimated age less than 870\,yr based on the present expansion rate of the supernova shock \citep{bietenholz08}.
{\bfc The \planck\ images in the upper panels of Figure~\ref{g21sed} show the SNR as a compact source at intermediate frequencies. The source is lost in confusion with unrelated galactic plane
structures at 30--44 GHz and at 217 GHz and higher frequencies. In addition to the aperture photometry as performed for all targets, we made a Gaussian fit at 70 GHz to the source to ensure the flux
refers to the compact source and not diffuse emission. The fitted Gaussian had a lower flux (2.7 Jy) than aperture photometry (4.3 Jy), but the residual from the Gaussian fit still shows the source and
is clearly an underestimate. Fitting with a more complicated functional form would yield a somewhat higher flux, so we are confident in the aperture photometry flux we report in Table~\ref{fluxtab}.

The lower panel of Figure~\ref{g21sed} shows the \planck\ flux densities together with radio data.}
The SED is relatively flat, more typical of a pulsar wind nebula than synchrotron emission from an old SNR
shock.
The microwave SED has been measured at two frequencies, and the \planck\ flux densities are in general agreement. A single
power law cannot fit the observations, as has been noted by \citet{salter89}. Instead, we show in Fig.~\ref{g21sed} a broken power law, which would be expected for a pulsar wind nebula, with a change in spectral index by
+0.5 above a break frequency \citep{reynolds09}. The data are consistent with a break frequency at 40\,GHz and a relatively flat, $\alpha=0.05$ spectral
index at lower frequencies.

\subsection{W\,44}

\begin{figure}
\includegraphics[width=9cm]{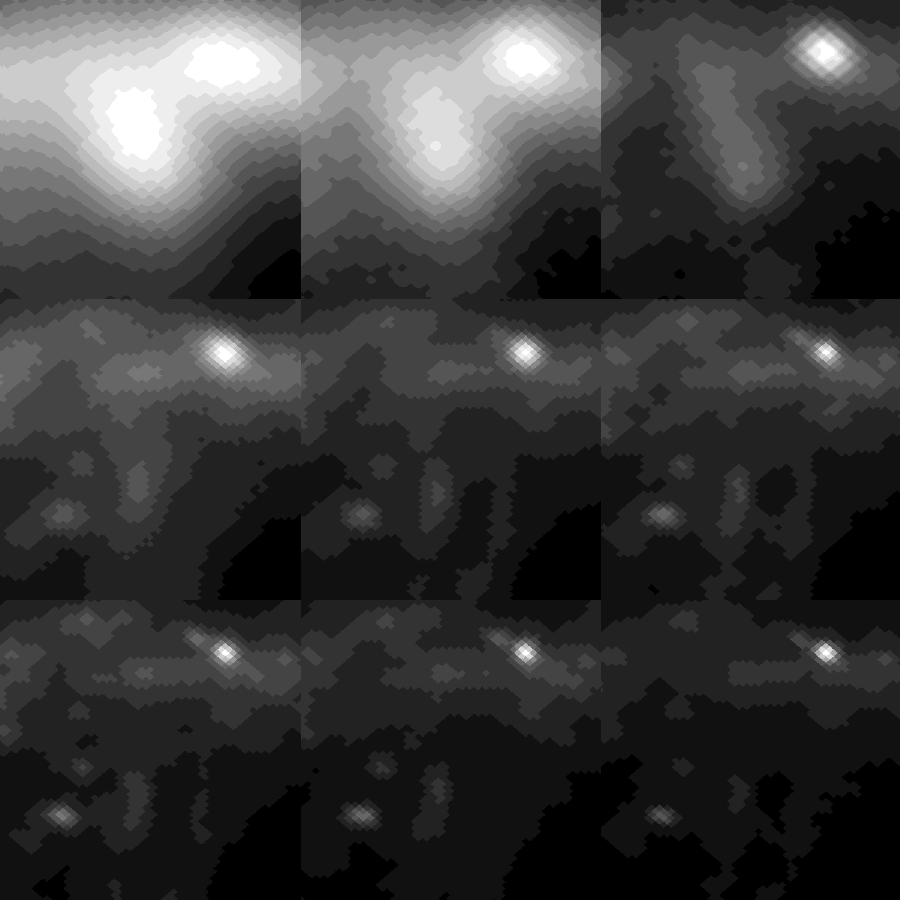}
\vskip 0.5em
\includegraphics[width=9cm]{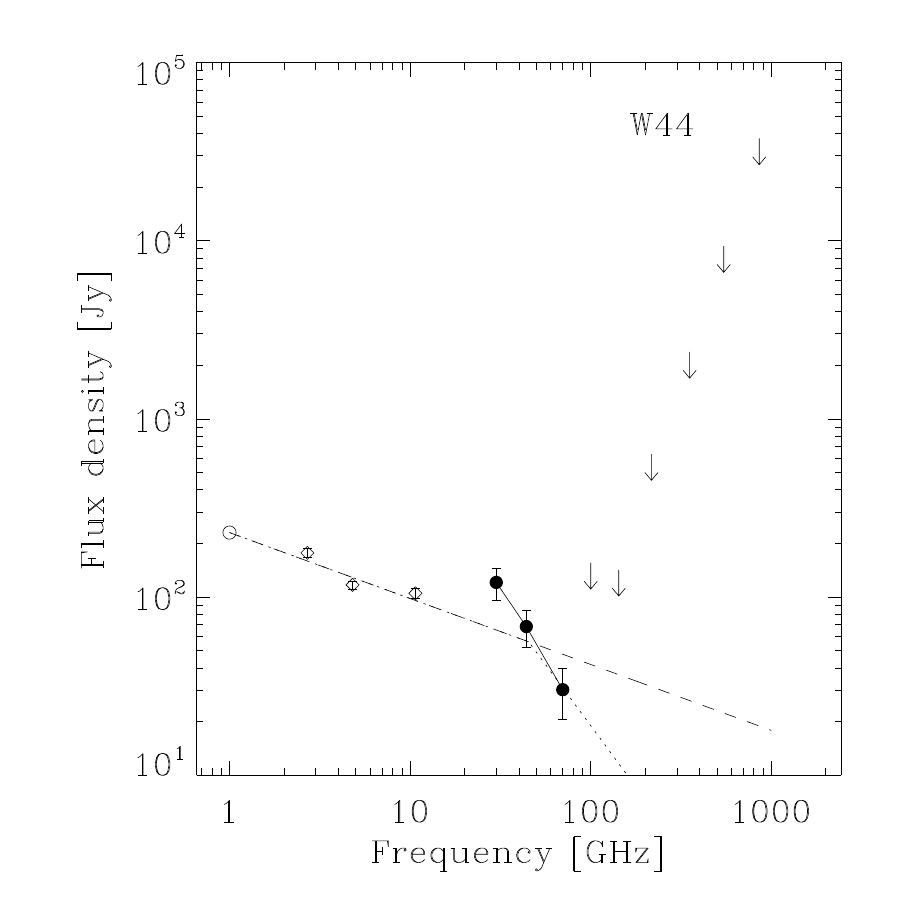}
\caption{
{\it Top\/}:  Images of the W\,44  environment at the nine \Planck\ frequencies (Table 1), increasing from 30\,GHz at top left to 857\,GHz at bottom right.  Each image is $100'$ on a side{\bfc , with galactic coordinate orientation.}
{\it Bottom\/}: Microwave SED of  W\,44. Filled circles are \Planck\ measurements from this paper, with 3$\sigma$ error bars. The open symbol at 1\,GHz is from the Green catalogue, and the dashed line emanating from it is a power law with spectral index from the Green catalogue. 
Open diamonds are radio flux densities {\bfc from\citet{velu76} at 2.7 GHz, \citet{sun11} at 5 GHz and \citet{kundu72} at 10.7 GHz.}
The dotted line illustrates a spectral break increasing the spectral index by 0.5 at 45\,GHz to match the \planck\ 70\,GHz flux density.
Downward arrows show the flux density measurements that are contaminated by unrelated foreground emission.
}
\label{w44sed}
\end{figure}

W\,44 is one of the brightest radio SNRs but is challenging for \planck\ because of its location in a very crowded portion of the Galactic plane. 
Figure~\ref{w44sed} shows the \planck\ images and flux densities.
The synchrotron emission is detected above the unrelated nearby regions at 30--70\,GHz.
Above 100\,GHz, there is a prominent structure in the \planck\ images that is at the {\bfc western}
(left-hand side in Fig.~\ref{w44sed})
border of the radio SNR but has a completely different 
morphology; we recognise this structure as being a compact \ion{H}{ii} region in the radio images, unrelated to the W\,44 SNR even if possibly due to a member of the same OB association as the progenitor. {\bfc At 70 GHz, the \ion{H}{ii} region flux is 22 Jy, comparable to the SNR; we ensured that there
is no contamination of the SNR flux by slightly adjusting the aperture radius to exclude the \ion{H}{ii} region.}
The \planck\ 70\,GHz flux density is somewhat lower than the radio power law, while the 30\,GHz flux density is higher. We consider the discrepancy at 30\,GHz flux density
as possibly due to confusion with unrelated large-scale emission from the Galactic plane. Figure~\ref{w44sed} includes a broken power law that could explain the lower 70\,GHz flux density, but the evidence from this single low flux density, especially given the severe confusion from unrelated sources in the field, does not conclusively demonstrate the existence of a spectral break for W\,44.

\subsection{CTB 80}

\begin{figure}
\includegraphics[width=9cm]{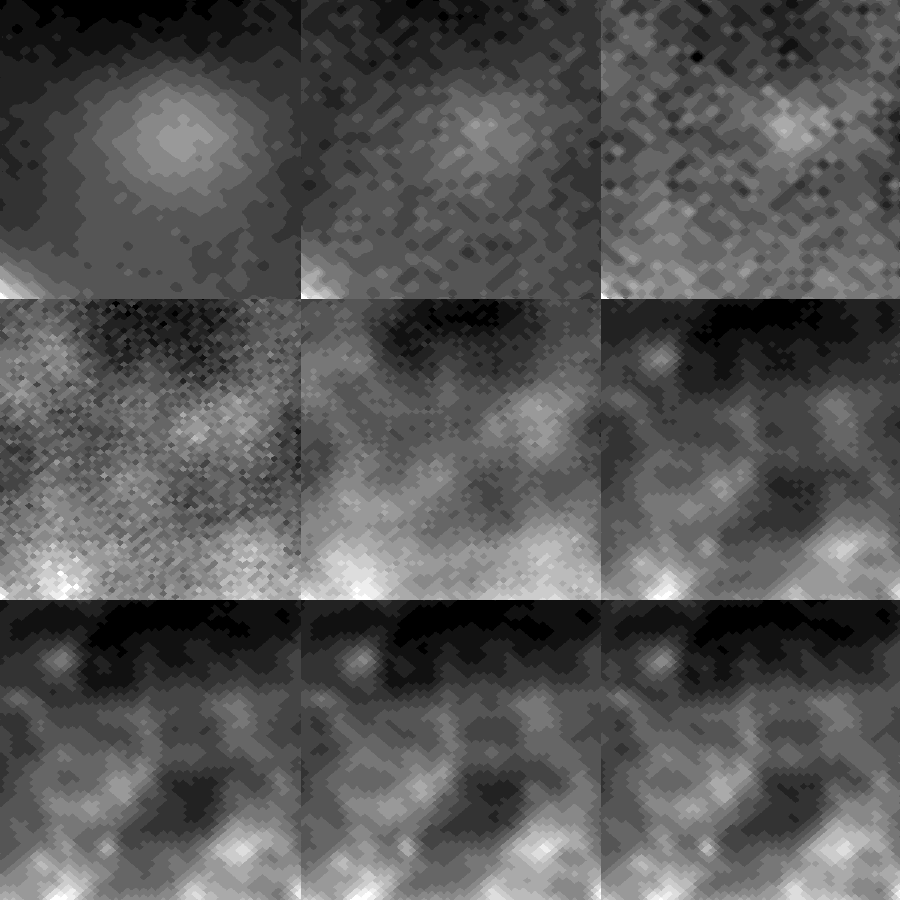}
\vskip 0.5em
\includegraphics[width=9cm]{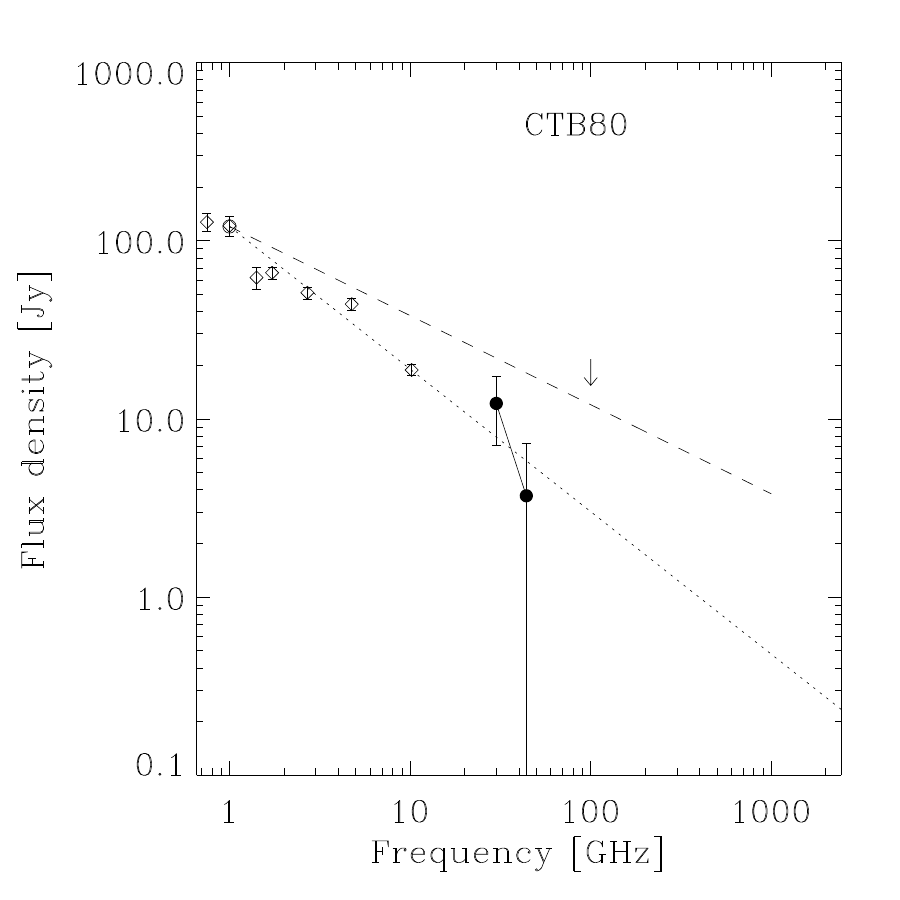}
\caption{
{\it Top\/}:  Images of the CTB\,80  environment at the nine \Planck\ frequencies (Table 1), increasing from 30\,GHz at top left to 857\,GHz at bottom right.  Each image is $100'$ on a side{\bfc , with galactic coordinate orientation.}
{\it Bottom\/}: Microwave SED of  CTB\,80. Filled circles are \Planck\ measurements from this paper, with 3$\sigma$ error bars. The open symbol at 1\,GHz is from the Green catalogue, and the dashed 
and dotted line emanating from it are power laws with spectral indices of 0.5 
{\bfcc{(Green catalogue) and 
0.8 (approximate fit for 1--100 GHz), respectively, for illustration.
Open diamonds are radio flux densities from 
\citet{velu76} at 0.75, 1.00 GHz;
}
\citet{montovani85} at 1.41, 1.72, 2.695 and 4.75 GHz; and \citet{sofue83} at 10.2 GHz.}
The downward arrow shows a flux density measurement that was contaminated by unrelated foreground emission.
}
\label{ctb80sed}
\end{figure}

CTB\,80 is powered by a pulsar that has traveled so far from the centre of explosion that it is now within the shell and injecting high-energy electrons directly into the swept-up ISM.
Figure~\ref{ctb80sed} shows that the SNR is only clearly detected at 30 and 44\,GHz, being lost in confusion with the unrelated ISM at higher frequencies. 
{\bfc The SNR is marginally resolved at low \planck\ frequencies, and the flux measurements include the entire region evident in radio images \citep{castelletti03}.}
The \Planck\ flux densities at 30 and 44\,GHz are consistent with a spectral index around $\alpha\simeq 0.8$, as shown in 
Fig.~\ref{ctb80sed}, continuing the trend that had previously been identified based on 10.2\,GHz  measurements \citep{sofue83}. The spectral index in the microwave is definitely
steeper than that seen from 0.4 to 1.4\,GHz, $\alpha=0.45\pm 0.03$ \citep{kothes06}, though the steepening is not as high as discussed above for pulsar wind nebulae where $\Delta\alpha=0.5$ can be due to significant cooling from synchrotron self-losses. 

\subsection{Cygnus Loop}

The Cygnus Loop is one of the largest SNRs on the sky, with an angular diameter of nearly 4\deg.
To measure the flux density, we use an on-source aperture of 120\arcm\ with a background annulus from 140\arcm\ to 170\arcm. On these large scales, the CMB 
fluctuations are a significant contamination, so we subtracted the CMB using the SMICA map \citep{planck2013-p06}.  
The SNR is well detected at 30\,GHz by \planck; a similar structure is evident in the 44 and 70\,GHz maps, but the
total flux density could not be accurately estimated at these frequencies due to uncertainty in the CMB subtraction.
The NW part of the remnant, which corresponds to NGC\,6992 optically, was listed in the ERCSC as an LFI source with ``no plausible match in existing radio catalogues.'' The actual match to the Cygnus Loop was made by \citet{ami12} as part of
 their effort to clarify the nature of such sources.

\begin{figure}
\includegraphics[width=9cm]{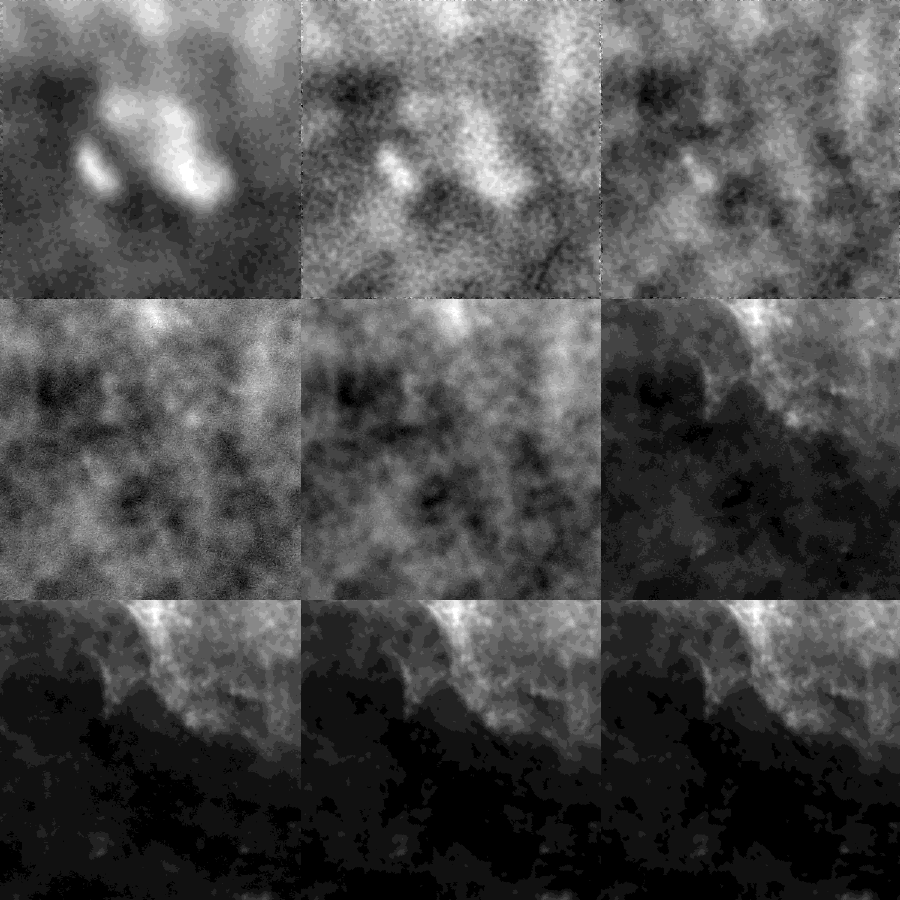}
\vskip 0.5em
\includegraphics[width=9cm]{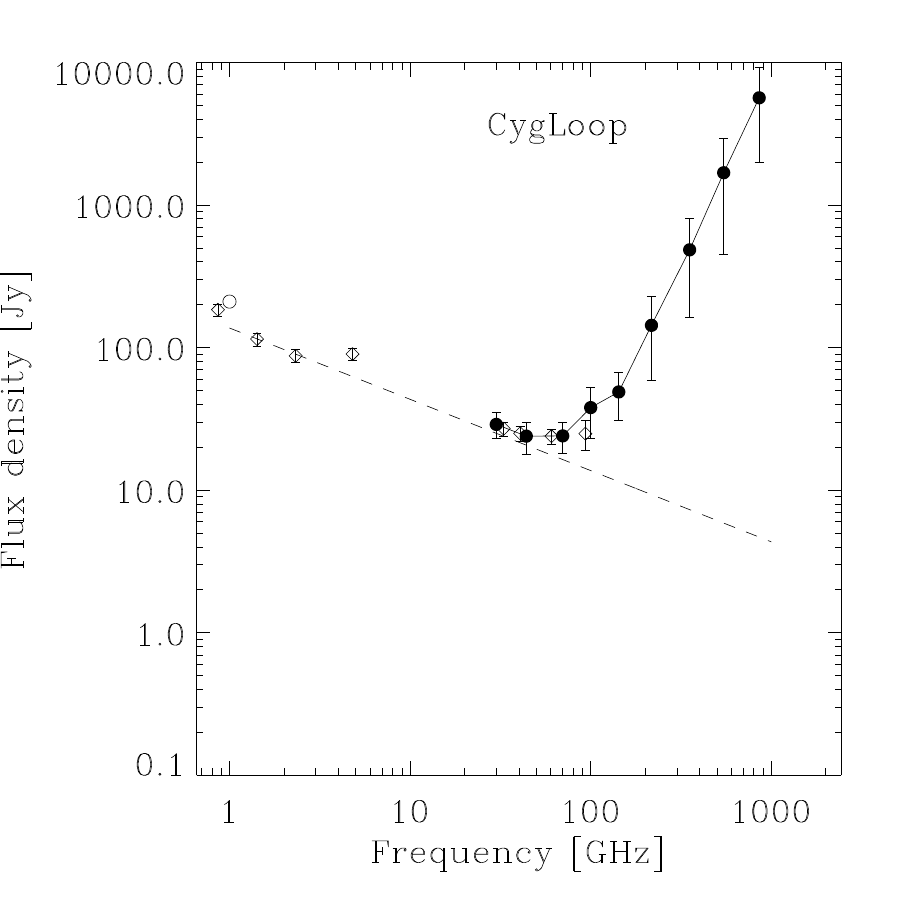}
\caption{
{\it Top\/}:  Images of the Cygnus Loop environment at the nine \Planck\ frequencies (Table 1), increasing from 30\,GHz at top left to 857\,GHz at bottom right. Each image is 435\arcm\ on a side, {\bfc with galactic coordinate orientation.}
The SNR is clearly visible in the lowest-frequency images, but a nearby star-forming region (to the north) confuses the SNR at frequencies above 100\,GHz.
{\it Bottom\/}: Spectral energy distribution of the Cygnus Loop including {\bfc fluxes measured by us from 
WMAP using the
same apertures as for {\it Planck}, and independent measurements from \citet{uyaniker04} and \citet{reich03}.
The dashed curve is a power law normalized through the radio fluxes.}
}
\label{cygloopsed}
\end{figure}

{\bfc For the purpose of illustrating the complete SED, we estimated the fluxes at all {\it Planck} frequencies.}
Figure~\ref{cygloopsed} shows the SED. The low-frequency emission is well matched by a power law with
spectral index $\alpha=0.46\pm 0.02$ from 1 to 60\,GHz. For comparison, a recent radio survey \citep{sun06,han12} measured the spectral index for the synchrotron emission from the Cygnus Loop and found a synchrotron index $\alpha=0.40\pm 0.06$, in excellent agreement with the \Planck\ results. There is no indication of a break in the power-law index all the way up to 60\,GHz (where dust emission becomes important). As discussed by \citet{han12}, previous indications of a break in the spectral index at much lower frequencies are disproved both by the 5\,GHz flux density they reported and now furthermore by the 30 and 44\,GHz flux densities reported here.

At high \planck\ frequencies, the images in Fig.~\ref{cygloopsed} show bright structured emission toward the upper right corner
(celestial E) and extending well outside the SNR; this unrelated emission is 
due to dust in molecular clouds illuminated by massive stars in the Cygnus-X region.
The bright, unrelated emission makes the measurement of high-frequency fluxes of the Cygnus Loop problematic; the
(heavily contaminated) aperture fluxes are well matched by a modified-blackbody fitted to the data with emissivity
index $\beta=1.46\pm 0.16$ and temperature $T_d=17.6\pm 1.9$ K, typical of molecular clouds.
The infrared emission from the SNR shell itself was detected at 60--100 $\mu$m by \citet{arendtcygnus}, who
derived a dust temperature of 31 K. The counterpart in the \planck\ images is not evident in Fig.~\ref{cygloopsed}.
Because the dust in the SNR shell is warmer than the ISM, we can use the 545 GHz emission as a tracer
of the unrelated emission and subtract it from the 857 GHz image to locate warmer dust. The subtracted image shows
faint emission that is clearly associated the shell with a surface brightness (after subtracting the scaled 545 GHz image) at 857 GHz of 0.2 MJy~sr$^{-1}$, where the 100 $\mu$m surface brightness is 8 MJy~sr$^{-1}$.  This surface brightness ratio can be explained
by a modified blackbody for dust with a temperature $T_d=31\pm 4$ K and an emissivity index $\simeq 1.5$, confirming
the temperature measurement by \citet{arendtcygnus}.

\subsection{HB\,21}

\begin{figure}
\includegraphics[width=9cm]{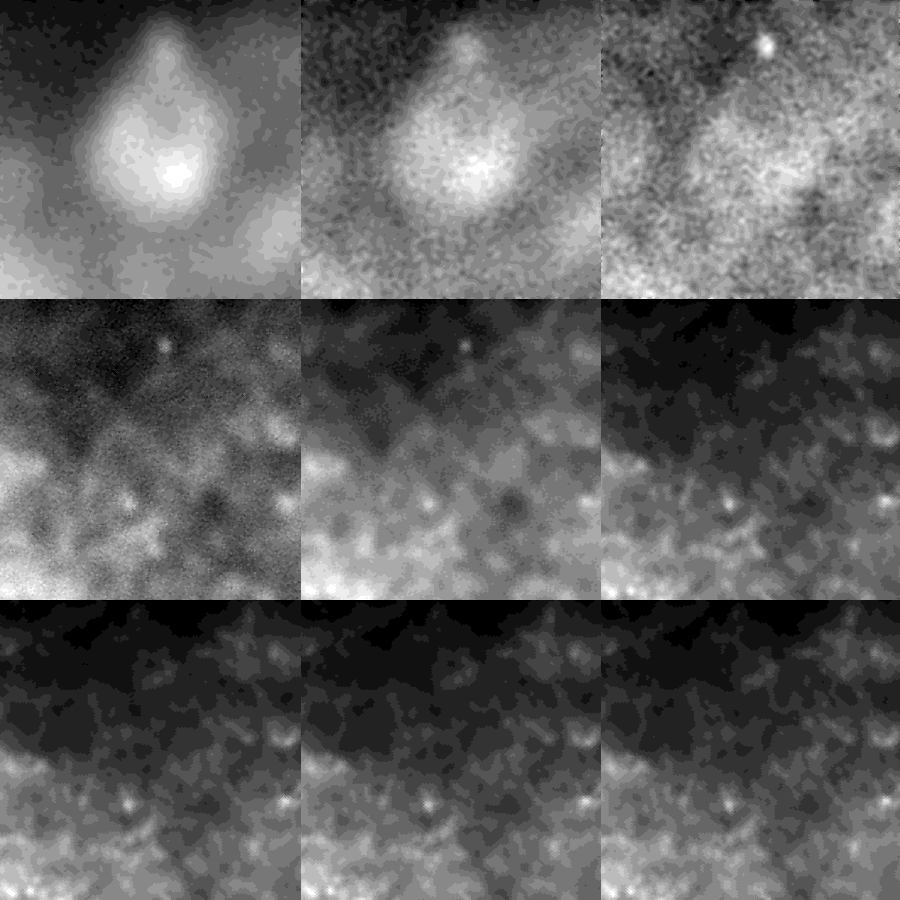}
\vskip 0.5em
\includegraphics[width=9cm]{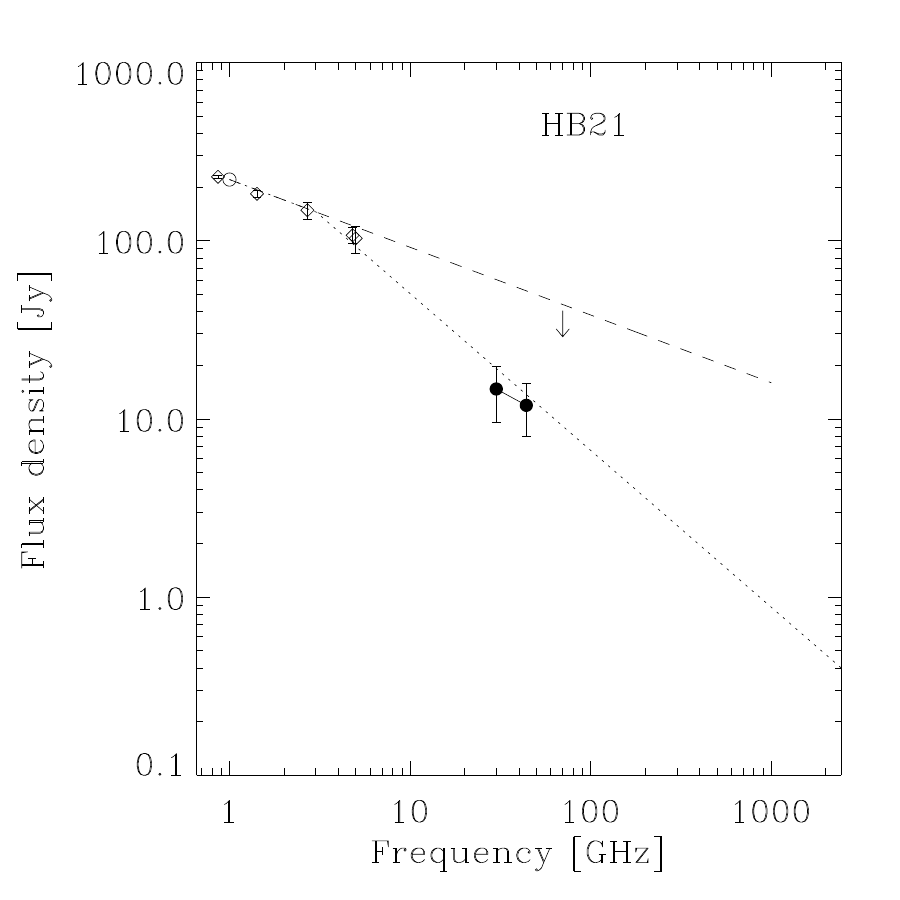}
\caption{
{\it Top\/}:  Images of the HB\,21  environment at the nine \Planck\ frequencies (Table 1), increasing from 30\,GHz at top left to 857\,GHz at bottom right.  Each image is {\bfc 233$'$ on a side, with galactic coordinate orientation.}
{\it Bottom\/}: Microwave SED of  HB\,21. Filled circles are \Planck\ measurements from this paper, with 3$\sigma$ error bars. The open symbol at 1\,GHz is from the Green catalogue, and the dashed line emanating from it is a power law with spectral index from the Green catalogue. 
The dotted line is a broken power-law model discussed in the text.
Open diamonds are radio flux densities {\bfc from  \citet{reich03}, \citet{willis73}, \citet{kothes06}, and \citet{gao11}}.
The downward arrow shows a flux density measurement that was contaminated by unrelated foreground emission.
}
\label{hb21sed}
\end{figure}

Figure~\ref{hb21sed} shows the microwave images and SED of the large SNR  HB\,21. The source nearly fills the displayed image
and is only clearly seen at 30 and 44\,GHz. A compact radio continuum source, 3C\,418, appears just outside the northern boundary of the SNR and has a flux density of approximately 2\,Jy at 30 and 44\,GHz.
The presence of this source at the boundary between the SNR and the background-subtraction annulus perturbs the flux density of the SNR by at most 1\,Jy, less than the quoted uncertainty.
The radio SED of HB\,21 is relatively shallow, with $\alpha=0.38$ up to 5\,GHz (Fig.~\ref{hb21sed}). The \Planck\ flux densities at 30 and 44\,GHz 
are significantly below the extrapolation of this power law. The target is clearly visible at both frequencies, and we confirmed that the flux densities are similar
in maps with and without CMB subtraction, so the low flux density does not appear to be caused by a cold spot in the CMB. 
\def\extra{
Measurement of the flux densities
for such large objects is difficult from ground-based telescopes, {\bfc  because they cannot cover large areas quickly enough to allow accurate total power 
measurements on large scales}. Measurements by  small telescopes over a range of frequencies up to 5\,GHz appear consistent \citep{kothes06}.
 Flux density measurement with \Planck\ should not be a problem on degree-sized scales; 
indeed this is the optimal range of design for the hardware and survey strategy. 
}

It therefore appears that the low flux densities in the microwave are due to a spectral break. 
For illustration, Fig.~\ref{hb21sed} shows a model where the spectral break is at 3\,GHz and the spectral index changes by $\Delta\alpha=0.5$ above that frequency.  
{\bfc An independent study of HB~21 found a spectral break at $5.9\pm 1.2$ GHz which is reasonably consistent with our results \citep{hewitt13}.}
This could occur if there have been significant energy losses since the time the high-energy particles were injected. This is probably the lowest frequency at 
which the break could occur and still be consistent with the radio flux densities. Even with this low value of $\nu_{\rm break}$, the {\Planck} flux densities are over predicted.
The spectral index change is likely to be greater than 0.5, which is possible for inhomogenous sources \citep{reynolds09}.  Given its detailed filamentary radio and optical morphology, HB\,21 certainly fits in this category.
The large size of the SNR indicates it is not young, so some significant cooling of the electron population is expected.

\subsection{Cas~A}

\begin{figure}
\includegraphics[width=9cm]{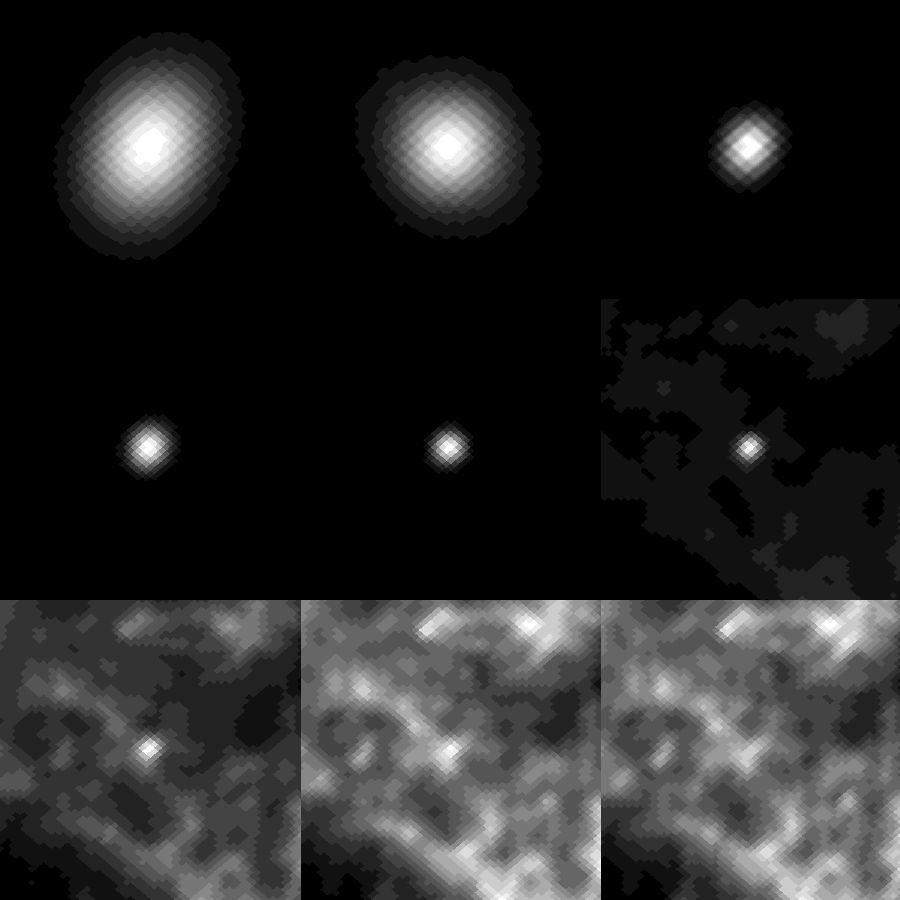}
\vskip 0.5em
\includegraphics[width=9cm]{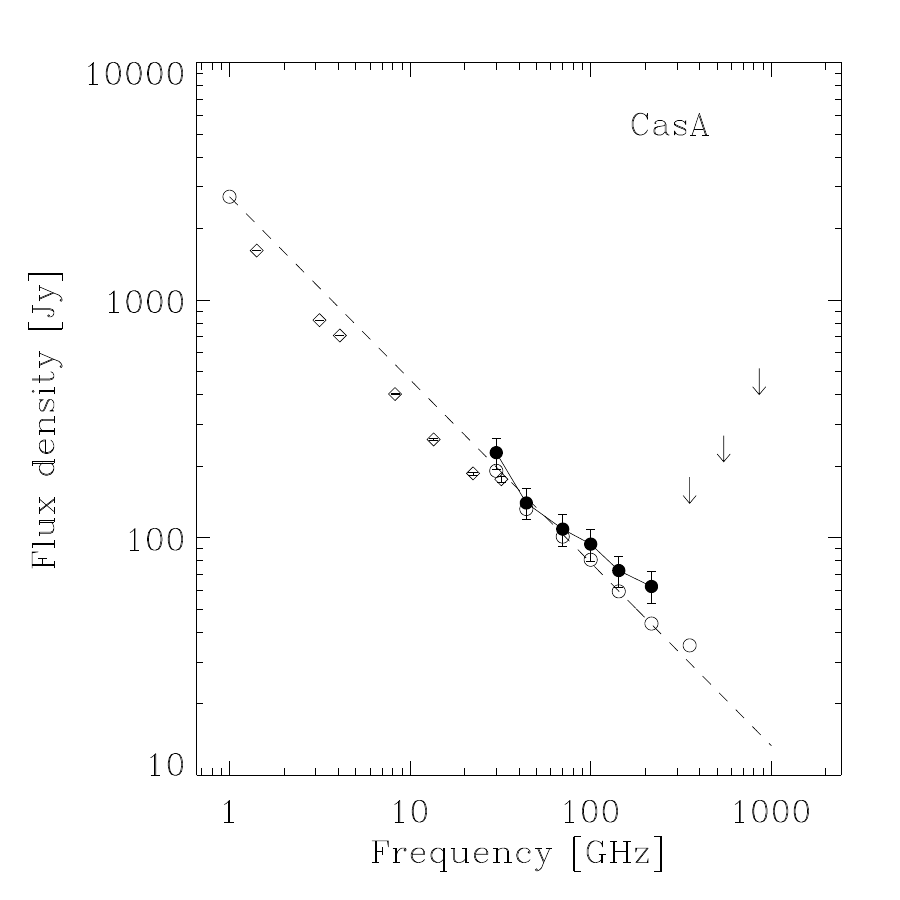}
\caption{
{\it Top\/}:  Images of the Cas~A environment at the nine \Planck\ frequencies (Table 1), increasing from 30\,GHz at top left to 857\,GHz at bottom right. Each image is $100\arcm$ on a side{\bfc, with galactic coordinate orientation.}
{\it Bottom\/}: Microwave SED of Cas~A. Filled circles are \Planck\ measurements from this paper, with 3$\sigma$ error bars. The open symbol at 1\,GHz is from the Green catalogue, and the dashed line emanating from it is a power law with spectral index from the Green catalogue. 
Open circles show the flux densities from the ERCSC,
{\bfc the compilation by \citet{baars} and \citet{mason99}}, after scaling for secular fading to the epoch of the \planck\ observations.
}
\label{casased}
\end{figure}

Cas~A is a very bright radio SNR, likely due to an historical supernova approximately 330\,yr ago \citep{ashworth80}.
In the \planck\ images (Fig.~\ref{casased}), Cas~A is a distinct compact source from 30--353\,GHz. Cas~A becomes confused with unrelated Galactic clouds at the highest
\planck\ frequencies (545 and 857\,GHz).
Inspecting the much higher-resolution images made with {\it Herschel}, it is evident that at
600\,GHz the cold interstellar dust would not be separable from the SNR \citep{barlow10}. Confusion with the bright and structured interstellar medium has made it difficult to assess the
amount of material directly associated with the SNR. Figure~\ref{casased} shows that from 30 to 143\,GHz, the \planck\ SED closely follows an
extrapolation of the radio power law. Cas~A was used as a calibration verification for the \planck\ Low-Frequency Instrument, and the flux densities at 30, 44, and 70\,GHz were shown to match
very well the radio synchrotron power law and measurements by WMAP \citep{planck2013-p02b}.

The flux densities at 217 and 353\,GHz appear higher than expected for synchrotron emission. We remeasured the excess at 353\,GHz using aperture photometry with
different background annuli, to check the possibility that unrelated Galactic emission is improperly subtracted and positively contaminates the flux density. 
At 353\,GHz, the flux density could be as low as 45\,Jy, which was obtained using a narrow aperture and background annulus centred on the source, or as high as 58\,Jy, obtained from a wide aperture
and background annulus, so we estimate a flux density and systematic uncertainty of $52\pm 7$ Jy. The flux density in the ERCSC is $35.2\pm 2.0$, which falls
close to the radio synchrotron extrapolation.  The excess flux density measured above the synchrotron prediction is $22\pm 7$, using the techniques in this paper, which is  significant. 
This excess could potentially be due to a coincidental peak in the unrelated foreground emission or to cool dust in the SNR, which is marginally resolved by \planck. Images at the
lowest frequency (600\,GHz) observed with  the {\it Herschel} Space Observatory \citep{barlow10} at much higher angular resolution than \Planck\ show that the
non-synchrotron microwave emission is a combination of both cold interstellar dust and freshly formed dust.

\subsection{Tycho}

\begin{figure}
\includegraphics[width=9cm]{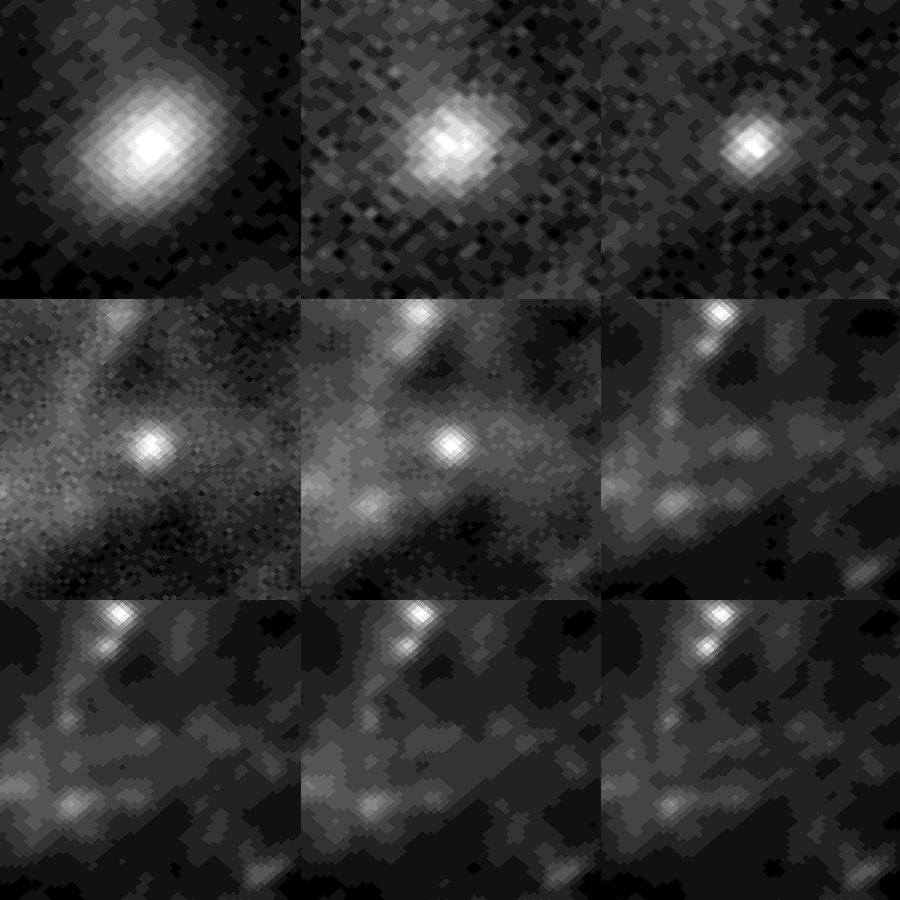}
\vskip 0.5em
\includegraphics[width=9cm]{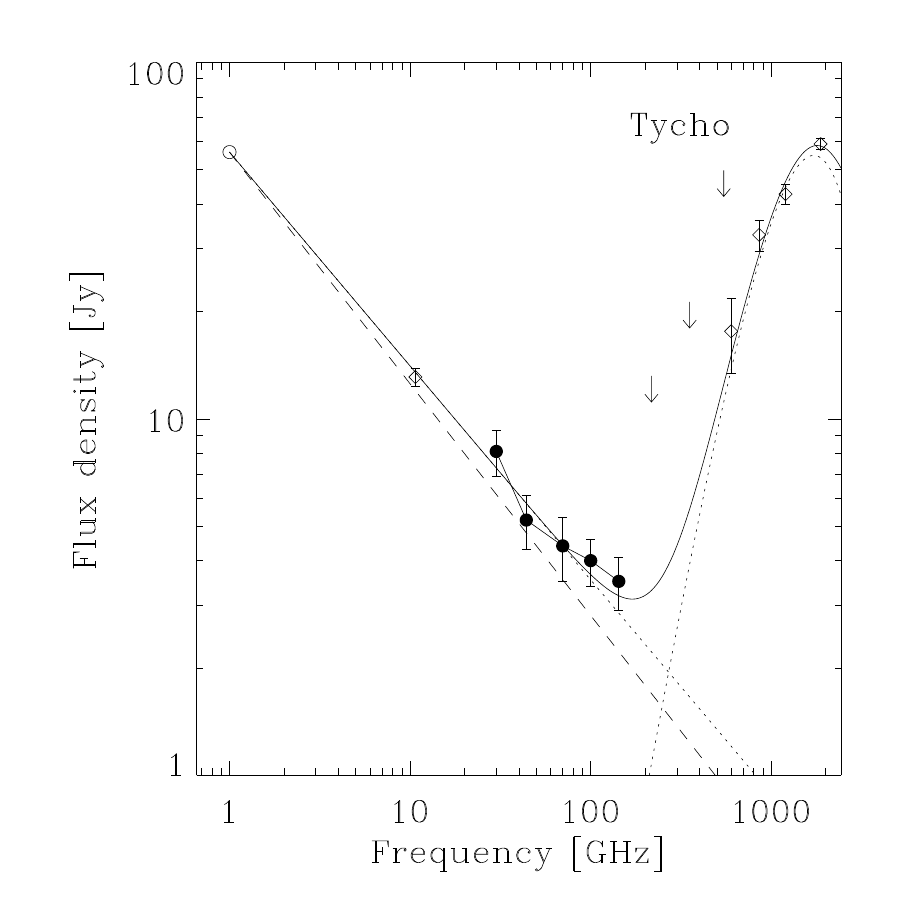}
\caption{
{\it Top\/}:  Images of the Tycho SNR  environment at the nine \Planck\ frequencies (Table 1), increasing from 30\,GHz at top left to 857\,GHz at bottom right.  Each image is 100\arcm\ on a side {\bfc, with galactic coordinate orientation.}
{\it Bottom\/}: Microwave SED of Tycho. Filled circles are \Planck\ measurements from this paper, with 3$\sigma$ error bars. The open symbol at 1\,GHz is from the Green catalogue, and the dashed line emanating from it is a power law with spectral index from the Green catalogue. Open diamonds are a 10.7\,GHz measurement from \citet{klein79} and {\it Herschel} flux density from \citet{gomez12}.
The dotted lines are a revised synchrotron power law with spectral index $\alpha=0.6$ at low frequency, based on the radio including 10.7\,GHz and the \planck\ 30--143\,GHz flux densities, and a modified blackbody with emissivity index $\beta=1$ and temperature 21\,K at high frequency, based on the {\it Herschel} data. Downward arrows are upper limits (99.5\% confidence) to the flux density 
from high-frequency \planck\ data. 
The solid line is a combination of the synchrotron and dust models.
}
\label{tychosed}
\end{figure}

The Tycho SNR comprises ejecta from a supernova explosion in 1572 AD. \planck\ detects the SNR from 30 to 545\,GHz and tentatively at 857\,GHz, as can be seen in the images in Fig.~\ref{tychosed}.
The SED shows at least two emission mechanisms. A continuation of the synchrotron spectrum dominates the 30--143\,GHz flux density, while a steeply-rising contribution, most likely due to dust, dominates above 143\,GHz. The SNR is well-detected by {\it Herschel} \citep{gomez12}, with its excellent spatial resolution. 
The {\it Herschel\/} image shows that cold dust emission is predominantly {\it outside\/} the boundary of the {\bfc ejecta-dominated portion of the} SNR.  While some is correlated with unrelated molecular clouds, much of 
this emission is attributed to swept-up ISM (as opposed to ejecta). From the {\it Herschel\/} study, the dust emission was fit with a cold component at temperature 21\,K. The emissivity index
of the modified blackbody that fits the {\it Herschel} and \planck\ SED is $\beta=1$. (The value of the emissivity index we report, $\beta=1$, is different from
the $\beta=1.5$ described by \citealt{gomez12} in the text of their paper, but the cold dust model plotted in their Fig.~12 appears closer to $\beta\simeq1$, in agreement with our Fig.~\ref{tychosed}.)

\subsection{3C\,58\label{c58sec}}

\begin{figure}
\includegraphics[width=9cm]{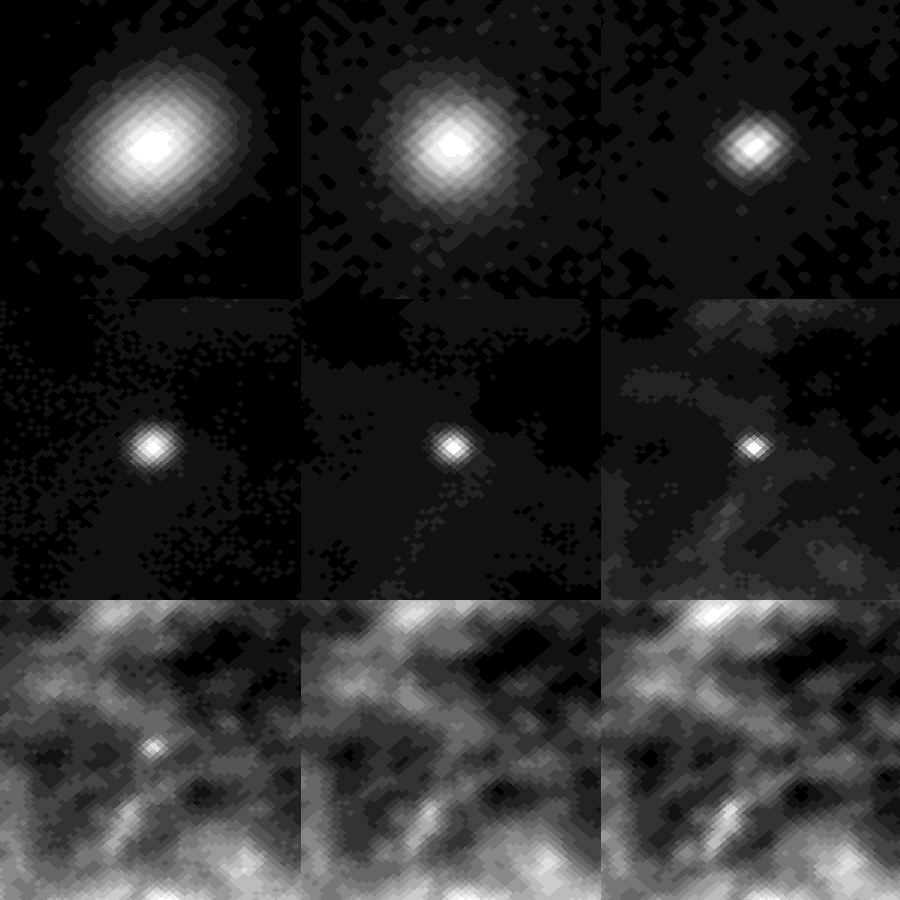}
\vskip 0.5em
\includegraphics[width=9cm]{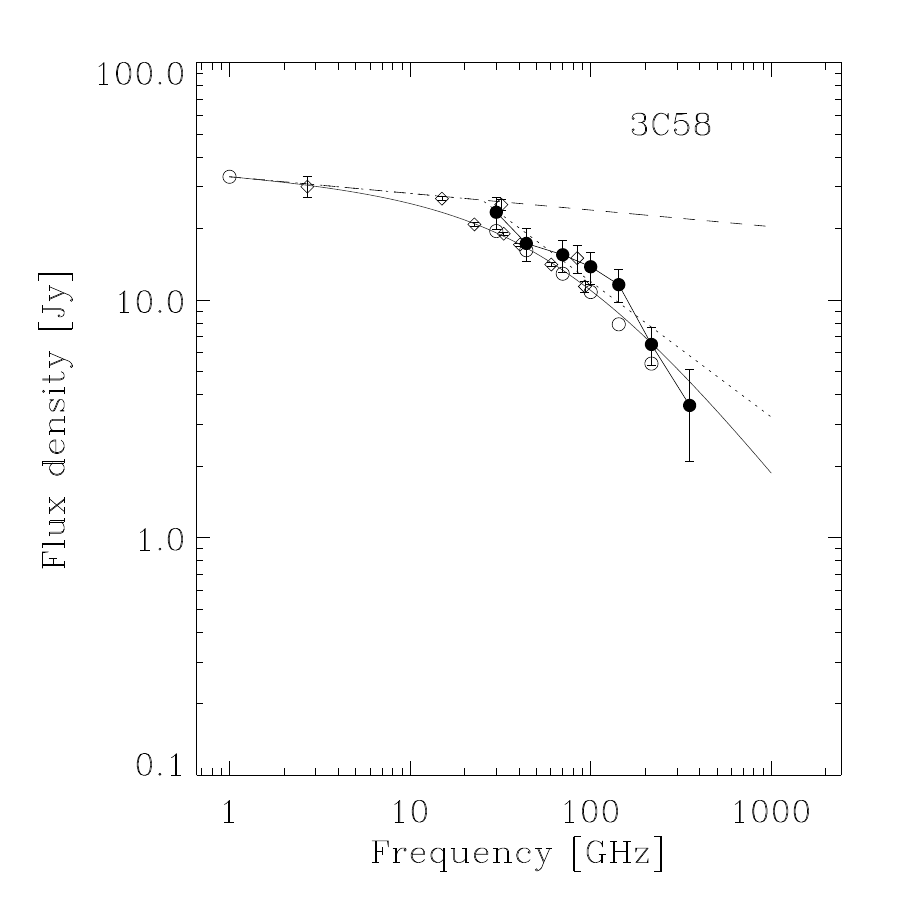}
\caption{
{\it Top\/}:  Images of the 3C\,58 environment at the nine \Planck\ frequencies (Table~1), increasing from 30\,GHz at top left to 857\,GHz at bottom right.  Each image is $100\arcm$ on a side,{\bfc  with galactic coordinate orientation.}
{\it Bottom\/}: Microwave SED of  3C\,58. Filled circles are \Planck\ measurements from this paper, with 3$\sigma$ error bars. The open symbol at 1\,GHz is from the Green catalogue, and the dashed line emanating from it is
a power law with spectral index from the Green catalogue. Open circles are from the \planck\ ERCSC.
Diamonds show published flux densities: \citet{salter89} measured $15\pm 2$ Jy at 84.2\,GHz; \citet{morsi87} measured $25.2\pm 1.4$ Jy at 32\,GHz; \citet{green75} measured $26.7\pm 0.5$ Jy at 15\,GHz; and 
\citet{green86} measured $30\pm 3$ Jy at 2.7\,GHz. Flux densities from \citet{weiland11} are also included as open diamonds, together with a solid line showing a fit to their data.
The dotted line is the broken power-law model for a 
pulsar wind nebula with energy injection spectrum matching the radio (low-frequency) data and 
a break at 25\,GHz due to synchrotron cooling (see Sect.~\ref{c58sec}).
}
\label{c58sed}
\end{figure}

3C\,58 is a compact source at \Planck\ frequencies, with a $9\arcm \times 5\arcm$ radio image size, so it becomes gradually resolved at frequencies above 100\,GHz. The adopted aperture photometry and correction procedure in this survey should recover the entire flux density of this SNR at all frequencies. Figure~\ref{c58sed} shows the SED. 
The flux density at 30\,GHz is just a bit lower than that predicted by extrapolating the 1\,GHz flux density with the radio spectral index. However, it is clear 
from the \Planck\ data that the brightness declines much more rapidly than predicted by an extrapolation of the radio SED.

The \Planck\ flux densities are in agreement with the low 84\,GHz flux density that had been previously measured by \citet{salter89}, who noted that their flux density measurement was $3\sigma$ below an extrapolation from lower frequencies. 
The \Planck\ data show that the decline in flux density, relative to an extrapolation from lower frequencies, begins around or before 30\,GHz and continues to at least 353\,GHz.

\begin{figure}
\includegraphics[width=10cm,height=10cm]{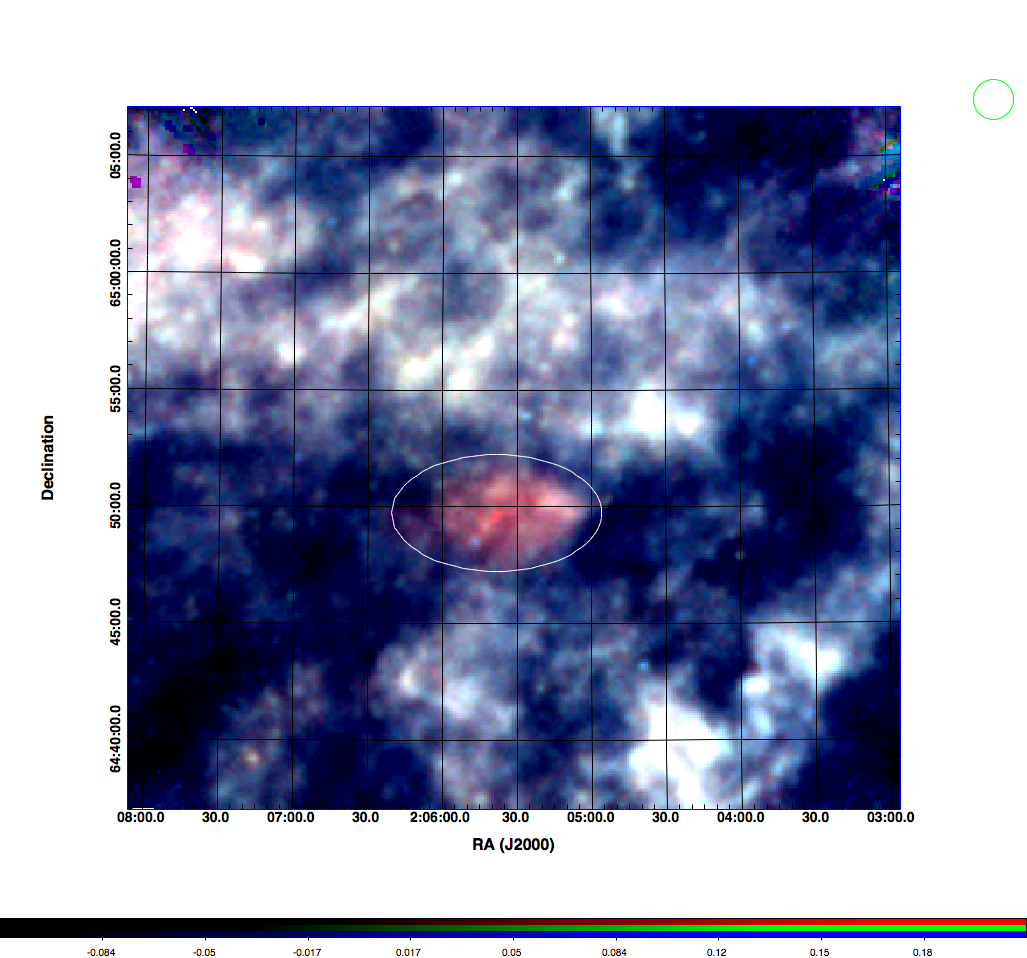}
\caption{
Multi-colour {\it Herschel}/SPIRE image of 3C\,58 created from images at 510\micron\ (587 GHz; red), 350\micron\ (857 GHz; green) and 250\micron\ (1200 GHz; blue). A white ellipse of $9'\times 5'$ diameter shows the size of the radio SNR. The diffuse interstellar medium is mostly white in this image, while the emission within the ellipse is distinctly red, being brightest at 510\micron\ while the diffuse interstellar medium is brightest at 250 $\mu$m. The red, compact source at the centre of the ellipse is coincident with PSR J0205+6449 that is thought to be the compact remnant of the progenitor star.
}
\label{c58herschel}
\end{figure}

There is a large gap in frequencies between the microwave and near-infrared detections of the pulsar wind nebula.
To fill in the SED, we reprocessed the {\it IRAS} survey data using the HIRES algorithm \citep{aumann90}, and we find a tentative detection of the SNR at 60\micron\ with a flux density of 0.4\,Jy.  At 100\micron\ the SNR is expected to be brighter, but is confused with the relatively-brighter diffuse interstellar medium.
In the far-infrared, we obtained the {\it Herschel}/SPIRE map, created for 3C\,58 as part of a large guaranteed-time key project led by M. Groenewegen, from the {\it Herschel} Science Archive. 
Figure~\ref{c58herschel} shows a colour combination of the SPIRE images. The SNR is clearly evident due to its distinct colours with respect to the diffuse ISM. In addition to the diffuse 
emission of the SNR, there is a compact peak in the emission at the location of the pulsar J0205+6449. We measure the flux density of the SNR to be $2.1\pm 0.4$\,Jy using an elliptical aperture of the same size as the radio image, and we measure the compact peak from the pulsar to be $0.05\pm 0.02$\,Jy using a small circular aperture centred on the pulsar with a background region surrounding the pulsar and within the SNR.
{\bfc Table~\ref{c58flux} summarizes the flux densities for 3C 58 from radio through infrared frequencies. For the {\it Planck} fluxes, because the source is compact, a conservative
uncertainty estimate of 10\% is included.}

{\bfc 
\begin{table}
\caption{Flux densities for 3C 58}
\label{c58flux}
\tiny
\setbox\tablebox=\vbox{
\newdimen\digitwidth
\setbox0=\hbox{\rm 0}
\digitwidth=\wd0
\catcode`*=\active
\def*{\kern\digitwidth}
\newdimen\signwidth
\setbox0=\hbox{+}
\signwidth=\wd0
\catcode`!=\active
\def!{\kern\signwidth}
%
\halign{\hbox to 0.7in{#}\tabskip=1em&
    \hfil#\hfil\tabskip=1em& 
    \hfil#\hfil\tabskip=1em& 
    \hfil#\hfil\tabskip=1em& 
    \hfil#\hfil\tabskip=0pt\cr
\noalign{\doubleline}
Frequency& Flux & Uncertainty & Reference\cr
[GHz]    & [Jy]   & [Jy]              & \cr
\noalign{\vskip 3pt\hrule\vskip 5pt}
****0.151  &       36.0 &    4.0 & \citet{green86}\cr
****0.327  &       33.9 &    3.0 & \citet{bietenholz08}\cr
****0.408  &       32.2 &    2.0 & \citet{kothes06}\cr
****1.42   &       31.9 &    1.0 & \citet{kothes06}\cr
****2.7    &       30** &      3** & \citet{green86}\cr
***15      &       26.7 &   0.5 & \citet{green75}\cr
***30 & 22.2 & 2.2 & this paper ({\it Planck})\cr
***32      &       25.2 &   1.4 & \citet{morsi87}\cr
***44 & 16.4 & 1.6 &this paper ({\it Planck})\cr
***70 & 14.2 & 1.4 &this paper ({\it Planck})\cr
***84.2   &       15** &     2.0 & \citet{salter89}\cr
**100 & 12.7 & 1.3 & this paper ({\it Planck})\cr
**143 & 10.8 & 1.1 & this paper ({\it Planck})\cr
**217 & *8.4 & 0.8 & this paper ({\it Planck})\cr
**353 & *4.8 & 0.7 & this paper ({\it Planck})\cr
**588 & *2.1 & 0.4 & this paper ({\it Herschel})\cr
*5000 & **0.37 & *0.15 & this paper ({\it IRAS})\cr
66666 & ***0.071 &  **0.028 & \citet{slane08}\cr
83333 & ***0.028 &  **0.012 & \citet{slane08}\cr
\noalign{\vskip 5pt\hrule\vskip 3pt}}}
\endPlancktable
\end{table}

}

3C\,58 is a pulsar wind nebula, like the Crab. The shallow spectral index ($S_\nu\propto \nu^{-\alpha}$ with $\alpha=0.07$) at low frequencies is due to injection of energetic particles into the nebula from the pulsar with an energy spectrum $dN/dE\propto E^{-s}$, with $s=2\alpha+1=1.14$. The electrons cool (become non-relativistic), which leads to a steepening of the spectrum, $\alpha\rightarrow \alpha+0.5$ \citep{reynolds09}, above a break frequency that depends upon the magnetic field and age. 
{\bfcc As an alternative to the idea of electron cooling, breaks in the synchrotron spectral index could also simply reflect breaks in the energy distribution of the relativistic particles injected into the pulsar wind nebula. Determining the location of the microwave spectral break was deemed `crucial, as are flux measurements in the sub-mm band' \citep{slane08}. The present paper provides both such measurements.}

\begin{figure}
\includegraphics[width=9cm]{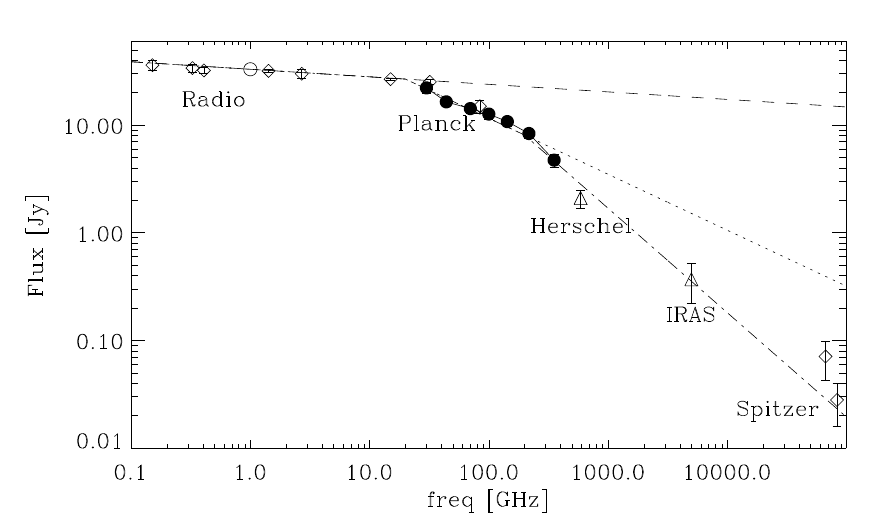}
\caption{
Microwave SED of  3C\,58 from the radio through infrared. Data are as in Fig.~\ref{c58sed} with the addition of lower-frequency radio flux densities \citep{greencatalog} and higher-frequency {\it IRAS} and {\it Spitzer} flux densities, and omission of the ERCSC flux densities.  The dashed line is the power-law fit to the radio flux densities;  the dotted line is the pulsar wind nebula model; and the dash-dotted line is that model with the addition of a second break with spectral index 0.92 above 200\,GHz.
}
\label{c58plot}
\end{figure}

The \Planck\ flux densities match the electron-cooling model for a break frequency around 20\,GHz. 
Figure~\ref{c58plot} shows a wider-frequency version of the SED for 3C\,58.
The existence of a break in the radio spectrum was previously indicated by a measurement at 84\,GHz \citep{salter89} that is now confirmed by the \Planck\ data 
with six independent measurements at frequencies from 30 to 217\,GHz. The {\it Wilkinson Microwave Anisotropy Probe} also clearly shows the decrease in flux density 
at microwave frequencies \citep{weiland11}, and the curved spectral model from their paper is at least as good a fit to the data as the broken power law.
At higher frequencies, the {\it IRAS} and {\it Spitzer} measurements are significantly over-predicted by the extrapolation of the microwave SED that matches the \Planck\ data.
\citet{green92} had already used the {\it IRAS} upper limit to show the synchrotron spectral index must steepen before reaching the far-infrared.
There may be, as indicated by \citet{slane08}, a second spectral break. We found that a good match to the radio through infrared SED can be made with a second
break frequency at 200\,GHz. The nature of this second break frequency is not yet understood.

\subsection{IC\,443\label{ic443sec}}

\begin{figure}
\includegraphics[width=9cm]{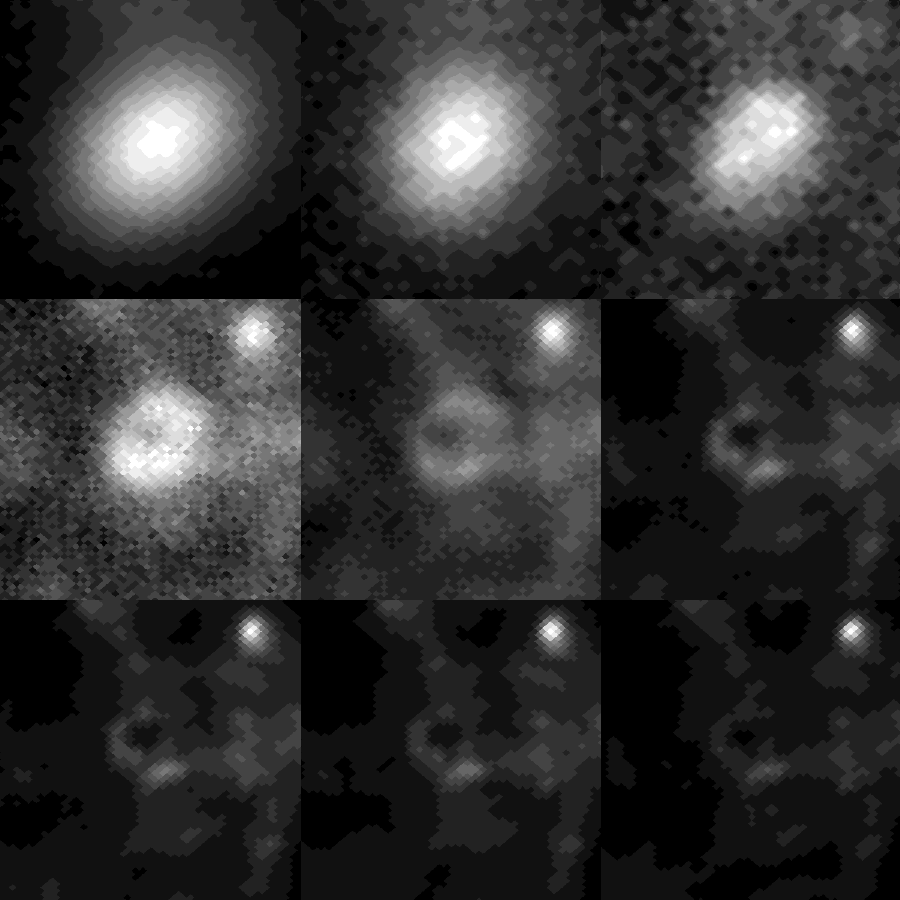}
\vskip 0.5em
\includegraphics[width=9cm]{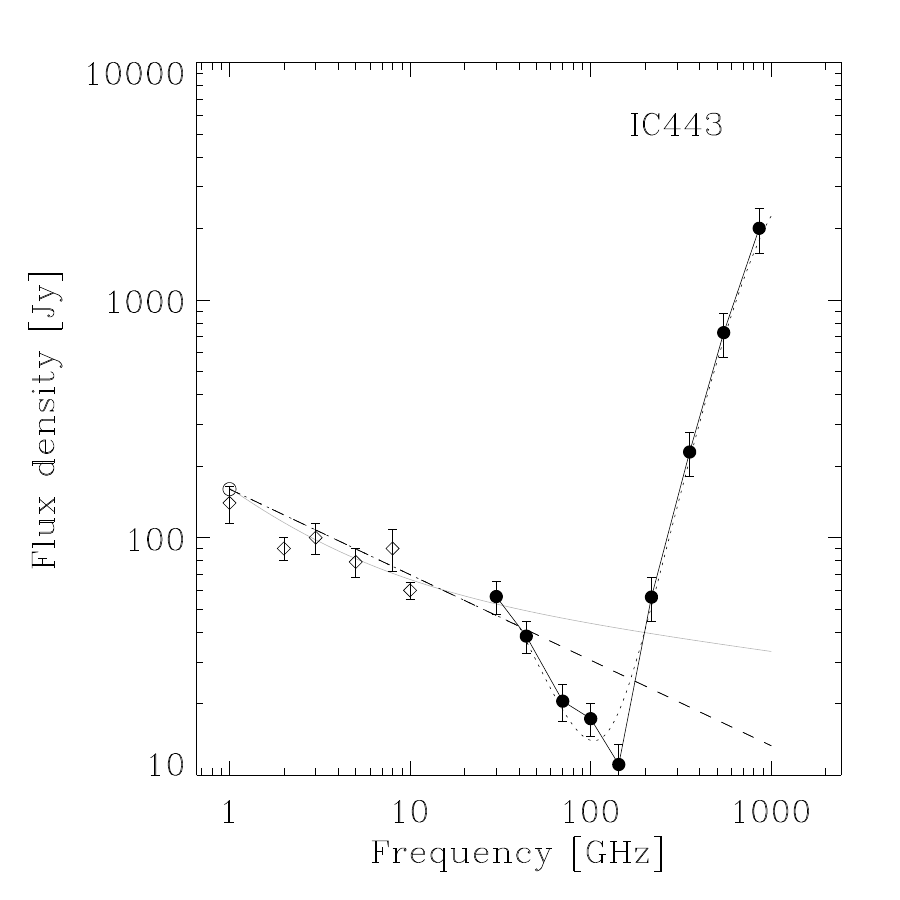}
\caption{
{\it Top\/}:  Images of the IC\,443  environment at the nine \Planck\ frequencies (Table 1), increasing from 30\,GHz at top left to 857\,GHz at bottom right. Each image is $100'$ on a side{\bfc, with galactic coordinate orientation.}
{\it Bottom\/}: Microwave SED of IC\,443. Filled circles are \Planck\ measurements from this paper, with 3$\sigma$ error bars. The open symbol at 1\,GHz is from the Green catalogue, and the dashed line emanating from it is
a power law with spectral index from the Green catalogue. 
Open diamonds are other radio flux densities from the compilation by \citet{castelletti11}.
The dotted line is a model (dust plus broken-power-law synchrotron) discussed in the text in \S\ref{ic443sec}.
The solid grey curve is an extrapolation from \citet{onic12}.
}
\label{ic443sed}
\end{figure}

The SNR IC\,443 is well known from radio to high-energy astrophysics, due to the interaction of a blast wave with both low and high-density material. The SNR is detected at all nine  
\Planck\ frequencies. 
The utility of the microwave flux densities is illustrated by comparing the observations of IC\,443 to extrapolations based only on lower-frequency data. In one recent paper,
there is a claim for an additional emission mechanism at 10\,GHz and above, with thermal free-free emission contributing at a level comparable to or higher than 
synchrotron radiation \citep{onic12}. The highest-frequency data point considered in that paper was at 8\,GHz. In Fig.~\ref{ic443sed} it is evident that the proposed
model including free-free emission is no better fit than a power-law scaling of the 1\,GHz flux density using the spectral index from the Green catalogue. The 10.7\,GHz and 30\,GHz flux densities are consistent with a single power law from 1--30\,GHz, with the single outlier at 8\,GHz being little more than 1$\sigma$ away.
Both the model of \citet{onic12} and the single power law predict the flux density at 30\,GHz perfectly, but they over predict the higher-frequency emission at 40 and143\,GHz by factors of more than
3 and 2, respectively.  Therefore, the shape of the IC\,443 SED requires a {\it dip} in emissivity in the microwave, rather than an excess due to free-free emission.
The decrease in flux density could be due to a break in the synchrotron power law from the injection mechanism of the energetic particles, or due to cooling losses by the
energetic particles.

At the higher \planck\ frequencies, another emission mechanism besides synchrotron radiation  dominates the brightness. 
The horseshoe-shaped eastern portion of the SNR is evident with similar morphology at 100 GHz and higher frequencies. The southern
part of the SNR, where shocks are impacting a molecular cloud \citep{rho01}, is physically distinct; this region
becomes significantly brighter at frequencies of 217 GHz and above.
While uncertainty in background subtraction 
makes an accurate flux difficult to measure, the fluxes were estimated all the way to 857 GHz for the purpose of Fig.~\ref{ic443sed}.
Based on the steep rise to higher frequency, and the similarity between the image at high and low frequencies, the higher-frequency emission is likely due to dust
grains that survive the shock. Contamination by unrelated Galactic plane emission is significant at these frequencies. Inspecting Fig.~\ref{ic443sed}, it is evident that the SNR
is well detected even at the highest \planck\ frequencies, because it is well resolved  and its spatial pattern can be recognised in the images. To determine whether the 857\,GHz flux density is due to the SNR or unrelated 
emission, we inspect the image along the ring of the SNR, finding an average brightness of 7\,K$_{\rm cmb}$ above the surrounding background,
which is equivalent to an average surface brightness of 21\,MJy~sr$^{-1}$. The solid angle of the ring is $2\pi \theta\Delta\theta$, with $\theta=80'$ being the radius of the SNR and $\Delta\theta\simeq 5'$ being the thickness of the ring. The total flux density from this rough estimate is 3000\,Jy, which is in order-of-magnitude agreement with the flux density shown in Fig.~\ref{ic443sed}. This affords
some quality verification of the measured flux density of the SNR dust emission.

The shape of the SED across radio and microwave frequencies can be reasonably approximated by a combination of synchrotron and dust emission.
A fit to the SED shows the dust emission has a temperature of $T=16$\,K with emissivity index $\beta=1.5$, 
where the dust emission depends on frequency as $\nu^{-\beta} B_\nu(T)$, and $B_\nu(T)$ is the Planck function
at temperature $T$.
The precise values are not unique and require combination with infrared data and
multiple temperature components given the complex mixture of dust in molecular, atomic, and shocked gas; dust dominates at frequencies above 140\,GHz.
The synchrotron emission follows a power law with spectral index of 0.36 from radio frequencies up to 40\,GHz, after which the spectral index steepens to 1.5.
The increase in spectral index is what makes IC\,443 relatively faint in the range 70--143\,GHz, compared to what might be expected from an extrapolation of
the radio power law. Possible causes for spectral breaks are discussed for other SNRs in sections that follow.

\subsection{Puppis A}

\begin{figure}
\includegraphics[width=9cm]{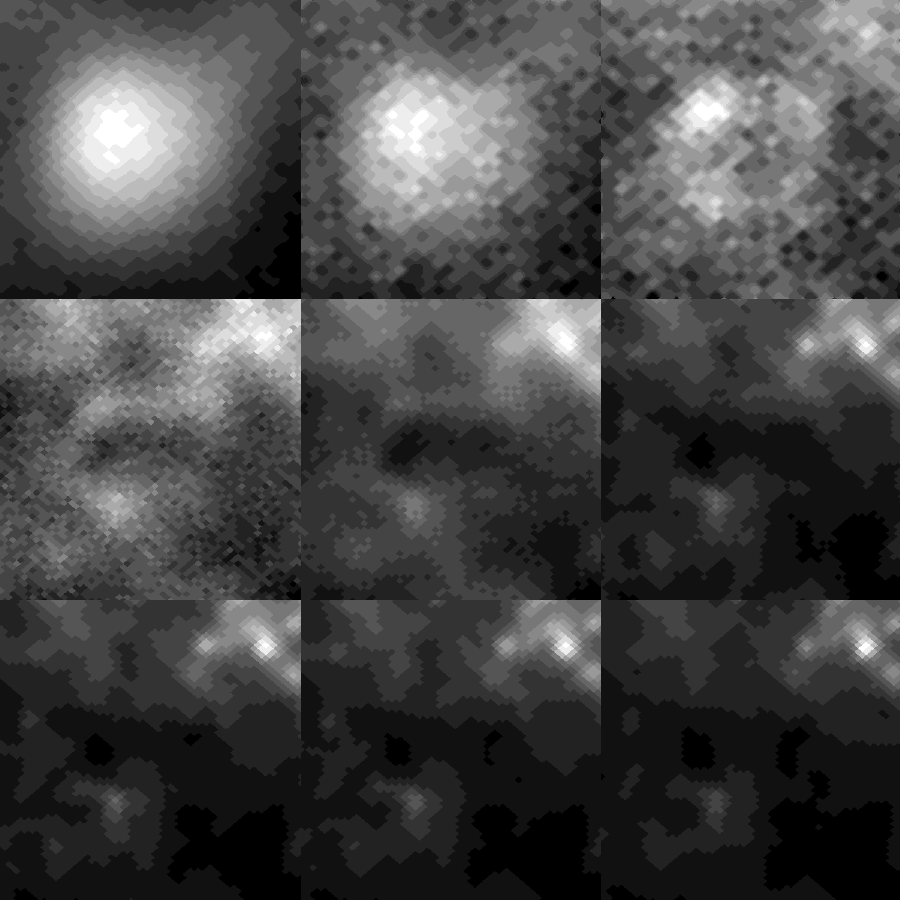}
\vskip 0.5em
\includegraphics[width=9cm]{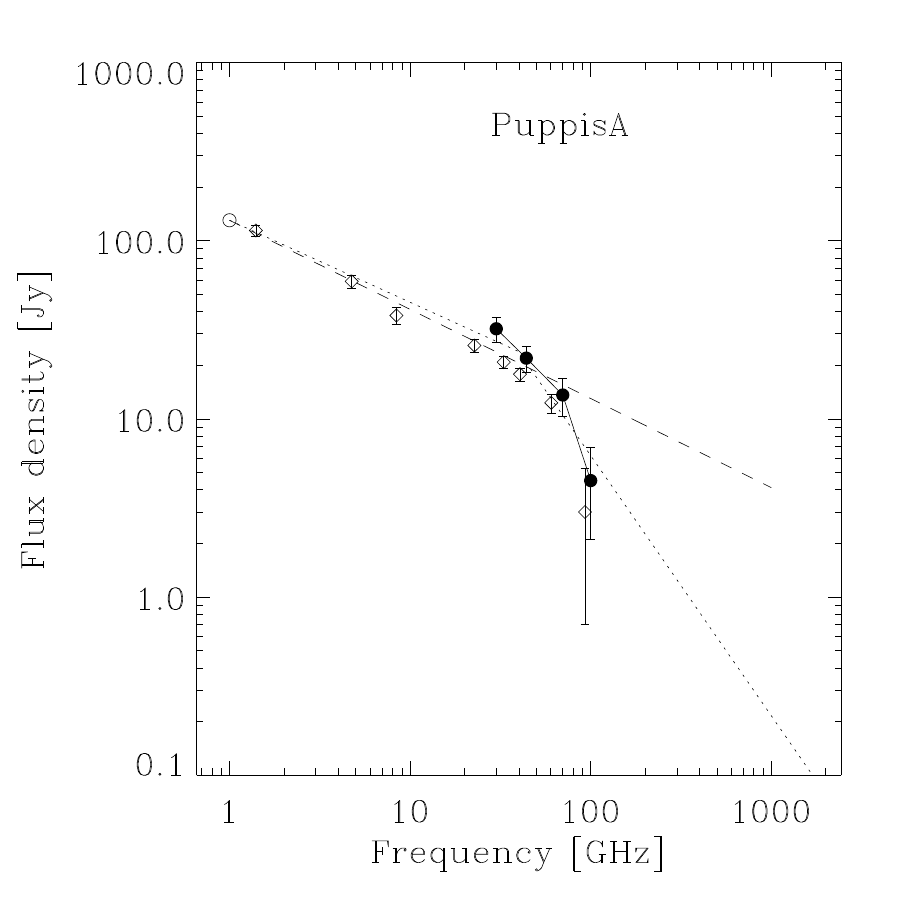}
\caption{
{\it Top\/}:  Images of the Puppis A  environment at the nine \Planck\ frequencies (Table 1), increasing from 30\,GHz at top left to 857\,GHz at bottom right.  Each image is $100'$ on a side{\bfc, with galactic coordinate orientation.}
{\it Bottom\/}: Microwave SED of  Puppis A. Filled circles are \Planck\ measurements from this paper, with 3$\sigma$ error bars. The open symbol at 1\,GHz is from the Green catalogue, and the dashed line emanating from it is a power law with spectral index from the Green catalogue. 
Open diamonds are radio flux densities from the \citet{milne93} and {\it WMAP} fluxes from
\citet{hewitt12}.
}
\label{puppisased}
\end{figure}

The  bright SNR Puppis A is evident in the \planck\ images at 30--143\,GHz. The flux densities measured from the \planck\ images at 30--70\,GHz roughly follow an extrapolation of the radio power law, suggesting the emission mechanism is synchrotron emission.
The 100\,GHz measurement is below the power law.
While this measurement is challenging due to contamination from unrelated emission evident in images at higher frequencies, the flux density at 100\,GHz does appear lower than the power law. A simple fit of a broken
power law in Fig.~\ref{puppisased} has a break frequency $\nu_{\rm break}=40$\,GHz above which the spectral index increases from 0.46 to 1.46.
{\bfc This fit is consistent with results from a study of WMAP observations \citep{hewitt12}.}

\subsection{Vela}

\begin{figure}
\includegraphics[width=9cm]{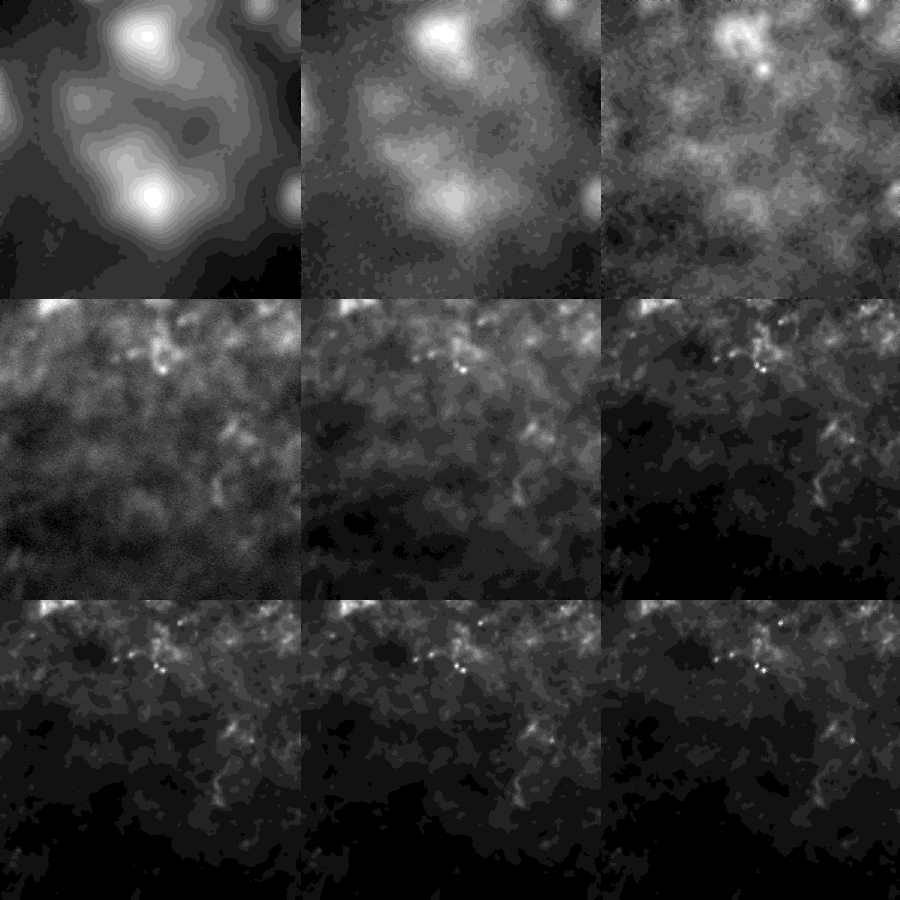}
\vskip 0.5em
\includegraphics[width=9cm]{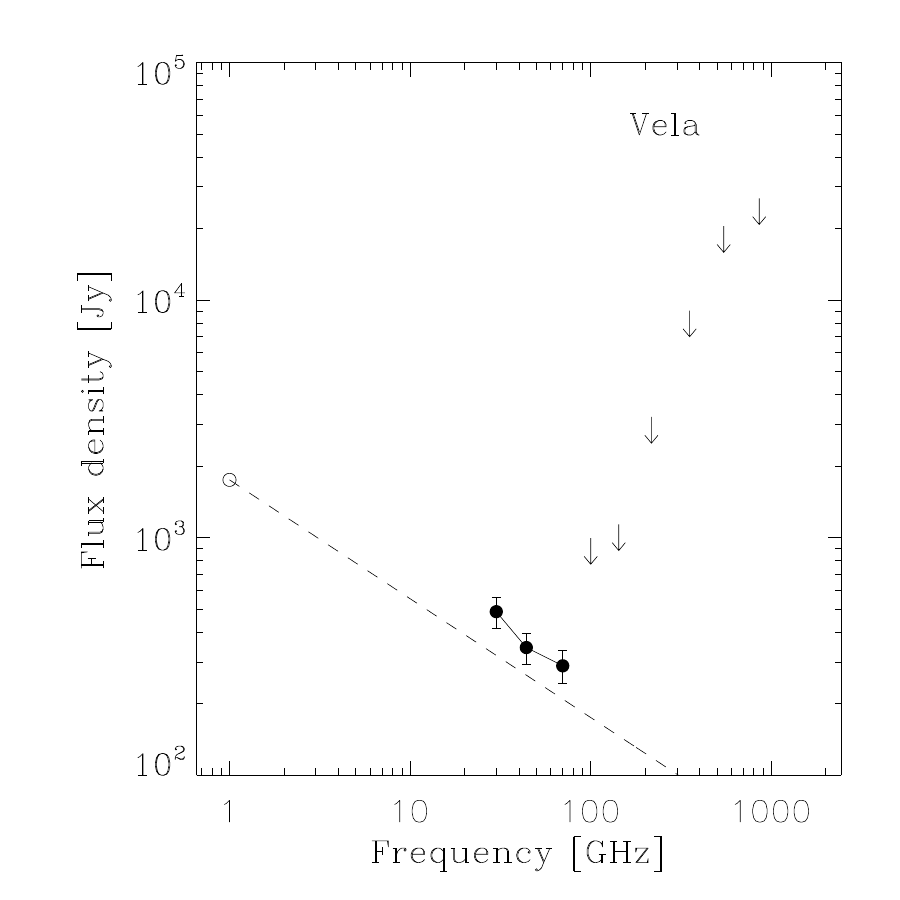}
\caption{
{\it Top\/}:  Images of the Vela  environment at the nine \Planck\ frequencies (Table 1), increasing from 30\,GHz at top left to 857\,GHz at bottom right.  Each image is $400'$ on a side{\bfc, with galactic coordinate orientation, centred $1^\circ$ north of Vela-X.}
{\it Bottom\/}: Microwave SED of  Vela. Filled circles are \Planck\ measurements from this paper, with 3$\sigma$ error bars. The open symbol at 1\,GHz is from the Green catalogue, and the dashed line emanating from it is a power law with spectral index from the Green catalogue. 
Downward arrows show the \planck\ high-frequency flux density measurements that are contaminated by unrelated foreground emission.
}
\label{velased}
\end{figure}

The very large Vela SNR is prominent in the lowest-frequency \planck\ images; 
{\bfc Figure~\ref{velased} shows a well-resolved shell even at 30 GHz. 
The centre of the image is shifted (by 1$^\circ$ upward in galactic latitude) from the Green catalogue position, so as to include the entire SNR shell. 
The object at the right-hand edge of the lower-frequency panels of Figure~\ref{velased} is actually the previous SNR from the survey, Puppis A. 

At frequencies above 70 GHz, the SNR is confused with unrelated emission from cold molecular clouds and cold cores. However, some 
synchrotron features remain visible at higher frequencies.
The relatively bright feature near the lower centre of the image is Vela-X, the bright nest part of the radio SNR. This feature can be traced all the way to 353 GHz, with
contrast steadily decreasing at higher and higher frequencies. 
Some dust-dominated features are visible at low frequencies. The bright feature at the centre-top of the low-frequency panels of Figure~\ref{velased} overlap
with the 
very prominent set of knots and filaments in the high-frequency images, with a steadily {\it decreasing} brightness for the knots and filaments. At frequencies
above 100 GHz, the dust features dominate here, while at 30--70 GHz the synchrotron from the SNR dominates.
}

The 30--70\,GHz \planck\ flux densities
follow an extrapolation of the radio power law with slightly higher spectral index, indicating the microwave emission mechanism is synchrotron,
with no evidence for a spectral break.

\subsection{PKS\,1209-51/52}

\begin{figure}
\includegraphics[width=9cm]{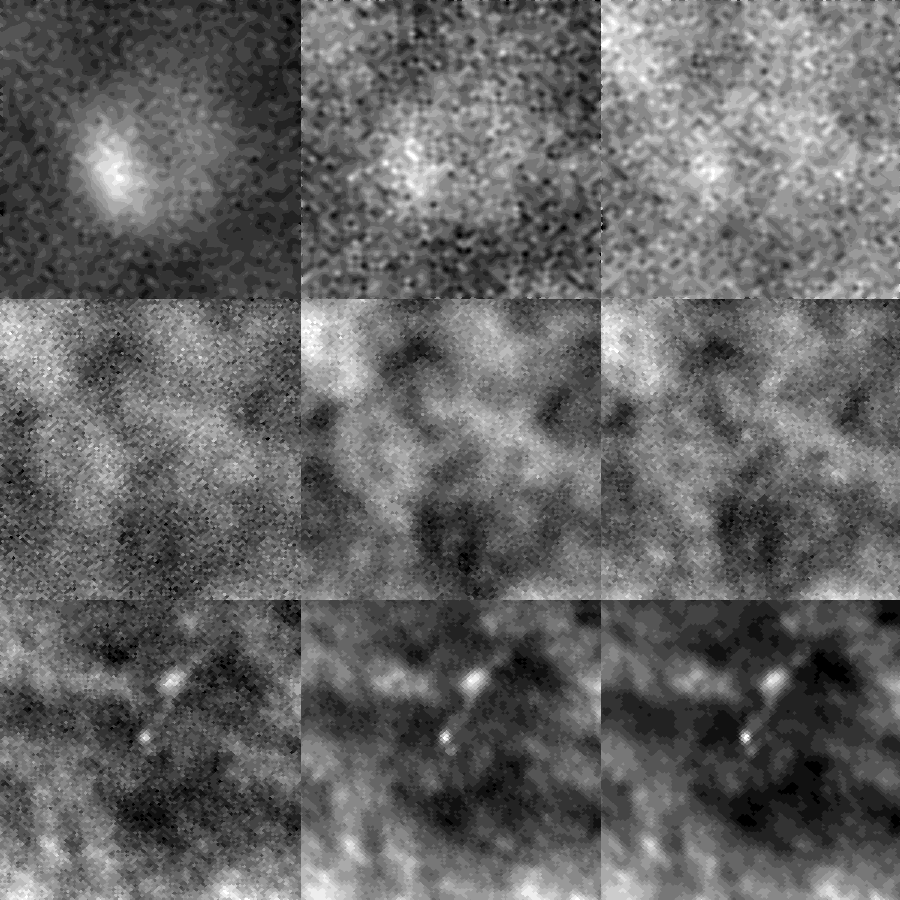}
\vskip 0.5em
\includegraphics[width=9cm]{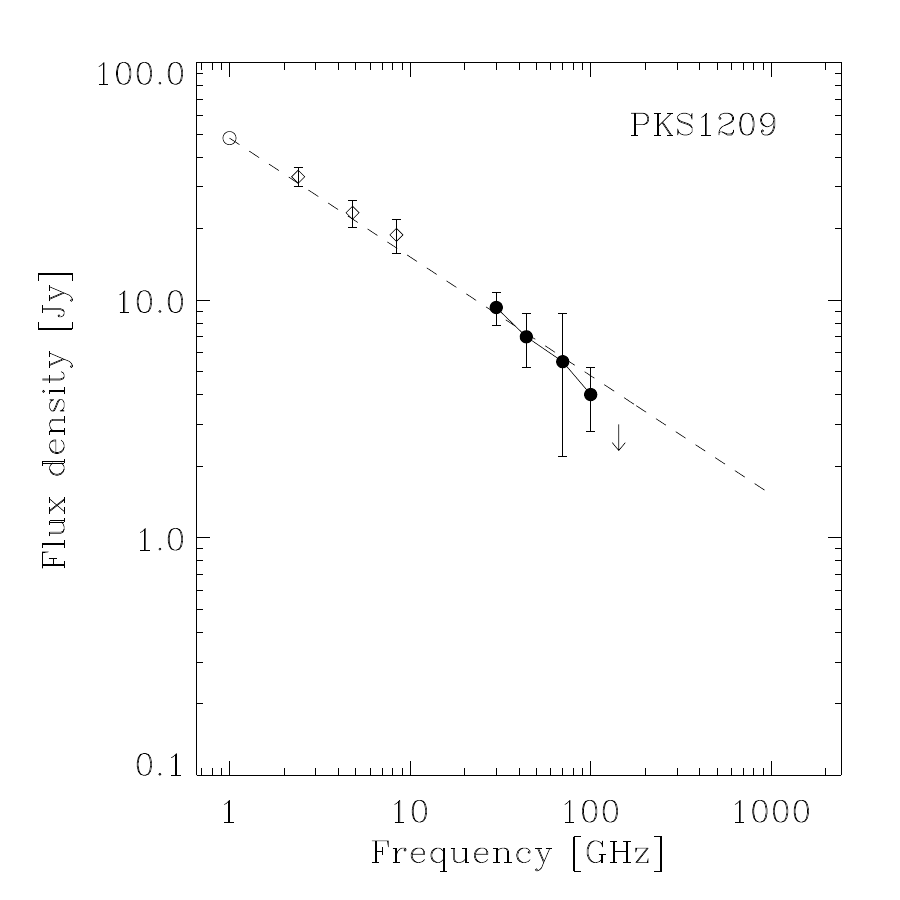}
\caption{
{\it Top\/}:  Images of the PKS\,1209-51/52  environment at the nine \Planck\ frequencies (Table 1), increasing from 30\,GHz at top left to 857\,GHz at bottom right.  Each image is {\bfc 180$'$ on a side, with galactic coordinate orientation.}
{\it Bottom\/}: Microwave SED of PKS\,1209-52. Filled circles are \Planck\ measurements from this paper, with 3$\sigma$ error bars. The open symbol at 1\,GHz is from the Green catalogue, and the dashed line emanating from it is a power law with spectral index from the Green catalogue. 
Open diamonds are radio flux densities from \citet{milne94}. 
The downward arrow shows a \planck\ high-frequency flux density measurement that was contaminated by unrelated foreground emission.
}
\label{pks1209sed}
\end{figure}

{\bfc The barrel-shaped \citep{kestevencaswell} } SNR PKS1209-51/52 is detected by \planck\ at low frequencies.
{\bfc The angular structure of the SNR overlaps in spatial scales with the CMB, so 
PKS1209-51/52 was masked in the CMB maps. 
Therefore, Figure~\ref{pks1209sed} shows the total intensity for this SNR, rather than the CMB-subtracted intensity. 
At
30--70 GHz, the SNR is clearly evident because it is significantly  brighter than the CMB and has the location and size
seen in lower-frequency radio images.
At 100--217 GHz, the SNR is lost in CMB fluctuations.
At 353--857 GHz, the region is dominated by interstellar dust emission. The object near the centre of the SNR in the high-frequency images is a reflection nebula, identified by \citet[object 381]{bran} on optical places as a $3'$ possible reflection
nebula; it was also noted as a far-infrared source without corresponding strong H~I emission \citep[object 9095]{rhk93}. There is no evidence for this object to be associated with the SNR nor the neutron star suspected to be the remnant of the progenitor \citep[X-ray source 1E 1207.4-5209]{vasisht97}, located $12'$ away. 
It is nonetheless remarkable that the
source is right at the centre of the SNR.

Figure~\ref{pks1209sed} shows the low-frequency emission seen by \planck\
continues the radio synchrotron spectrum closely up to 70 GHz, with no evidence for a spectral break.
}

\subsection{RCW\,86}

\begin{figure}
\includegraphics[width=9cm]{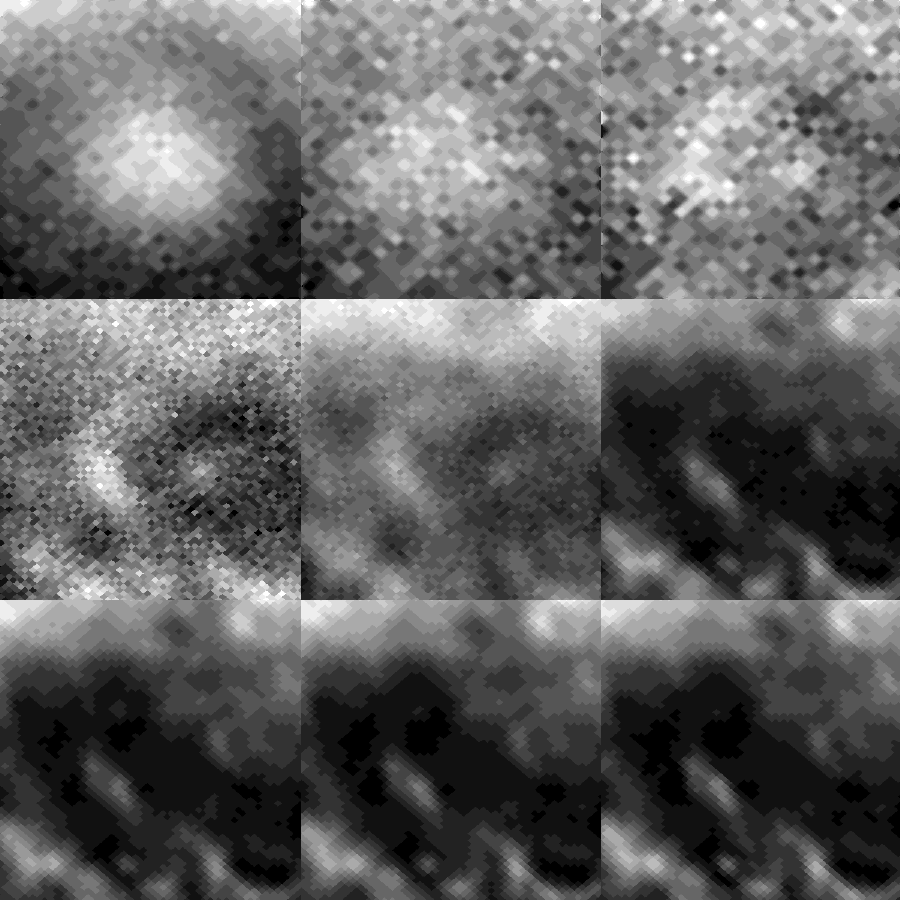}
\vskip 0.5em
\includegraphics[width=9cm]{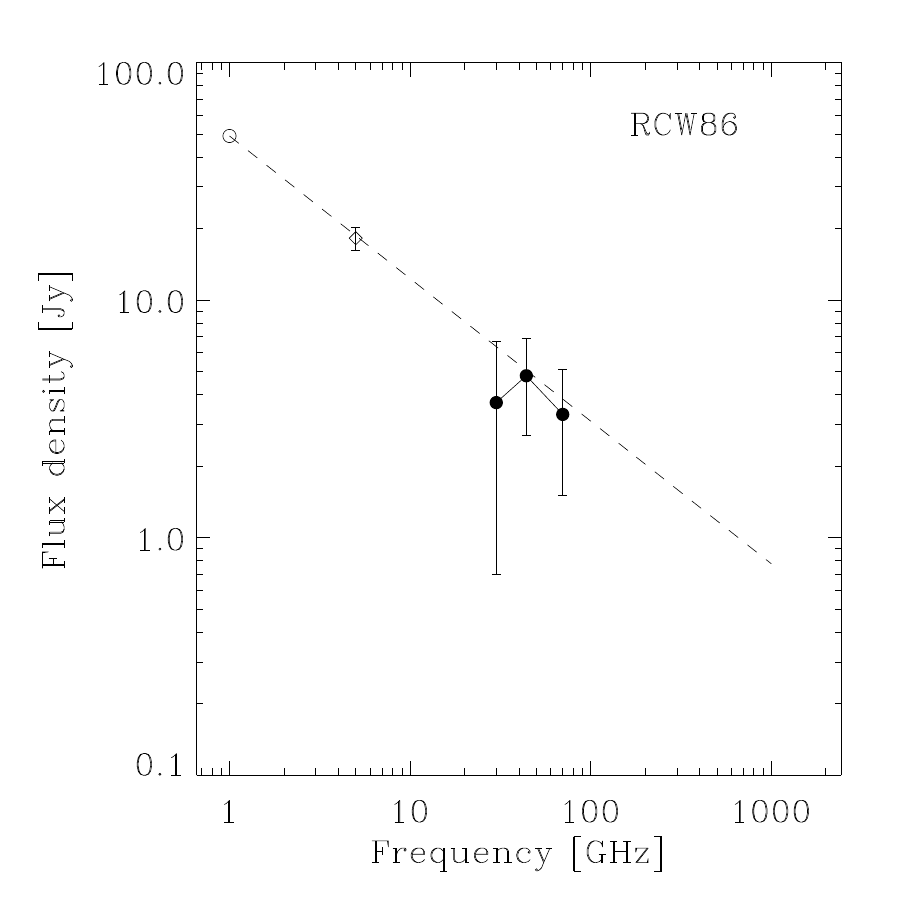}
\caption{
{\it Top\/}:  Images of the RCW\,86  environment at the nine \Planck\ frequencies (Table 1), increasing from 30\,GHz at top left to 857\,GHz at bottom right.  Each image is 100\arcm\ on a side, with galactic coordinate orientation.
{\it Bottom\/}: Microwave SED of RCW\,86. Filled circles are \Planck\ measurements from this paper, with 3$\sigma$ error bars. The open symbol at 1\,GHz is from the Green catalogue, and the dashed line emanating from it is a power law with spectral index from the Green catalogue. 
The open diamonds is the 5 GHz flux density from \citet{caswell75}.
}
\label{rcw86sed}
\end{figure}

The RCW\,86 supernova remnant, possibly that of a Type~I SN within a stellar wind bubble \citep{williams11}, is evident in the low-frequency 30--70\,GHz \planck\ images.
At higher frequencies the synchrotron emission from the SNR is confused with other emission. 
{\bfc The feature near the left-centre of the highest 6 frequency images of RCW~86 in Fig.~\ref{rcw86sed} is a dark molecular cloud, DB 315.7-2.4 \citep{dutrabica}. 
The \planck\ emission from this location is due to dust, with brightness steadily increasing with frequency over the \planck\ domain. The cloud is located near the edge of the SNR, and the CO velocity \citep[-37 km~s$^{-1}$;][]{otrupeck00} is approximately as expected for a cloud at the distance estimate for the SNR (2.3 kpc). Any relation between the cold molecular cloud and the SNR is only plausible; there is no direct evidence for interaction. In any event, the dust in this cloud makes
it impossible to measure the synchrotron brightness at 100 GHz, and the 70 GHz flux has higher uncertainty.}

{\bfc For the SNR synchrotron emission,} the \planck\ flux densities, shown in Fig.~\ref{rcw86sed}, are consistent within $1\sigma$ with an extrapolation of the radio synchrotron power law. The existence of X-ray synchrotron emission \citep{rho02} suggests that the energy distribution of relativistic electrons continues to high energy.

\subsection{MSH\,15-5{\it 6}}

\begin{figure}
\includegraphics[width=9cm]{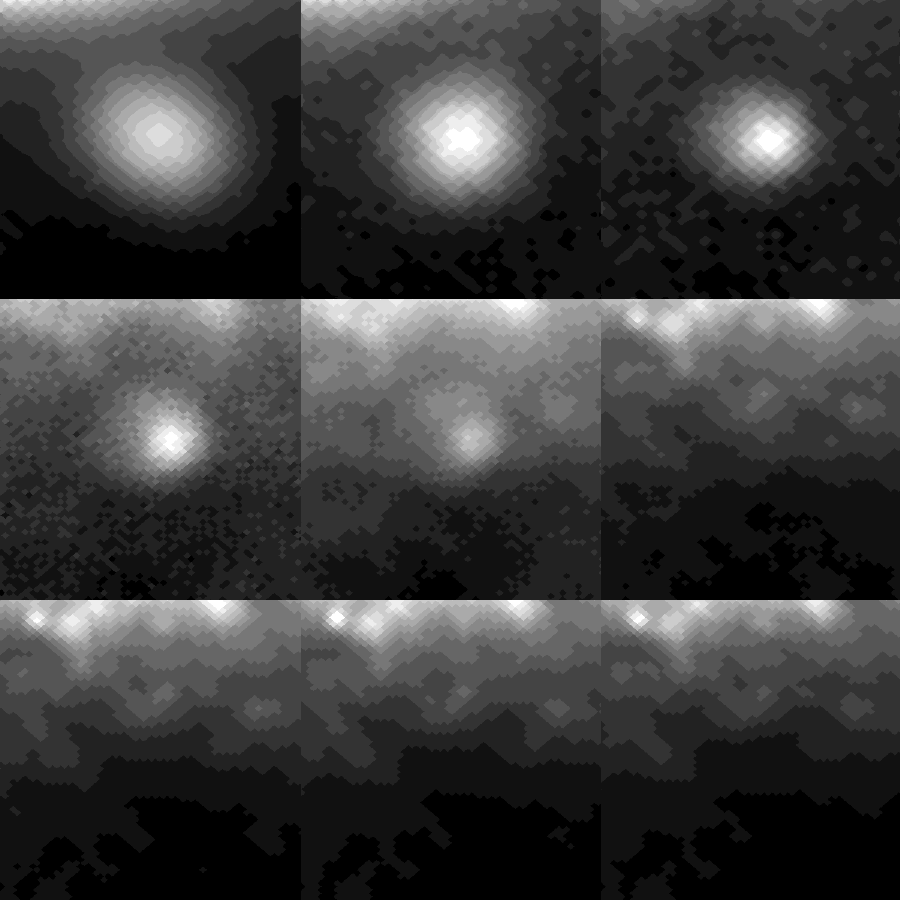}
\vskip 0.5em
\includegraphics[width=9cm]{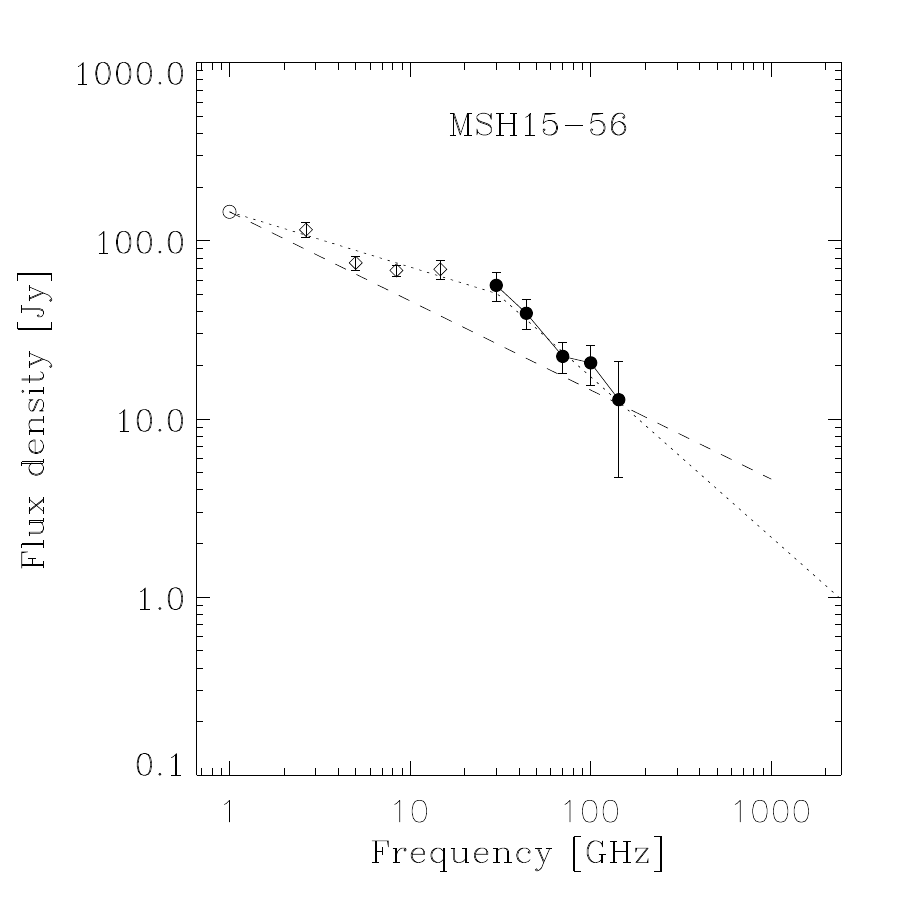}
\caption{
{\it Top\/}:  Images of the MSH\,15-5{\it 6}  environment at the nine \Planck\ frequencies (Table 1), increasing from 30\,GHz at top left to 857\,GHz at bottom right.  Each image is 100\arcm\ on a side, with galactic coordinate orientation.
{\it Bottom\/}: Microwave SED of MSH\,15-5{\it 6}. Filled circles are \Planck\ measurements from this paper, with 3$\sigma$ error bars. The open symbol at 1\,GHz is from the Green catalogue, and the dashed line emanating from it is a power law with spectral index from the Green catalogue. 
Open diamonds are radio flux densities from the  {\bfc \citet{dickel00}  and \citet{milne79} 
that constrain the slope through the {\it Planck} data.} 
}
\label{msh1556sed}
\end{figure}

The radio-bright SNR MSH\,15-5{\it 6} is well detected in the first five \planck\ frequencies, 30--143\,GHz. 
The SNR is a `composite,' with a steep-spectrum shell and a brighter, flat-spectrum plerionic core.
The \planck\ flux densities do not follow a single power law matching the published radio flux densities. Just connecting the 1\,GHz radio flux densities to the \planck\ flux densities, the spectral index is in the range 0.3--0.5.
The \planck\ flux densities themselves follow a steeper power law than can match the radio flux densities, and suggest a break in the spectral index. 
Figure~\ref{msh1556sed} shows a broken-power-law fit, where the spectral index steepens from 0.31 to 0.9 at 30\,GHz.
Low-frequency radio observations show that the plerionic core of the SNR has a flatter spectral
index than the shell, while higher-frequency flux densities of the core alone from 4.8 to 8.6\,GHz  
have a spectral index of 0.85 \citep{dickel00}.
{\bfc The relatively flat spectral index at radio frequencies and up to about 30 GHz may indicate injection of fresh electrons, as in
the Crab and 3C\,58, which are driven by pulsar wind nebulae. However, there has been, to date, no pulsar detected in MSH\,15-5{\it 6}, and the
spectral index is not as flat as in the known, young pulsar wind nebulae.
The apparent break in spectral index to a stepper slope above 30 GHz suggests possible energy loss of the highest-energy particles.}

\subsection{SN\,1006}

\begin{figure}
\includegraphics[width=9cm]{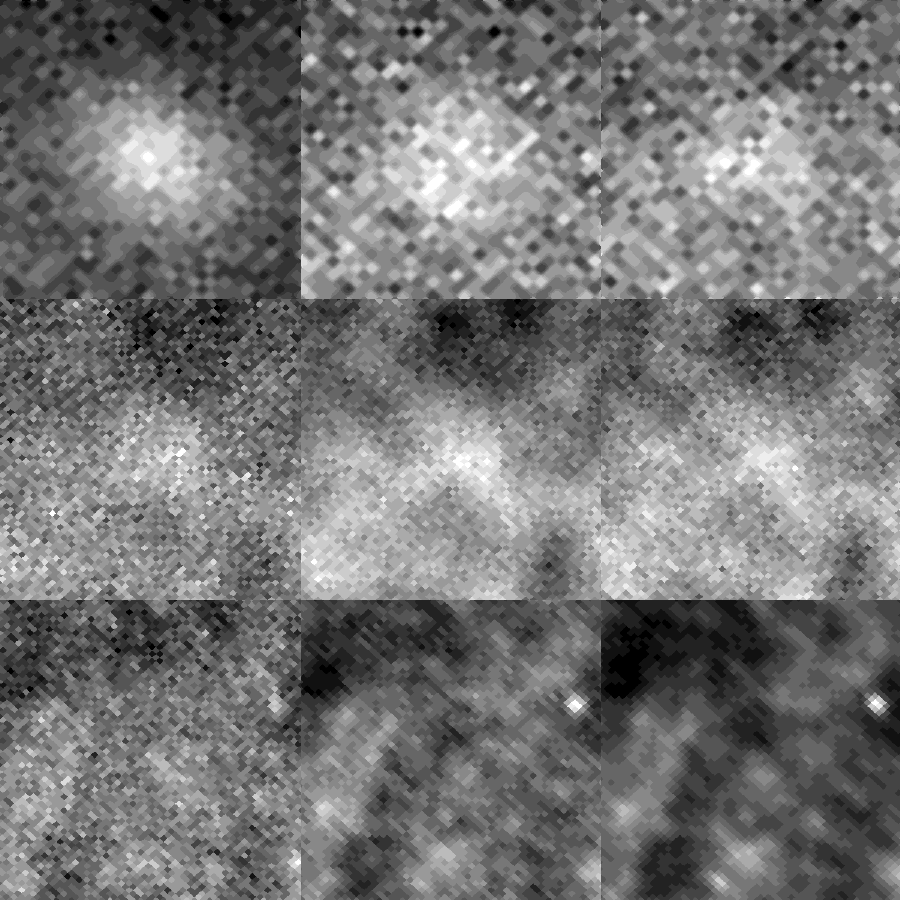}
\vskip 0.5em
\includegraphics[width=9cm]{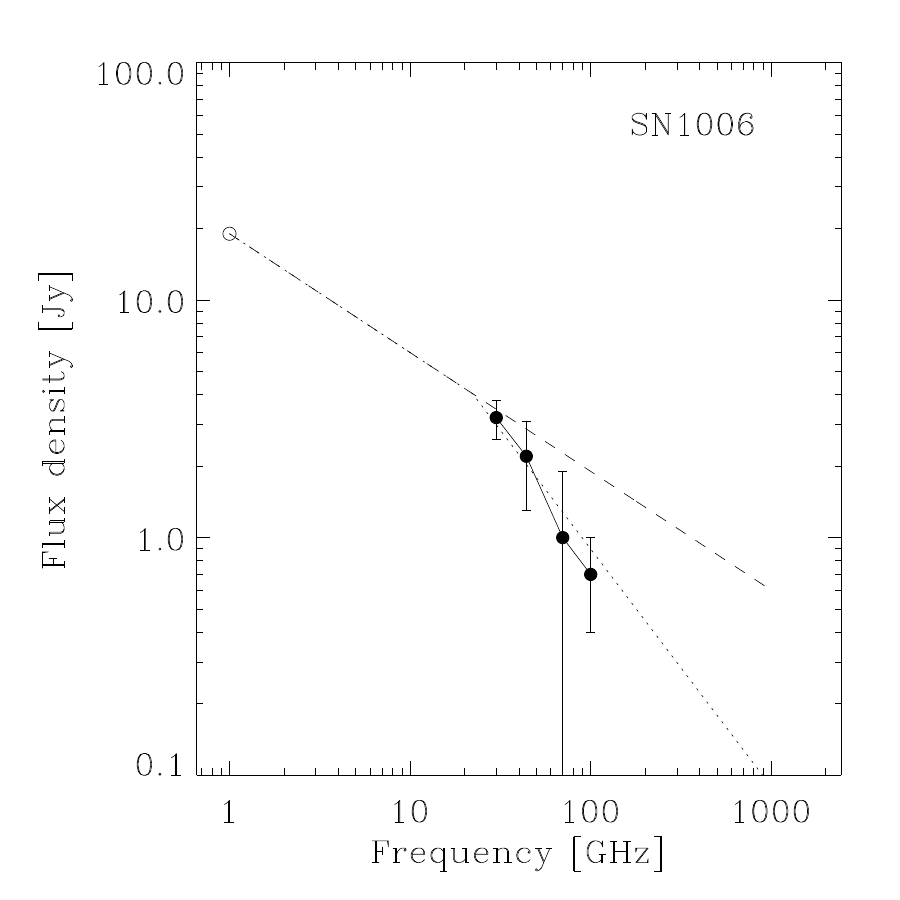}
\caption{
{\it Top\/}:  Images of the SN\,1006 environment at the nine \Planck\ frequencies (Table 1), increasing from 30\,GHz at top left to 857\,GHz at bottom right.  Each image is 100\arcm\ on a side, with galactic coordinate orientation.
{\it Bottom\/}: Microwave SED of SN\,1006. Filled circles are \Planck\ measurements from this paper, with 3$\sigma$ error bars. The open symbol at 1\,GHz is from the Green catalogue, and the dashed line emanating from it is a power law with spectral index from the Green catalogue. 
The dotted line is a broken power-law fit discussed in the text.
}
\label{sn1006sed}
\end{figure}

SN\,1006 is well-detected at 30--44\,GHz. At higher frequencies it becomes faint and possibly confused with unrelated emission. However, the field is not
as crowded as it is for most other SNRs, and we suspect that the observed decrement in flux density below the extrapolation of the radio power law
at 70 and 100\,GHz may be due to a real break in the spectral index. If so, then the frequency of that break is in the range $20 < \nu_{\rm break}<30$\,GHz.
For illustration, Fig.~\ref{sn1006sed} shows a broken power-law fit with $\nu_{\rm break}=22$\,GHz, above which the spectral index steepens from $\alpha=0.5$ to 
$\alpha=1$ as predicted for synchrotron losses. The \planck\ data appear to match this model well.

\clearpage

\section{Conclusions}

The flux densities of \ndetected\ known Galactic supernova remnants were measured from the  \planck\ microwave all-sky survey with the following conclusions.
We find new evidence for spectral index breaks in G21.5-0.9, HB\,21, MSH\,15-5{\it 6}, SN\,1006, and we confirm the previously detected spectral
break in 3C\,58, including a new detection with {\it Herschel\/}. 

{\bfc
Table~\ref{indextab} summarizes the new spectral indices required to fit the radio through microwave SED
of SNRs. These values correspond to the dashed lines in the SEDs
for each SNR in this paper. For each SNR in this paper for which the {\it Planck} data indicated in a spectrum noticeably different from the radio power-law extrapolation, the frequency of the spectral break ($\nu_{\rm break}$) and the spectral
index at lower and higher frequencies ($\alpha_1$ and $\alpha_2$, respectively) are listed. The actual SEDs
should be consulted before using the spectral
index values by themselves, because they are only applicable over the region shown, and they are only 
mathematical approximations to what is more likely a continuous distribution of energies with evolving losses.
}

The breaks in spectral index are consistent with synchrotron losses of electrons injected by a central source.
We extend the radio synchrotron spectrum for young SNRs Cas~A and Tycho with no evidence for extra emission mechanisms.
The distinction in properties between those supernova remnants that do or do not show a break in their power-law spectral index is not readily evident. 
The supernova remnants with spectral breaks include examples that range from bright to faint and young to mature, and they also include examples both with and without stellar remnants.  
A combination of cosmic-ray acceleration by the shocks and the pulsars, deceleration in denser environments, and ageing may lead to the variation in synchrotron shapes.

\begin{table}
\caption{\bfc  Sychrotron spectral indices}
\label{indextab}
\tiny
\setbox\tablebox=\vbox{
\newdimen\digitwidth
\setbox0=\hbox{\rm 0}
\digitwidth=\wd0
\catcode`*=\active
\def*{\kern\digitwidth}
\newdimen\signwidth
\setbox0=\hbox{+}
\signwidth=\wd0
\catcode`!=\active
\def!{\kern\signwidth}
\halign{\hbox to 0.7in{#\leaderfil}\tabskip=2em&
    \hfil#\hfil\tabskip=0em& 
    \hfil#\hfil\tabskip=1em& 
    \hfil#\hfil\tabskip=1em& 
    \hfil#\hfil\tabskip=0pt\cr
\noalign{\doubleline}
\omit &  \hfil Spectral & index\hfil & $\nu_{\rm break}$ \cr \cline{2-3} 
 SNR & $\alpha_1$ & $\alpha_2$& [GHz]\cr
\noalign{\vskip 3pt\hrule\vskip 5pt}
G21.5-0.9   &   0.05 &  0.55 & 45\cr
W 44          & 0.37 & 1.37 & 45 \cr
CTB 80      & 0.80 & ... & none\cr
HB 21         & 0.38 & 0.88 & 3\cr
3C 58          &   0.07 &  0.57 & 25\cr
IC 443        & 0.36 & 1.56 & 40 \cr
Puppis A    & 0.46 & 1.46 & 40 \cr
MSH 15-5{\it 6} &  0.31 & 0.9 & 30\cr
SN 1006     & 0.50 & 1.0 & 22 \cr
\noalign{\vskip 5pt\hrule\vskip 3pt}}}
\endPlancktable
\end{table}

\begin{acknowledgements}

The Planck Collaboration acknowledges the support of: ESA; CNES, and
CNRS/INSU-IN2P3-INP (France); ASI, CNR, and INAF (Italy); NASA and DoE
(USA); STFC and UKSA (UK); CSIC, MINECO, JA and RES (Spain); Tekes, AoF,
and CSC (Finland); DLR and MPG (Germany); CSA (Canada); DTU Space
(Denmark); SER/SSO (Switzerland); RCN (Norway); SFI (Ireland);
FCT/MCTES (Portugal); ERC and PRACE (EU). A description of the Planck
Collaboration and a list of its members, indicating which technical
or scientific activities they have been involved in, can be found at
\url{http://www.cosmos.esa.int/web/planck/planck-collaboration}.

\end{acknowledgements}

\bibliographystyle{aa}

\bibliography{Planck_bib,Psnr_bib}

\appendix
\section{Aperture photometry details}
More details on the aperture photometry are provided in Table~\ref{apphottab}.
For each flux density measurement, the following quantities are listed.

\noindent (1) The supernova remnant (SNR) name, in the same order as in Tables~\ref{sourcetab} and
\ref{fluxtab}.

\noindent (2) The observing frequency in GHz.

\noindent (3)  `Flux' is the aperture-corrected, background-subtracted flux of the source with its uncertainty. The uncertainty is a quadrature combination of the statistical uncertainty and the flux calibration uncertainty. The flux calibration uncertainty is the flux multiplied by the mission calibration uncertainty from Table~\ref{planckprop}. The statistical uncertainty is the rms of the pixel values in the background annulus, $I_{\rm rms}$, multiplied by the number of pixels in
the on-source aperture and divided by the square-root of the number of pixels in the off-source annulus,
then scaled by the same aperture correction as was applied to the flux.

\noindent (4) $\theta_{\rm ap}$ is the aperture radius, within which the source flux was summed.

\noindent (5) $f_{\rm A}$ is the aperture correction factor, from equation~\ref{eq:fap}. 

\noindent (6) $F_{\rm bg}$ is the flux of the background that was subtracted from the on-source sum of the source flux. This was calculated by taking the median brightness within the background annulus, then multiplying by the area of the on-source aperture. To aid direct comparison we also applied the same aperture correction as was applied to the source flux. Because of the line lines of sight through the Galactic disk at the locations of most of the SNRs, the backgrounds are high, and $F_{\rm bg}$ is often
higher than the flux of the SNR. In itself, this is not a fundamental limitation, because a smooth background can be much brighter than the target without affecting the photometry.

\noindent (7) $I_{\rm peak}$ is the peak intensity in the on-source aperture, in the units of the map for which the photometry was being performed. For frequencies 30--353 GHz, the units are mK in thermodynamic CBM brightness temperature units. For 545--857 GHz, the units are MJy~sr$^{-1}$.

\noindent (8) $I_{\rm bg}$ is the median intensity in the background annulus, in the same units 
units as $I_{\rm peak}$.

\noindent (9) $I_{\rm rms}$ is the standard deviation (rms) of the pixel values in the background annulus, in the same units as $I_{\rm peak}$. 
The peak, per-pixel signal-to-noise for each target can be estimated as
$I_{\rm peak}/I_{\rm rms}$ because the target causes the brightest pixel in the on-source aperture. This
gives a quantitative measure of how readily visible the target is, at the angular scale of the {\it Planck} beam. 

\begin{table*}
\caption{Aperture photometry details for \Planck\ flux density measurements of supernova remnants.}
\label{apphottab}
\centering
\begin{tabular}{ l r c c c c c c c}
\hline\hline

(1)    & (2)  & (3) & (4) & (5) & (6) & (7)& (8) & (9)\\
Source & Freq. & Flux & $\theta_{\rm ap}$ & $f_{\rm A}$ & $F_{\rm bg}$ & $I_{\rm peak}$ & $I_{\rm bg}$ & $I_{\rm rms}$ \\
       & [GHz] & [Jy] & [$'$]              &            & [Jy] & $^{\rm a}$ & $^{\rm a}$& $^{\rm a}$ \\
\hline
      G21.5-0.9 &  70 & $   4.3 \pm    0.6 $ &   10.2 &    1.4 &   7.1 &   3.1 &    1.4 &    0.4 \\
      G21.5-0.9 & 100 & $   2.5 \pm    0.5 $ &    8.1 &    1.4 &  17.6 &   4.5 &    3.1 &    0.5 \\
      G21.5-0.9 & 143 & $   2.9 \pm    0.4 $ &    6.0 &    1.4 &  15.4 &   4.5 &    2.9 &    0.3 \\
            W44 &  30 & $ 121.0 \pm    8.2 $ &   36.1 &    1.4 & 149.2 &  28.6 &   11.6 &    5.8 \\
            W44 &  44 & $  67.2 \pm    5.2 $ &   33.2 &    1.4 & 103.1 &  11.0 &    4.5 &    2.2 \\
            W44 &  70 & $  29.5 \pm    3.3 $ &   28.0 &    1.0 &  64.1 &   5.3 &    2.3 &    1.6 \\
          CTB80 &  30 & $  14.5 \pm    1.7 $ &   64.9 &    1.0 &  39.3 &   3.3 &    1.3 &    2.2 \\
          CTB80 &  44 & $   8.4 \pm    1.2 $ &   63.3 &    1.0 &  25.8 &   1.3 &    0.4 &    1.0 \\
        CygLoop &  30 & $  22.2 \pm    1.6 $ &  174.3 &    1.0 &  61.1 &   1.3 &    0.3 &    0.4 \\
           HB21 &  30 & $  15.4 \pm    1.7 $ &   93.3 &    1.0 & 110.6 &   3.6 &    1.8 &    0.8 \\
           HB21 &  44 & $  10.0 \pm    1.2 $ &   92.2 &    1.0 &  74.8 &   1.3 &    0.6 &    0.3 \\
           CasA &  30 & $ 227.2 \pm   11.4 $ &   25.0 &    1.4 &  15.8 &  76.8 &    2.6 &    1.5 \\
           CasA &  44 & $ 139.7 \pm    7.0 $ &   20.6 &    1.4 &  12.1 &  33.6 &    1.4 &    0.7 \\
           CasA &  70 & $ 108.9 \pm    5.5 $ &   10.4 &    1.4 &   4.4 &  40.8 &    0.8 &    0.7 \\
           CasA & 100 & $  93.9 \pm    4.7 $ &    8.4 &    1.4 &   5.7 &  34.5 &    0.9 &    0.3 \\
           CasA & 143 & $  72.7 \pm    3.7 $ &    6.5 &    1.4 &   7.2 &  28.7 &    1.2 &    0.3 \\
           CasA & 217 & $  62.2 \pm    3.2 $ &    5.1 &    1.4 &  21.4 &  34.1 &    4.3 &    0.6 \\
           CasA & 353 & $  59.9 \pm    3.9 $ &    5.0 &    1.4 &  82.5 &  72.8 &   28.8 &    3.3 \\
          Tycho &  30 & $   8.4 \pm    0.4 $ &   25.5 &    1.4 &  10.0 &   4.3 &    1.6 &    0.2 \\
          Tycho &  44 & $   5.4 \pm    0.3 $ &   21.1 &    1.4 &   5.3 &   2.0 &    0.6 &    0.1 \\
          Tycho &  70 & $   4.5 \pm    0.3 $ &   11.4 &    1.4 &   2.9 &   1.9 &    0.4 &    0.1 \\
          Tycho & 100 & $   3.9 \pm    0.2 $ &    9.6 &    1.4 &   6.0 &   1.9 &    0.7 &    0.1 \\
          Tycho & 143 & $   3.5 \pm    0.2 $ &    8.0 &    1.4 &   9.1 &   2.0 &    1.0 &    0.1 \\
           3C58 &  30 & $  23.7 \pm    1.2 $ &   25.7 &    1.4 &   5.0 &   8.2 &    0.8 &    0.2 \\
           3C58 &  44 & $  17.5 \pm    0.9 $ &   21.3 &    1.4 &   3.2 &   4.3 &    0.3 &    0.1 \\
           3C58 &  70 & $  15.3 \pm    0.8 $ &   11.9 &    1.4 &   2.3 &   5.5 &    0.3 &    0.1 \\
           3C58 & 100 & $  13.8 \pm    0.7 $ &   10.1 &    1.4 &   3.4 &   4.7 &    0.4 &    0.1 \\
           3C58 & 143 & $  11.6 \pm    0.6 $ &    8.6 &    1.4 &   5.9 &   4.2 &    0.6 &    0.1 \\
           3C58 & 217 & $   6.5 \pm    0.4 $ &    7.6 &    1.0 &  11.9 &   5.1 &    1.6 &    0.1 \\
           3C58 & 353 & $   3.6 \pm    0.5 $ &    7.5 &    1.0 &  45.1 &  14.4 &   10.4 &    0.6 \\
           Crab &  30 & $ 425.4 \pm   21.3 $ &   25.3 &    1.4 &   5.9 & 138.8 &    0.9 &    2.5 \\
           Crab &  44 & $ 313.8 \pm   15.7 $ &   20.9 &    1.4 &  10.0 &  75.4 &    1.1 &    1.3 \\
           Crab &  70 & $ 286.7 \pm   14.4 $ &   11.1 &    1.4 &   3.1 & 107.0 &    0.5 &    1.3 \\
           Crab & 100 & $ 278.5 \pm   13.9 $ &    9.2 &    1.4 &   0.8 &  95.8 &    0.1 &    0.4 \\
           Crab & 143 & $ 255.3 \pm   12.8 $ &    7.4 &    1.4 &   2.5 &  89.0 &    0.3 &    0.2 \\
           Crab & 217 & $ 221.5 \pm   11.1 $ &    6.3 &    1.4 &   5.8 & 108.5 &    0.9 &    0.1 \\
           Crab & 353 & $ 197.8 \pm    9.9 $ &    6.2 &    1.4 &  27.6 & 163.1 &    6.6 &    0.3 \\
           Crab & 545 & $ 153.7 \pm    7.7 $ &    6.2 &    1.4 &  89.1 &  46.0 &    6.3 &    0.3 \\
           Crab & 857 & $ 136.9 \pm    7.2 $ &    6.1 &    1.4 & 249.6 &  57.1 &   17.8 &    0.8 \\
          IC443 &  30 & $  56.0 \pm    2.8 $ &   38.9 &    1.4 &  13.9 &   9.8 &    0.9 &    0.8 \\
          IC443 &  44 & $  38.1 \pm    2.0 $ &   36.2 &    1.4 &  11.0 &   3.9 &    0.4 &    0.3 \\
          IC443 &  70 & $  19.5 \pm    1.1 $ &   31.5 &    1.0 &  10.9 &   2.1 &    0.3 &    0.2 \\
          IC443 & 100 & $  16.8 \pm    0.9 $ &   30.9 &    1.0 &  26.1 &   1.4 &    0.4 &    0.2 \\
          IC443 & 143 & $  11.4 \pm    0.9 $ &   30.5 &    1.0 &  62.3 &   1.3 &    0.7 &    0.2 \\
          IC443 & 217 & $  56.2 \pm    4.0 $ &   30.2 &    1.0 & 234.8 &   5.5 &    2.0 &    0.8 \\
          IC443 & 353 & $ 230.3 \pm   16.2 $ &   30.2 &    1.0 & 995.4 &  35.7 &   13.8 &    5.1 \\
          IC443 & 545 & $ 728.8 \pm   51.4 $ &   30.2 &    1.0 &3138.8 &  33.3 &   13.0 &    5.1 \\
          IC443 & 857 & $2003.5 \pm  144.2 $ &   30.2 &    1.0 &8697.4 &  92.1 &   36.0 &   15.6 \\
        PuppisA &  30 & $  31.5 \pm    1.7 $ &   51.4 &    1.0 &  28.6 &   5.9 &    1.5 &    0.7 \\
        PuppisA &  44 & $  20.6 \pm    1.2 $ &   49.3 &    1.0 &  22.3 &   2.2 &    0.6 &    0.3 \\
        PuppisA &  70 & $  13.3 \pm    1.0 $ &   46.0 &    1.0 &  30.3 &   1.2 &    0.4 &    0.2 \\
        PuppisA & 100 & $  11.7 \pm    1.1 $ &   45.6 &    1.0 &  71.6 &   1.3 &    0.5 &    0.2 \\
           Vela &  30 & $ 470.6 \pm   23.7 $ &  192.8 &    1.0 & 407.9 &   8.3 &    1.5 &    4.7 \\
           Vela &  44 & $ 313.6 \pm   15.9 $ &  192.3 &    1.0 & 323.7 &   3.0 &    0.6 &    2.2 \\
           Vela &  70 & $ 235.4 \pm   12.4 $ &  191.5 &    1.0 & 455.0 &   1.8 &    0.4 &    1.5 \\
        PKS1209 &  30 & $   9.7 \pm    0.5 $ &   71.9 &    1.0 &  12.9 &   1.7 &    0.3 &    0.1 \\
        PKS1209 &  44 & $   8.2 \pm    0.6 $ &   70.5 &    1.0 &  12.5 &   0.8 &    0.2 &    0.1 \\
        PKS1209 &  70 & $   7.4 \pm    0.9 $ &   68.2 &    1.0 &  24.3 &   0.6 &    0.1 &    0.1 \\
        PKS1209 & 100 & $   9.7 \pm    0.8 $ &   67.9 &    1.0 &  45.4 &   0.5 &    0.2 &    0.1 \\
          RCW86 &  30 & $   3.7 \pm    0.9 $ &   40.1 &    1.4 &  29.6 &   2.8 &    1.9 &    0.9 \\
          RCW86 &  44 & $   4.0 \pm    0.6 $ &   37.4 &    1.4 &  15.2 &   1.1 &    0.5 &    0.3 \\
          RCW86 &  70 & $   2.0 \pm    0.5 $ &   33.0 &    1.0 &   7.6 &   0.5 &    0.2 &    0.1 \\
       MSH15-56 &  30 & $  55.5 \pm    3.5 $ &   37.7 &    1.4 &  34.8 &  14.6 &    2.5 &    6.5 \\
       MSH15-56 &  44 & $  38.7 \pm    2.5 $ &   35.0 &    1.4 &  19.8 &   6.1 &    0.8 &    1.9 \\
       MSH15-56 &  70 & $  22.7 \pm    1.6 $ &   30.1 &    1.0 &  11.7 &   4.1 &    0.4 &    0.5 \\
       MSH15-56 & 100 & $  18.4 \pm    1.7 $ &   29.5 &    1.0 &  26.6 &   2.7 &    0.5 &    0.6 \\
       MSH15-56 & 143 & $  12.8 \pm    2.7 $ &   29.0 &    1.0 &  76.1 &   2.1 &    0.9 &    0.7 \\
         SN1006 &  30 & $   3.6 \pm    0.2 $ &   33.4 &    1.4 &   3.0 &   1.2 &    0.3 &    0.1 \\
         SN1006 &  44 & $   3.3 \pm    0.3 $ &   30.3 &    1.4 &   1.2 &   0.6 &    0.1 &    0.1 \\
         SN1006 &  70 & $   2.6 \pm    0.3 $ &   24.5 &    1.0 &   1.0 &   0.5 &    0.0 &    0.1 \\
         SN1006 & 100 & $   2.7 \pm    0.2 $ &   23.7 &    1.0 &   1.4 &   0.4 &    0.0 &    0.1 \\

                 \hline
\end{tabular} 
\tablefoot{$^{\rm a}$Units for the intensities are mK for frequencies 30--353 GHz, and MJy~sr$^{-1}$ for 
frequencies 545--857 GHz.}
\end{table*}

\raggedright
\end{document}